\documentclass[aip,jcp,reprint,noshowkeys,superscriptaddress]{revtex4-1}
\usepackage{graphicx,dcolumn,xcolor,microtype,multirow,amscd,amsmath,amssymb,amsfonts,physics,wrapfig,txfonts,tikz,siunitx}
\usepackage[version=4]{mhchem}

\newcommand{\ie}{\textit{i.e.}}

\usepackage[normalem]{ulem}
\newcommand{\titou}[1]{\textcolor{black}{#1}}

\newcommand{\QP}{\textsc{quantum package}}

\newcommand{\hT}{\Hat{T}}
\newcommand{\hH}{\Hat{H}}

\newcommand{\bH}{\Bar{H}}

\newcommand{\EFCI}{E_\text{FCI}}

\newcommand{\EVCC}{E_\text{VCC}}
\newcommand{\ETCC}{E_\text{TCC}}

\newcommand{\PsiCC}{\Psi_\text{CC}}
\newcommand{\PsiO}{\Psi_0}

\newcommand{\cre}[1]{a_{#1}^\dagger} %Creation operator
\newcommand{\ani}[1]{a_{#1}} %Annihilation operator
\newcommand{\ta}[2]{t_{#1}^{#2}} %Cluster amplitudes
\newcommand{\f}[2]{f_{#1}^{#2}} %Fock operator matrix elements

\usepackage[
colorlinks=true,
linkcolor=blue,
        citecolor=blue,
    breaklinks=true
	]{hyperref}
\begin{document}

\newcommand{\LCPQ}{Laboratoire de Chimie et Physique Quantiques (UMR 5626), Universit\'e de Toulouse, CNRS, UPS, France}

\title{Variational coupled cluster for ground and excited states}

\author{Antoine Marie}
\affiliation{\LCPQ}
\author{F\'abris Kossoski}
\affiliation{\LCPQ}
\author{Pierre-Fran\c{c}ois Loos}
\email{loos@irsamc.ups-tlse.fr}
\affiliation{\LCPQ}

% Abstract
\begin{abstract}
In single-reference coupled-cluster (CC) methods, one has to solve a set of non-linear polynomial equations in order to determine the so-called amplitudes which are then used to compute the energy and other properties.
Although it is of common practice to converge to the (lowest-energy) ground-state solution, it is also possible, thanks to tailored algorithms, to access higher-energy roots of these equations which may or may not correspond to genuine excited states.
Here, we explore the structure of the energy landscape of variational CC (VCC) and we compare it with its (projected) traditional version (TCC) in the case where the excitation operator is restricted to paired double excitations (pCCD).
By investigating two model systems (the symmetric stretching of the linear \ce{H4} molecule and the continuous deformation of the square \ce{H4} molecule into a rectangular arrangement) in the presence of weak and strong correlations, the performance of VpCCD and TpCCD are gauged against their configuration interaction (CI) equivalent, known as doubly-occupied CI (DOCI), for reference Slater determinants made of ground- or excited-state Hartree-Fock orbitals or state-specific orbitals optimized directly at the VpCCD level.
The influence of spatial symmetry breaking is also investigated.
\end{abstract}

% Title
\maketitle

%%%%%%%%%%%%%%%%%%%%%%%%%%%%%%%%%%%%%%%%%%%%%%%%%%
\section{Coupled Cluster and Strong Correlation}
\label{sec:SRCC}
%%%%%%%%%%%%%%%%%%%%%%%%%%%%%%%%%%%%%%%%%%%%%%%%%%

Single-reference (SR) coupled-cluster (CC) methods offers a reliable description of weakly correlated systems through a well-defined hierarchy of systematically improvable models. \cite{Cizek_1966,Paldus_1972,Crawford_2000,Bartlett_2007,Shavitt_2009}
On top of this hierarchy stands full CC (FCC), which is equivalent to full configuration interaction (FCI), and consequently provides, at a very expensive computational cost, the exact wave function and energy of the system in a given basis set. 
Fortunately, more affordable methods have been designed and the popular CCSD(T) method, which includes singles, doubles and non-iterative triples, is nowadays considered as the gold standard of quantum chemistry for ground-state energies and properties. \cite{Purvis_1982,Raghavachari_1989} 
Despite its success for weakly correlated systems, it is now widely known that CCSD(T) flagrantly breaks down in the presence of strong correlation as one cannot efficiently describe such systems with a single (reference) Slater determinant. 
This has motivated quantum chemists to design multi-reference CC (MRCC) methods. \cite{Jeziorski_1981,Mahapatra_1998,Mahapatra_1999,Lyakh_2012,Kohn_2013} 
However, it is fair to say that these methods are computationally demanding and still far from being black-box.

Because SRCC works so well for weak correlation, it would be convenient to be able to treat strong correlation within the very same framework. 
This is further motivated by the fact that one can compensate the poor quality of the reference wave function by simply increasing the maximum excitation degree of the CC expansion. 
However, this is inevitably associated with a rapid growth of the computational cost, and hence one cannot always afford this brute-force strategy.
The development of SR-based methods for strong correlation is ongoing and some of them (usually based on the ``addition-by-subtraction'' principle) have shown promising results. 
A non-exhaustive list includes pair coupled-cluster doubles, \cite{Limacher_2013,Limacher_2014,Henderson_2014a,Henderson_2014b,Stein_2014,Shepherd_2016,Boguslawski_2017a,Boguslawski_2017b,Johnson_2017,Boguslawski_2019} singlet-paired CCD, \cite{Bulik_2015,Gomez_2016} the distinguishable cluster methods, \cite{Kats_2013,Kats_2014,Kats_2015,Kats_2016,Kats_2018,Kats_2019,Kats_2019a,Rishi_2016,Rishi_2019,Rishi_2019a} CCD-based variants involving a well-defined subset of diagrams, \cite{Scuseria_2008,Peng_2013,Scuseria_2013,Shepherd_2014,Shepherd_2014a} the $n$CC hierarchy, \cite{Bartlett_2006,Musial_2007} and parametrized CCSD. \cite{Huntington_2010}
Each of these methods sheds new light on the failures of SRCC to treat static correlation. 
For the sake of brevity, we omit the single-reference prefix hereafter.

The CC wave function $\ket{\PsiCC}$ is obtained by applying a wave operator onto a single Slater determinant reference $\ket{\PsiO}$ as
\begin{equation}
  \ket{\PsiCC} = e^{\hT} \ket{\PsiO}.
\end{equation}
In CC theory, the wave operator is defined as the exponential of the cluster operator
\begin{equation}
  \hT = \sum_{k=1}^N \hT_k,
\end{equation}
which is the sum of the $k$th-degree excitation operator up to $k=N$ (where $N$ is the number of electrons). In second quantized form, we have
\begin{equation}
  \label{eq:excitationOp}
  \hT_k = \frac{1}{(k!)^2} \sum_{ij\dots}\sum_{ab\dots} \ta{ij\dots}{ab\dots} \cre{a}\cre{b} \dots \ani{j}\ani{i},
\end{equation}
$\ani{i}$ and $\cre{a}$ being the second quantization annihilation and creation operators, respectively, which annihilates (creates) an electron in the spin-orbital $i$ ($a$). The cluster amplitudes $\ta{ij\dots}{ab\dots}$ are the quantities of interest in order to compute the CC energy (see below).

Throughout the paper, $p$, $q$, $r$, and $s$ denote general spin-orbitals, $i$, $j$, $k$, and $l$ refer to occupied spin-orbitals (hole states) and $a$, $b$, $c$, and $d$ to unoccupied spin-orbitals (particle states).

In quantum mechanics, one convenient way to determine the parameters of a wave function ans\"atz is to minimize the energy with respect to its parameters. 
The Rayleigh-Ritz variational principle ensures that the energy thus obtained is an upper bound to the exact ground-state energy.
Following this strategy, the variational CC (VCC) energy \cite{Bartlett_1988,Kutzelnigg_1991,Szalay_1995,Kutzelnigg_1998,Kutzelnigg_2010,Cooper_2010,Knowles_2010,Robinson_2011,Harsha_2018}
\begin{equation}
	\label{eq:EVCC}
	\EVCC 
	=  \frac{\mel{\PsiO}{e^{\hT^{\dag}} \hH e^{\hT}}{\PsiO}}{\mel{\PsiO}{e^{\hT^{\dag}} e^{\hT}}{\PsiO}},
\end{equation}
is thus minimized with respect to the cluster amplitudes which ensures
\begin{equation}
  \label{eq:varPcp}
  \min_{\ta{ij\dots}{ab\dots}}\EVCC \ge \EFCI.
\end{equation}
Unfortunately, independently of the excitation rank of $\hT$, this procedure is not tractable in practice.
Indeed, because the series expansion of the exponential in Eq.~\eqref{eq:EVCC} does not truncate before the $N$th-order term, VCC has an inherent exponential scaling with respect to system size.

Usually, one sacrifices the attractive upper bound property of the variational principle in exchange for computational tractability. 
To do so, the similarity-transformed Schr\"odinger equation
\begin{equation}
  \label{eq:schroEq}
  e^{-\hT} \hH e^{\hT} \ket{\PsiO} = \bH \ket{\PsiO} = E \ket{\PsiO}
\end{equation}
is projected onto the reference determinant $\ket{\PsiO}$, which gives
\begin{equation}
  \label{eq:TCCnrj}
  \ETCC = \mel{\PsiO}{\bH}{\PsiO}.
\end{equation}
This energy can be seen as the expectation value of a similarity-transformed Hamiltonian $\bH = e^{-\hT} \hH e^{\hT}$ for the reference determinant $\ket{\PsiO}$. 
One can expand $\bH$ thanks to the Baker-Campbell-Hausdorff formula and show that this series naturally truncates after the fourth-order term. 
This truncation is due to the two-electron nature of the Hamiltonian and is responsible for the affordable polynomial scaling of this method (contrary to the exponential cost of the variational approach).
In such a case, the cluster amplitudes are no longer determined by minimization of the VCC energy functional \eqref{eq:EVCC} but via the amplitude equations
\begin{equation}
	\label{eq:T2_eq}
	\mel*{\Psi_{ij\dots}^{ab\dots}}{\bH}{\PsiO} = 0,
\end{equation}
which are the projection of the similarity-transformed Schr\"odinger equation \eqref{eq:schroEq} onto excited determinants.
In Eq.~\eqref{eq:T2_eq}, the determinant $\ket*{\Psi_{ij\dots}^{ab\dots}}$ is obtained by promoting the electrons occupying the orbitals $i,j,\dots$ in $\ket{\PsiO}$ to the vacant orbitals $a,b,\dots$. 
One usually refers to this type of methods as traditional CC (TCC).

As reported in Refs.~\onlinecite{VanVoorhis_2000,Cooper_2010,Evangelista_2011}, VCC has been shown to give correct results in situations where TCC fails.
These benchmark studies evidenced that the breakdowns of TCC cannot be explained solely by its single-reference nature,
as part of the problem actually originates from its non-variational character.
Unfortunately, because of the exponential scaling of VCC, it is computationally cumbersome and cannot be applied in practice except for small molecules in small basis sets. 
This drawback has motivated the search for approximate methods that retain the advantages of VCC but at a polynomial cost. 
Because VCC inherits its exponential scaling from the lack of truncation of its energy functional [see Eq.~\eqref{eq:EVCC}], some authors have designed ingenious truncation schemes. \cite{Bartlett_1988, Kutzelnigg_1991} 
The quasi-variational CC (QVCC) method from Knowles' group has been designed along these lines. \cite{Robinson_2011,Robinson_2012,Robinson_2012a,Robinson_2012b,Robinson_2012c} 
This method, which is an improvement of the former linked pair functional, \cite{Knowles_2010} can be seen as an infinity summation of a given subset of diagrams of the VCC energy functional. 
This method has most of the desirable properties of an approximate VCC theory [see Ref.~\onlinecite{Robinson_2012} for an exhaustive discussion of these properties] but is not an upper bound to the exact energy. 
Yet, QVCC has been proved to be much more robust than TCC in cases where the latter exhibits non-variational collapse below the FCI energy like, for example, in the symmetric bond stretching of the nitrogen and acetylene molecules. \cite{Robinson_2012a,Robinson_2012b,Robinson_2012c}

Using VCC instead of TCC has its advantages but its computational complexity is very nettlesome. 
It would be simpler if one could describe strong correlation while retaining the projective way of solving the equations and its associated polynomial cost.  
Surprisingly, restricting the cluster operator to paired double excitations (pCCD), which is a simplification with respect to CC with doubles (CCD), \cite{Pople_1976} can give qualitatively good results for strongly correlated systems. \cite{Limacher_2013,Limacher_2014,Henderson_2014a,Henderson_2014b,Stein_2014,Gomez_2016,Shepherd_2016,Johnson_2017,Boguslawski_2017a,Boguslawski_2017b,Boguslawski_2019} This can be understood thanks to the concept of seniority number which is defined as the number of unpaired electrons in a determinant. \cite{Ring_1980} 
Indeed, the seniority-zero subspace (\ie, the set of all closed-shell determinants) has proven to give a good description of static correlation. \cite{Bytautas_2011} Unfortunately, doubly-occupied configuration interaction (DOCI), which is a CI calculation in the seniority-zero subspace, inherits the exponential scaling of its FCI parent. \cite{Allen_1962,Smith_1965,Veillard_1967,Weinhold_1967,Couty_1997,Kollmar_2003,Bytautas_2011}
However, benchmark results \cite{Henderson_2014a,Henderson_2014b,Henderson_2015,Shepherd_2016} have shown that pCCD provides ground-state energies which are almost indistinguishable from the DOCI ones but at a mean-field computational cost, hence providing a tractable way to qualitatively describe strongly correlated systems. 
Note that pCCD is equivalent to the antisymmetric product of 1-reference orbital geminals (AP1roG) \cite{Limacher_2013,Limacher_2014,Tecmer_2014,Boguslawski_2014a,Boguslawski_2014b,Boguslawski_2014c,Tecmer_2015,Boguslawski_2015,Boguslawski_2016a,Boguslawski_2016b,Johnson_2017,Fecteau_2020,Johnson_2020} which has been designed as a computationally tractable approximation to the antisymmetric product of geminals (APG), \cite{Coleman_1963,Coleman_1965} a method that has been recently further explored by the group of Scuseria. \cite{Henderson_2019,Khamoshi_2019,Henderson_2020,Dutta_2020,Khamoshi_2021,Dutta_2021}

Because the seniority-zero subspace is not invariant to orbital rotations, one must energetically optimize the orbitals to obtain the optimal pairing scheme, (\ie, the orbital set that minimizes the energy in the seniority-zero subspace). \cite{Bytautas_2011} 
In Ref.~\onlinecite{Limacher_2013}, Limacher \textit{et al.}~determined this pairing scheme by optimizing the orbitals at the DOCI level and then using these orbitals for their geminal wave function methods. 
Later, Henderson \textit{et al.}~designed an orbital-optimized pCCD (oo-pCCD) procedure which provides a more straightforward route to obtain this optimal pairing scheme. \cite{Henderson_2014a}

%%%%%%%%%%%%%%%%%%%%%%%%%%%%%%%%%%%%%%%%%%%%%%%%%%
\section{Coupled Cluster and Excited States}
\label{sec:ES}
%%%%%%%%%%%%%%%%%%%%%%%%%%%%%%%%%%%%%%%%%%%%%%%%%%

%=====================================
\subsection{TCC for excited states}
\label{sec:TCC4ES}
%=====================================
Excited-state energies and properties can be computed within the TCC paradigm through the well-established equation-of-motion (EOM) formalism. \cite{Rowe_1968,Monkhorst_1977,Koch_1990,Stanton_1993,Koch_1994} 
In EOM-CC, one applies a suitably chosen (linear) excitation operator on a ground-state CC wave function to compute excited states.
This procedure can be conveniently recast as a non-Hermitian eigenvalue problem involving the similarity-transformed Hamiltonian $\bar{H}$ in a space of excited determinants. \cite{Shavitt_2009}
Like in ground-state TCC, one can systematically expand the excitation space to form a well-defined hierarchy of EOM methods. 
As an example, EOM-CCSD restricts the set of excited determinants to singles and doubles.
EOM-CCSD is known to accurately describe single excitations \cite{Loos_2018b,Loos_2020c} but dramatically fails to describe double excitations because of the lack of triples and higher excitations.\cite{Loos_2019c,Loos_2020d} 
This shortcoming can be corrected by the inclusion of these higher excitations but this is not without a steep increase of the computational cost. \cite{Kucharski_1991,Christiansen_1995b,Kucharski_2001,Kowalski_2001,Hirata_2000,Hirata_2004}

Albeit being by far the most popular excited-state formalism, EOM is not the only route to excited states within CC theory. 
Indeed, the amplitude equations \eqref{eq:T2_eq} constitute a set of non-linear polynomial equations and consequently possess many solutions. 
These solutions, sometimes labeled as ``non-standard'', can be non-physical or correspond to genuine excited states. \cite{Piecuch_2000}
Therefore, performing a first standard ground-state CC calculation and a second one converging towards a given excited state provides an alternative way to obtain excitation energies. \cite{Adamowicz_1985,Lee_2019}
Lee \textit{et al.}~refers to this type of methods as $\Delta$CC \cite{Lee_2019} by analogy with the $\Delta$SCF methods where one basically follows the same procedure but at the self-consistent field (SCF) level. 
Indeed, the use of Hartree-Fock (HF) or Kohn-Sham higher-energy solutions corresponding to excited states is becoming more and more popular and new algorithms designed to target such solutions, like the maximum overlap method (MOM) \cite{Gilbert_2008,Barca_2014,Barca_2018a,Barca_2018b} or more involved variants, \cite{Thom_2008,Zhao_2016a,Ye_2017,Shea_2018,Thompson_2018,Ye_2019,Tran_2019,Burton_2019c,Zhao_2020,Hait_2020,Hait_2020b,Levi_2020a,Levi_2020b,Dong_2020,Hait_2021} are being actively developed. 
Besides providing a qualitatively good description of excited states, \cite{Hait_2021} these solutions can also be very helpful for $\Delta$CC methods, as we shall illustrate below (see also Ref.~\onlinecite{Lee_2019}).

The set of orbitals used, particularly the orbitals that constitute $\ket{\PsiO}$, strongly influences the performance of $\Delta$CC methods.
Importantly, the use of state-specific orbitals plays the role of a magnifying glass and facilitates the convergence towards a given CC solution by enlarging the associated basin of attraction.
In addition to the orbital set, two other factors influence in a significant way the solutions that can be reached: the guess amplitudes for the CC equations and the algorithm employed for solving these equations. 
Even if the chosen orbitals can enlarge or shrink the basin of attraction of a given solution, one still has to pick an appropriate starting point within this basin to be able to converge to the desired solution. 
Moreover, the type of iterative algorithms (usually based on Newton-Raphson method and/or supplemented by Pulay's DIIS method \cite{Pulay_1980,Pulay_1982,Scuseria_1986}) must also be carefully chosen so as to target, for example, saddle points or maxima instead of minima. 
For example, as shown in Ref.~\onlinecite{Kossoski_2021}, the usual CC iterative algorithm is inappropriate to converge towards excited states.

Because of the non-linearity of the CC equations [see Eq.~\eqref{eq:T2_eq}], the number of solutions can be higher that the physically meaningful number. 
However, claiming that a given solution corresponds to a genuine electronic state (or not) is a rather tricky task as the overall picture behind the structure of the CC solutions is still far from being thoroughly understood. 
Zivkovic and Monkhorst were the first to tackle this outstanding problem with their seminal work on the existence conditions of the higher roots of the CC equations. \cite{Zivkovic_1977,Zivkovic_1978}
However, their model was too simplistic and most of the pathological solutions that they found or predicted were due to this unrealistic model as argued later by Jankowski \textit{et al.}~who investigated the CCD solutions of $^{1}A_1$ symmetry in the \ce{H4} molecule. \cite{Jankowski_1994,Jankowski_1994a,Jankowski_1995} 
Still, they evidenced that some non-standard solutions may be non-physical. 
They also showed that the CC solution structure highly depends on the reference. \cite{Jankowski_1995} 
Few years later, the introduction of the homotopy method (which gives all the solutions of a set of non-linear equations) in the CC paradigm enabled the first systematic study on the structure of the CC energy landscape. \cite{Kowalski_1998,Kowalski_1998a} 
In particular, these studies showed that, in practice, the number of CC solutions is much lower than the theoretical upper bound known as B\'ezout's number. 
We refer the interested reader to the series of papers by Jankowski \textit{et al.} \cite{Jankowski_1999,Jankowski_1999a,Jankowski_1999b,Jankowski_1999c} and the book chapter of Piecuch and Kowalski \cite{Piecuch_2000} for an extensive discussion about the homotopy method and the higher-energy solutions of the CC equations.
We should also mention that the homotopy method has been employed to locate the CC solutions of the PPP model for some cyclic polyenes, \cite{Podeszwa_2002,Podeszwa_2003} as well as in the context of MRCC and the Bloch equation formalism. \cite{Paldus_1993,Kowalski_2000,Kowalski_2000a}

More recent studies have further improved our understanding of the CC energy landscape from which multiple solutions emerge. \cite{Mayhall_2010,Lee_2019} 
As pointed out by Mayhall in their study of the CCSD solutions in the \ce{NiH} molecule, the problem of the CC solution structure still needs to be addressed for more realistic systems. \cite{Mayhall_2010} 
Lee \textit{et al.}~showed that $\Delta$CC can provide fairly accurate double excitation and double core-hole energies. \cite{Lee_2019} 
Recently, we have pursued along these lines by analyzing the non-standard solutions of the pCCD equations. 
We have shown that the agreement between pCCD and DOCI holds for excited states on the condition that state-specific optimized orbitals are employed. \cite{Kossoski_2021}
Moreover, Ref.~\onlinecite{Kossoski_2021} brought some answers to Mayhall's open question as we have shown that $\Delta$oo-pCCD provides double excitation energies that are comparable in terms of accuracy to the more expensive EOM-CCSDT method \cite{Noga_1987,Scuseria_1988b,Kucharski_2001,Kowalski_2001,Kowalski_2001b} for a set of small (yet realistic) molecules.
It is worth mentioning again that all the studies mentioned above deal with TCC methods.
\titou{Note also that Ref.~\onlinecite{Johnson_2017} discusses, in particular, alternative ways to find multiple excited states for AP1roG (and related methods).}

%=====================================
\subsection{VCC for excited states}
\label{sec:VCC4ES}
%=====================================

For the sake of clarity, from here on, we restrict ourselves to VCCD (\ie, $\hT = \hT_2$) but the procedure presented below is general and can be applied to higher-order variants.
To the best of our knowledge, the present study is the first one to investigate excited states at the VCC level.

Because saddle points and maxima of the HF energy functional represent excited states, one can genuinely wonder if the same holds for the VCC energy functional \eqref{eq:EVCC}.
Thus, we seek for its stationary points, \ie, the different sets of cluster amplitudes $\boldsymbol{t}$ with elements $\ta{ij}{ab}$ satisfying
\begin{equation}
  \label{eq:dEVCC}
  \pdv{\EVCC}{\ta{ij}{ab}} = r_{ij}^{ab} = 0,
\end{equation}
where the VCCD residuals $r_{ij}^{ab}$ are the elements of the tensor $\boldsymbol{r}$. 
The ground-state variational solution obtained via the minimization of Eq.~\eqref{eq:EVCC} is also a solution of the more general equations \eqref{eq:dEVCC} which provide all the stationary solutions of the VCCD equations.
In this study, we restrict ourselves to solutions with real cluster amplitudes.
The explicit expressions of the residual equations under this assumption are derived in Appendix~\ref{app:appendixA}.
Of course, stationary points of the VCC energy functional associated with complex cluster amplitudes may also exist. 
Indeed, the hermiticity of $e^{\hT^{\dagger}}\hH e^{\hT}$ ensures that $E_\text{VCC}$ is real for any set of amplitudes. \cite{Kutzelnigg_1991} 

Because VCC has an inherent exponential scaling, one can take advantage of the more convenient FCI representation to implement VCC algorithms. \cite{VanVoorhis_2000,Cooper_2010} 
Following Van Voorhis and Head-Gordon, we represent the (unnormalized) CC wave function as a CI vector (\ie, in the Slater determinant basis) by $N$ successive applications of the cluster operator $\hT$ on the reference wave function:
\begin{equation}
  \label{eq:CCtoCI}
  \begin{split}
    &\ket{\PsiCC} = e^{\hT}\ket{\PsiO} \\
    &= \ket{\PsiO} + \hT \qty(\ket{\PsiO} + \frac{\hT}{2}\qty( \dots \qty(\ket{\PsiO} + \frac{\hT}{N-1} \qty(\qty(1 + \frac{\hT}{N})\ket{\PsiO}))\dots)).
  \end{split}
\end{equation}
Using this CI representation, the action of second quantized operators on the CC wave function is quite straightforward, and one can evaluate the energy \eqref{eq:EVCC} and the residuals \eqref{eq:derivationVCCampEq} by simple matrix products.
Note that the coefficients of the resulting CC wave function [see Eq.~\ref{eq:CCtoCI}] are equal to the cluster analysis of the CI coefficients. \cite{Cizek_1969,Monkhorst_1977,Lehtola_2017,Magoulas_2021}

In their VCCD benchmark study, Van Voorhis and Head-Gordon \cite{VanVoorhis_2000} relied on the standard TCCD iterative procedure (where one computes an approximate diagonal Jacobian matrix based on the difference of the Fock matrix elements $f_p^q$) to solve Eq.~\eqref{eq:dEVCC}:
\begin{equation}
  \label{eq:updateAmpVCC}
  \ta{ij}{ab} \leftarrow \ta{ij}{ab} - \frac{r_{ij}^{ab}}{\f{a}{a} + \f{b}{b} - \f{i}{i} - \f{j}{j}}.
\end{equation}
However, this approximate form of the Jacobian matrix cannot be employed to target excited states as it systematically converges towards the ground state or eventually diverges (see Ref.~\onlinecite{Kossoski_2021} for an exhaustive discussion on this point). 
\titou{If one aims at excited states, one should be aware that they generally are saddle points of the energy landscape.
However, local minima could also correspond to physical excited states but it was not the case for the two model systems considered here.}
\titou{Therefore, to target saddle points, one should take into account the curvature of the energy landscape. 
To do so, we consider the Jacobian matrix $\boldsymbol{J}$ with elements}
\begin{equation}
  \label{eq:jacobian}
  J_{ij,kl}^{ab,cd} = \pdv{r_{ij}^{ab}}{t_{kl}^{cd}},
\end{equation}
which is then used to update the amplitudes according to the usual Newton-Raphson algorithm, \ie,
\begin{equation}
  \label{eq:updateAmpVCCDiagHess}
  \boldsymbol{t} \leftarrow \boldsymbol{t} - \boldsymbol{J}^{-1} \cdot \boldsymbol{r}.
\end{equation}
The general expression of the Jacobian matrix elements is given in Appendix \ref{app:appendixA}.
\titou{Note that this updating scheme of the amplitudes [see Eq.~\eqref{eq:updateAmpVCCDiagHess}] is more computationally demanding than the usual one [see Eq.~\eqref{eq:updateAmpVCC}] as it requires to compute the entire Jacobian matrix and invert it.}
\titou{We should nonetheless mention that alternative algorithms are available to target such solutions. 
For example, approximate Newton-Raphson schemes which preserve the information about the energy curvature contained in the exact Jacobian matrix, or large-scale iterative solvers where one does not need to construct the full Jacobian, could also be employed.}

In difficult cases, it can be useful to damp the Newton-Raphson steps. 
However, one has to ensure that the structure of the Jacobian matrix is preserved during this process.
This can be done by diagonalizing the Jacobian and adding a positive/negative constant to the positive/negative eigenvalues,
similarly to what we have recently done for orbital optimization at the pCCD level. \cite{Kossoski_2021}

To fully specify our algorithm, we still need to choose our reference $\ket{\PsiO}$ as well as the starting values of the cluster amplitudes. 
In this study, we rely on both ground- and excited-state HF wave functions as references in order to study the influence of state-specific references.
The orbitals employed to construct these excited-state HF wave functions have been obtained using initial MOM (IMOM). \cite{Gilbert_2008,Barca_2014,Barca_2018a,Barca_2018b}
State-specific orbitals optimized at the correlated level are also considered, as discussed below.
Regarding the starting values of the cluster amplitudes $\boldsymbol{t}$, once again we have taken advantage of the FCI representation by obtaining these via a cluster analysis of the corresponding CI eigenvectors. \cite{Monkhorst_1977,Lehtola_2017}

%=====================================
\subsection{Orbital optimization for excited states}
\label{sec:OO4ES}
%=====================================

The solutions obtained via this iterative process [see Eq.~\eqref{eq:updateAmpVCCDiagHess}] are stationary points of the VCCD energy functional with respect to the cluster amplitudes but not with respect to the orbital coefficients. 
Indeed, the orbitals have usually been obtained at the HF level and no longer represent a stationary point when electron correlation is introduced. 
The next step is thus to optimize the orbitals at the corresponding correlated level to find solutions that are stationary with respect to both the cluster amplitudes and the orbital coefficients.

As usually done, \cite{Scuseria_1987,Bozkaya_2011} we introduce a unitary operator $e^{\hat{\kappa}}$ into the VCCD energy functional,
\begin{equation}
  \label{eq:orbVCC}
  \EVCC(\hat{\kappa}) = \frac{\mel{\PsiO}{e^{\hT^{\dag}} e^{-\hat{\kappa}} \hH e^{\hat{\kappa}} e^{\hT}}{\PsiO}}{\mel{\PsiO}{e^{\hT^{\dag}} e^{\hT}}{\PsiO}},
\end{equation}
to account for orbital rotations. Now, Eq.~\eqref{eq:orbVCC} can be minimized with respect to the cluster amplitudes $t_{ij}^{ab}$ and to the orbital rotation parameters $\kappa_{pq}$ of the one-electron anti-Hermitian operator $\hat{\kappa}$. 
For a given set of cluster amplitudes, we search for the stationary points with respect to the orbital rotation parameters using the second-order Newton-Raphson method. 
We then expand the VCC energy around $\boldsymbol{\kappa} = \boldsymbol{0}$,
\begin{equation}
	\label{eq:energy_expansion}
	\EVCC(\boldsymbol{\kappa}) \approx \EVCC(\boldsymbol{0}) + \boldsymbol{g} \cdot \boldsymbol{\kappa} + \frac{1}{2} \boldsymbol{\kappa^{\dag}} \cdot \boldsymbol{H} \cdot \boldsymbol{\kappa},
\end{equation}
where $\boldsymbol{g}$ is the orbital gradient and $\boldsymbol{H}$ is the orbital Hessian, both evaluated at $\boldsymbol{\kappa}=\boldsymbol{0}$, \ie,
\begin{align}
  \label{eq:orbGradHess}
	g_{pq} & = \left. \pdv{\EVCC(\boldsymbol{\kappa})}{ \kappa_{pq}} \right|_{\boldsymbol{\kappa} = \boldsymbol{0}},
	&
	H_{pq,rs} & = \left. \pdv{\EVCC(\boldsymbol{\kappa})}{ \kappa_{pq}}{ \kappa_{rs}} \right|_{\boldsymbol{\kappa} = \boldsymbol{0}}.
\end{align}
The orbitals are then updated following the usual Newton-Raphson step
\begin{equation}
  \label{eq:updateCoeff}
  \boldsymbol{C} \leftarrow  \boldsymbol{C} \cdot e^{-\boldsymbol{H}^{-1} \cdot \boldsymbol{g}},
\end{equation}
where $\boldsymbol{C}$ is the orbital coefficient matrix.
Then, one finds the solution of Eq.~\eqref{eq:dEVCC} for this new set of orbitals and the procedure is repeated until convergence.
To compute the gradient and the Hessian, one must compute the one- and two-body density matrices, \cite{Henderson_2014a} with respective elements 
\begin{subequations}
\begin{align}
  \label{eq:onebody}
  \gamma_{pq} &= \sum_{\sigma} \frac{\mel{\PsiCC}{\cre{q_{\sigma}}\ani{p_{\sigma}}}{\PsiCC}}{\braket{\PsiCC}{\PsiCC}}, \\
  \label{eq:twobody}
  \Gamma_{pq,rs} &= \sum_{\sigma \sigma'} \frac{\mel{\PsiCC}{\cre{s_{\sigma}}\cre{r_{\sigma'}}\ani{q_{\sigma'}}\ani{p_{\sigma}}}{\PsiCC}}{\braket{\PsiCC}{\PsiCC}},
\end{align}
\end{subequations}
where the orbital index refers to spatial orbitals, and $\sigma$ and $\sigma'$ to spin indexes.
Once again, we take advantage of the CI representation of the VCCD wave function to compute these quantities. 
We express the string of second quantized operators in Eqs.~\eqref{eq:onebody} and \eqref{eq:twobody} as a matrix in the Slater determinant basis, and then evaluate the elements of the one- and two-body density matrices by simple matrix products.

In the present study, we restrict the cluster operator to a pair double excitation operator
\begin{equation}
	\hat{T} = \sum_{ia} t_{ii}^{aa} P_a^{\dag} P_i,
\end{equation}
(with $P_q^{\dag} = a_{q\uparrow}^{\dag} a_{q\downarrow}^{\dag}$) and investigate the properties of ground and excited states at the traditional pCCD (TpCCD) and variational pCCD (VpCCD) levels.
This choice is motivated by the two following arguments. 
Firstly, our aim is to compare the VpCCD solution structure with its TpCCD counterpart (which has received our attention recently \cite{Kossoski_2021}) in order to provide new insights into the multiple solutions of the VCC equations.
Secondly, this restriction of the cluster operator significantly lowers both the computational cost and the complexity of the energy landscape, hence simplifying the present analysis. 
The VpCCD equations are easily obtained from their VCCD analogs (see Appendix~\ref{app:appendixB} for their explicit expressions). 
We refer the interested reader to Ref.~\onlinecite{Henderson_2014a} for a complete list of equations and an exhaustive discussion of the orbital optimization algorithm in the case of ground-state TpCCD and to Ref.~\onlinecite{Kossoski_2021} for the case of excited-state TpCCD.

In the following, taking the symmetric dissociation of the linear \ce{H4} molecule as a first case study, ground- and excited-state energies obtained at the TpCCD and VpCCD levels are compared to DOCI for three different sets of orbitals: ground-state HF orbitals, state-specific HF orbitals and state-specific orbitals optimized at the VpCCD level.
In a second stage, we look at the various TpCCD, VpCCD and DOCI electronic states in the presence of strong correlation (\ie, near degeneracies) by examining the continuous deformation of \ce{H4} from a square to a rectangular arrangement.

%%%%%%%%%%%%%%%%%%%%%%%%%%%%%%%% 
\section{Computational details}
\label{sec:compdet}
%%%%%%%%%%%%%%%%%%%%%%%%%%%%%%%%

The computational methods investigated here (HF, MOM, TpCCD, VpCCD, DOCI, and FCI) have been implemented as standalone \textsc{mathematica} modules, \cite{Mathematica} which makes them easily interconnectable and modifiable depending on the actual purpose.
These are provided in an accompanying notebook available for download from Zenodo at \href{http://doi.org/10.5281/zenodo.4971904}{http://doi.org/10.5281/zenodo.4971904}.
All the calculations have been performed in the restricted formalism.
The only required input is the one- and two-electron integrals which are usually computed with a third-party software like {\QP}.\cite{Garniron_2017b,Garniron_2018,Garniron_2019}
The convergence threshold (based on the DIIS commutator) was set to $10^{-10}$ a.u.~for the restricted HF (RHF) calculations, while the convergence thresholds (based on the maximum absolute value of the gradient) for the cluster amplitude and orbital optimization procedures were both set to $10^{-6}$ a.u.

%%%%%%%%%%%%%%%%%%%%%%%%%%%%%%%%
\section{Results and discussion}
\label{sec:res}
%%%%%%%%%%%%%%%%%%%%%%%%%%%%%%%%

%========================================================================
\subsection{Influence of the orbital set: the linear \ce{H4} molecule}
\label{subsec:linearH4}
%========================================================================

%%% FIG 1 %%%
\begin{figure*}
  \includegraphics[width=0.58\textwidth]{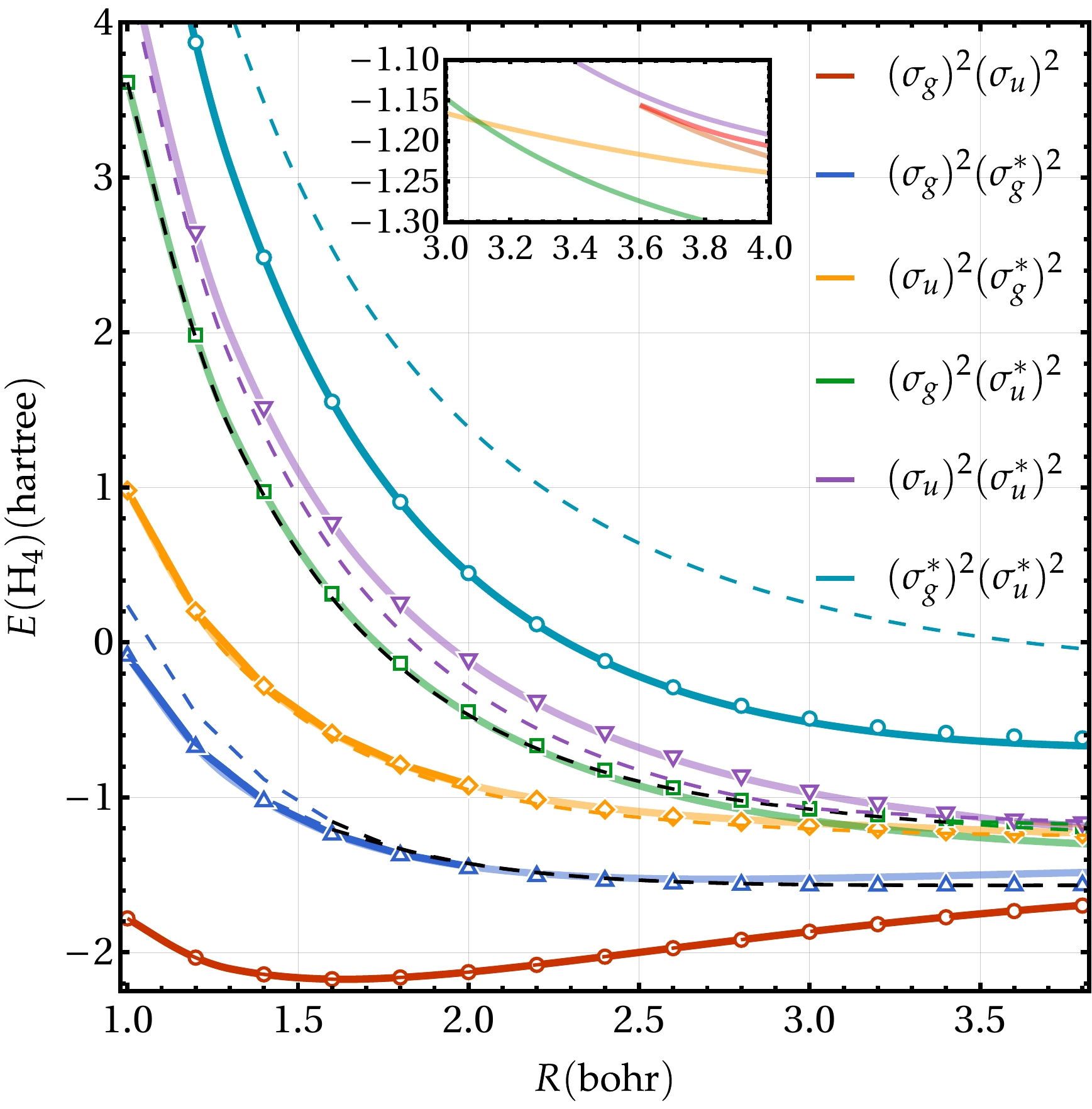}
  \includegraphics[width=0.31\textwidth]{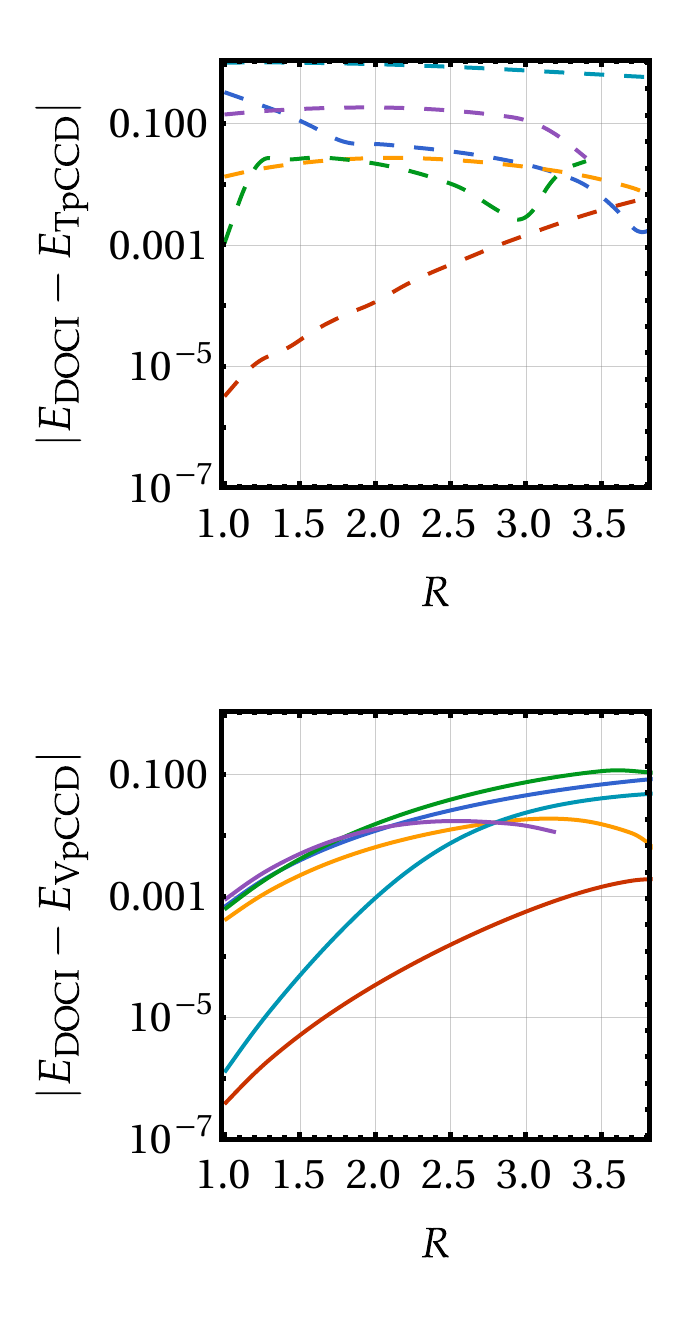}
  \caption{
    Left: Energies (in hartree) of the linear \ce{H4} molecule in the STO-6G basis set as functions of the bond length $R$ (in bohr) for various methods using the ground-state RHF determinant as reference wave function: DOCI (markers), VpCCD (solid), and TpCCD (dashed). 
    The real part of the complex TpCCD solutions are represented as dashed black lines.
	Right: Energy differences between DOCI and TpCCD (VpCCD) as functions of $R$ in the top (bottom) panel.
	}
  \label{fig:H4RHF}
\end{figure*}
%%% %%% %%% %%% 

%%% FIG 2 %%%
\begin{figure}
	\includegraphics[width=\linewidth]{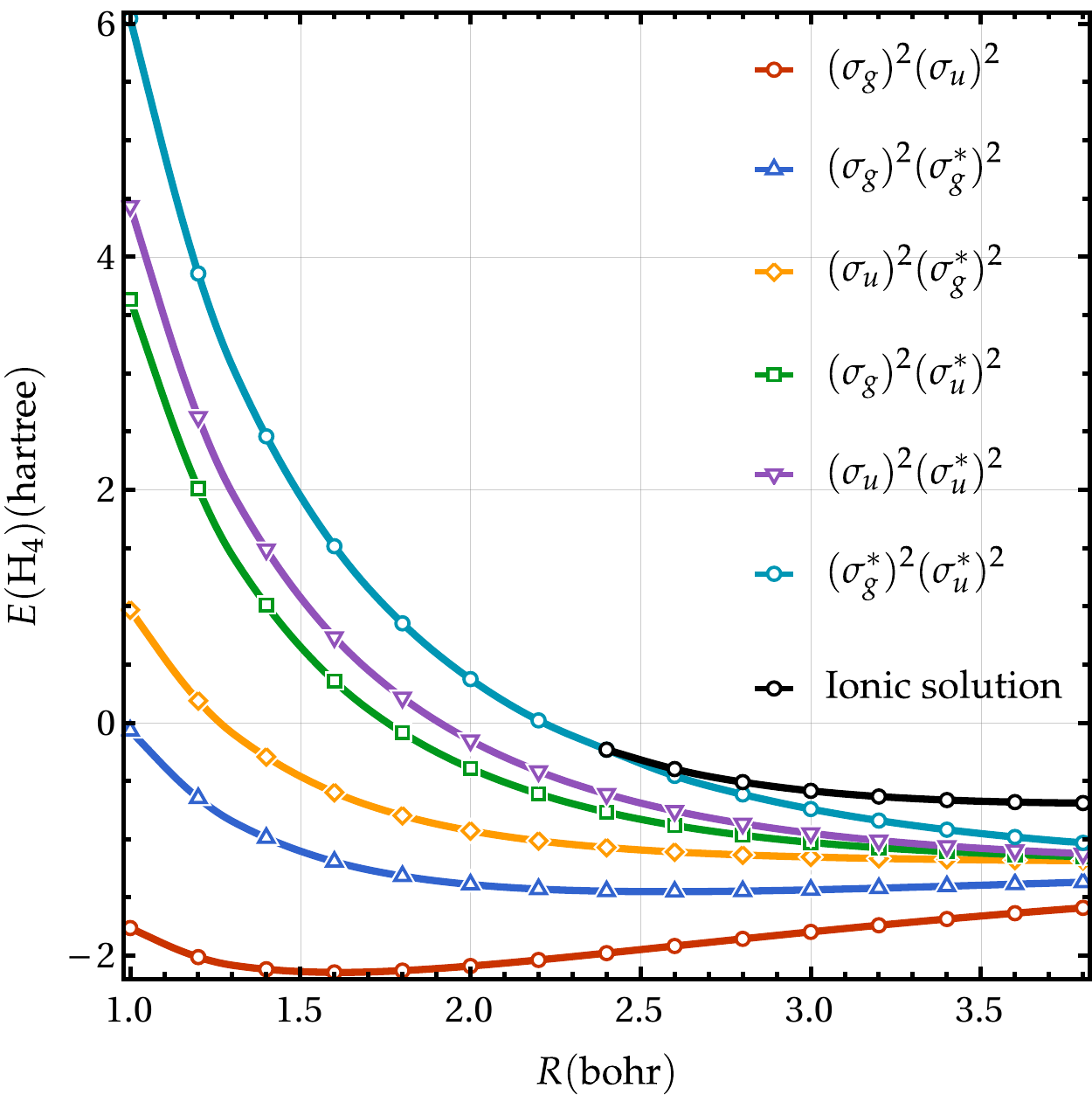}
	\caption{
	Energies (in hartree) of various RHF solutions as functions of the bond length $R$ (in bohr) for the linear \ce{H4} molecule in the STO-6G basis set.
	\label{fig:H4MOM}}
\end{figure}
%%% %%% %%% %%%

%%% FIG 3 %%%
\begin{figure*}
  \includegraphics[width=0.58\textwidth]{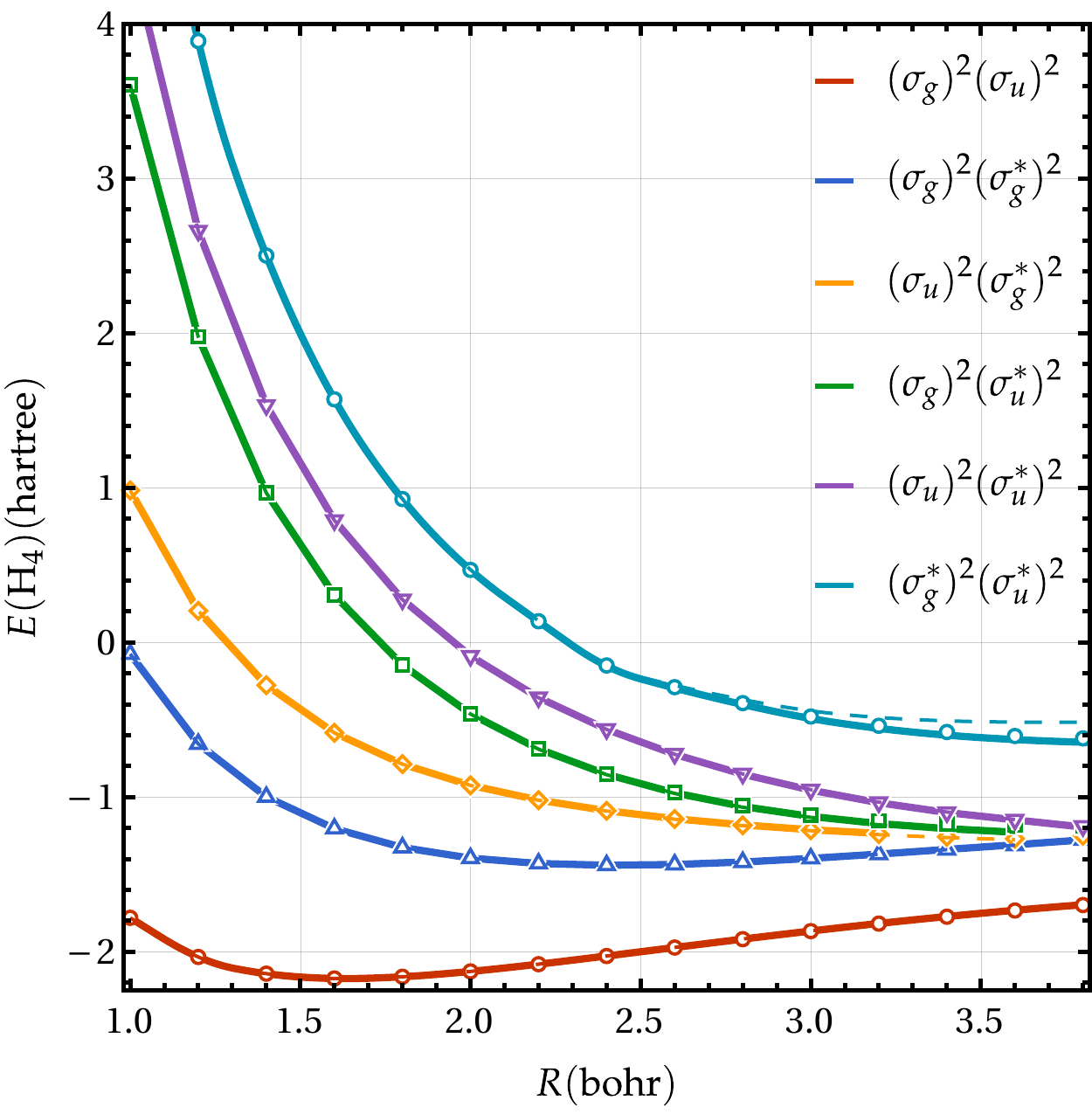}
  \includegraphics[width=0.31\textwidth]{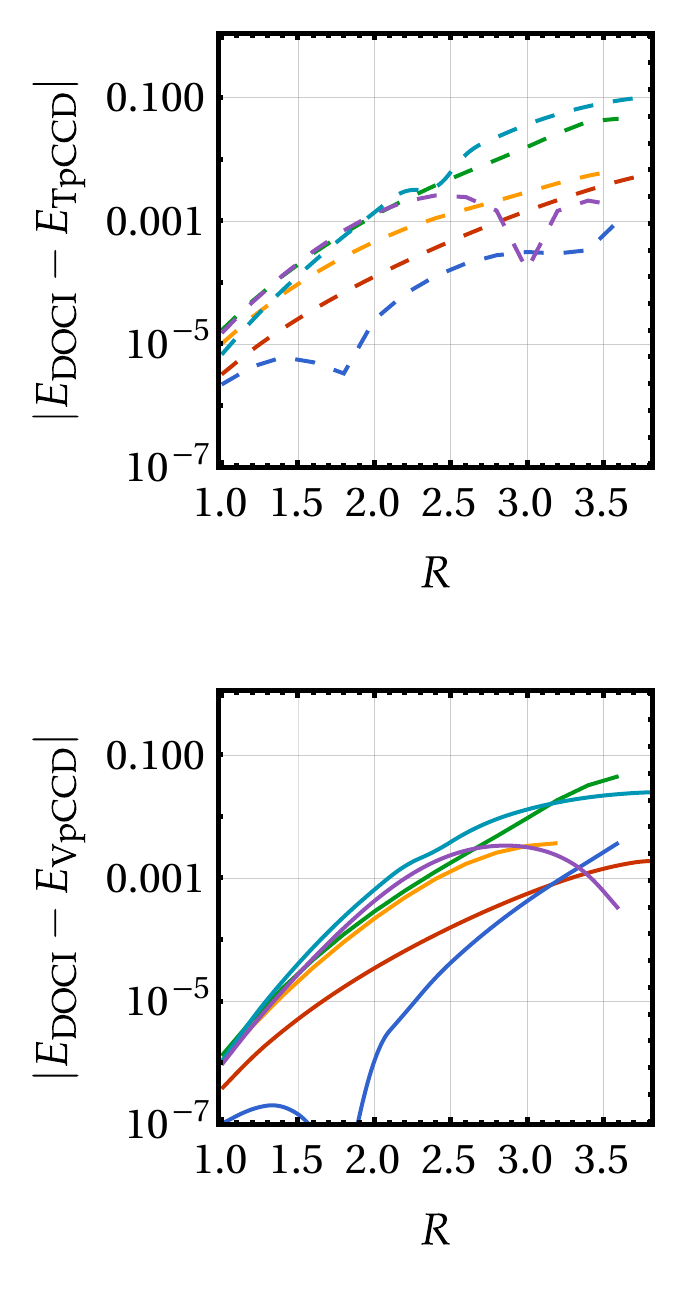}
  \caption{
    Left: Energies (in hartree) of the linear \ce{H4} molecule in the STO-6G basis set for various methods using state-specific RHF determinants as functions of the bond length $R$ (in bohr): DOCI (dots), VpCCD (solid), and TpCCD (dashed). 
    Right: Energy differences between DOCI and TpCCD (VpCCD) as functions of $R$ in the top (bottom) panel.}
  \label{fig:H4Correlated}
\end{figure*}
%%% %%% %%% %%%

%%% FIG 4 %%%
\begin{figure*}
  \includegraphics[width=0.58\textwidth]{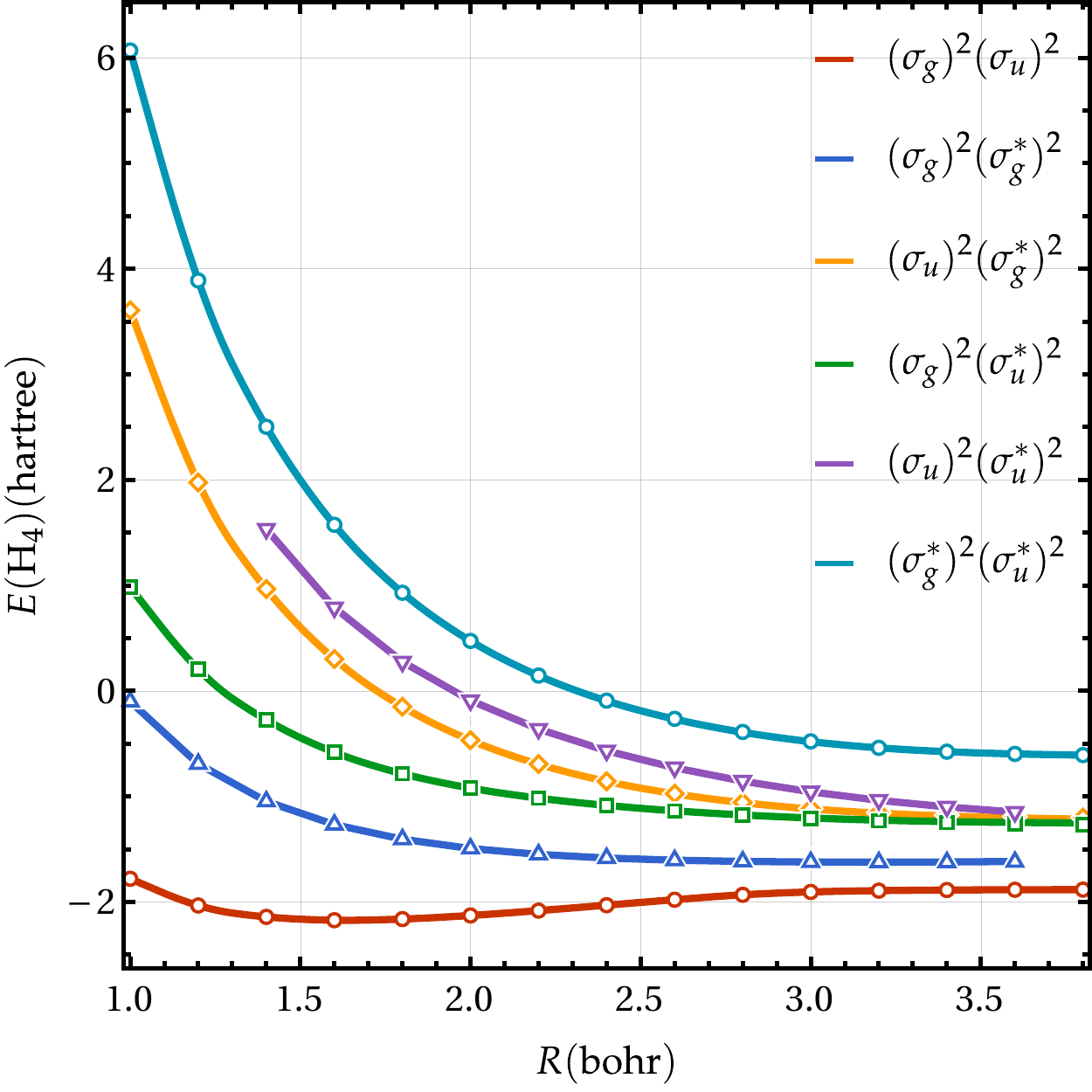}
  \includegraphics[width=0.31\textwidth]{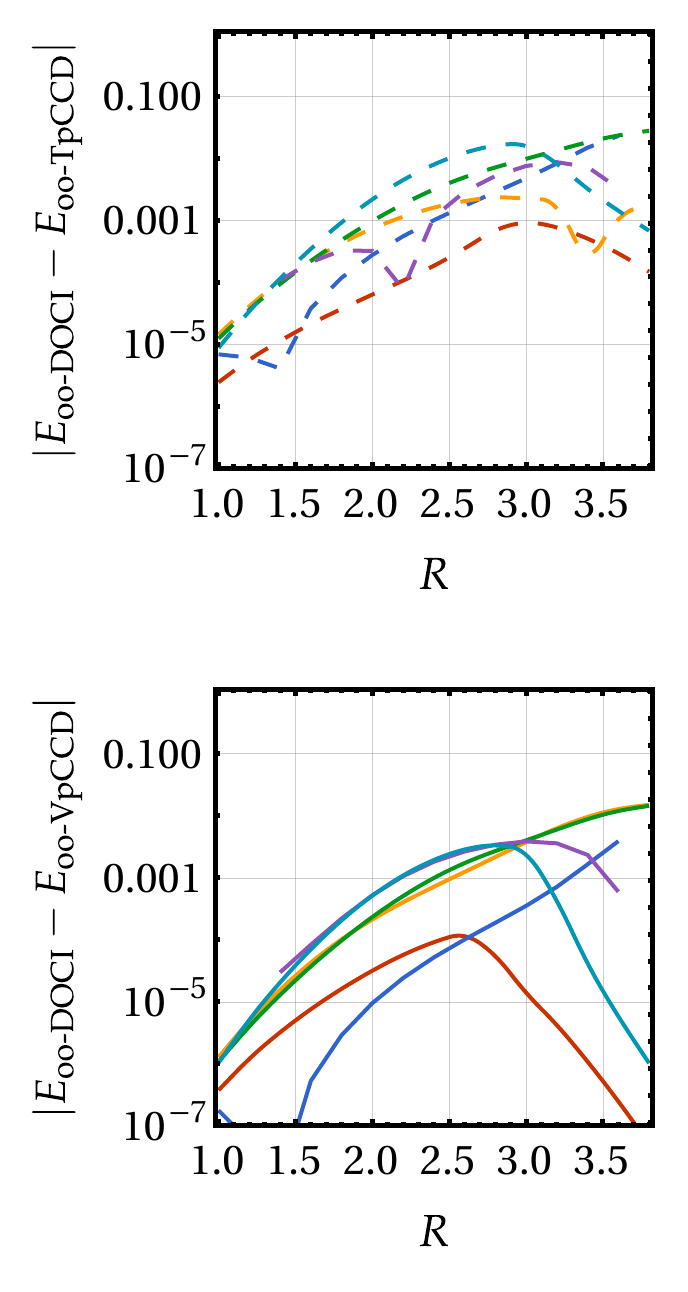}
  \caption{
  Left: Energies (in hartree) of the linear \ce{H4} model in a STO-6G basis set for the orbital-optimized VpCCD method (solid) and DOCI using the same orbitals (dots) as functions of the bond length $R$ (in bohr). 
  Right: Energy differences between oo-DOCI and oo-TpCCD (oo-VpCCD) as functions of $R$ in the top (bottom) panel.}
    \label{fig:H4ooCorrelated}
\end{figure*}
%%% %%% %%% %%%

As a first example, we consider the symmetric stretching of the linear \ce{H4} molecule in a minimal basis (STO-6G \cite{Hehre_1969}). 
This corresponds to a system with 4 electrons in 4 spatial orbitals with respective symmetries $\sigma_g$, $\sigma_u$, $\sigma_g^*$, and $\sigma_u^*$ (in ascending energies).
Linear chains of hydrogens are prototypical examples of left-right correlation and, therefore, have been widely studied in order to probe electronic structure methods in presence of such correlation. \cite{Hachmann_2006,Al-Saidi_2007,Sinitskiy_2010,Bytautas_2011,Stella_2011,Robinson_2012c,Limacher_2013,Kats_2013,Henderson_2014a,Motta_2017,Motta_2020,Vollhard_2020,Giner_2020}
Hereafter, the distance between two consecutive hydrogens is denoted by $R$.
The first stage of this study consists in investigating the quality of the TpCCD and VpCCD ground- and excited-state energies in the case where the reference wave function is chosen as the ground-state RHF determinant, a choice that obviously induces a bias towards the ground state.
The VpCCD energies (solid lines) are plotted alongside the DOCI ones (markers) in the left-hand-side of Fig.~\ref{fig:H4RHF}.
Thanks to the simplicity of this model, one can access, via \textsc{mathematica}'s implementation of the Jenkins-Traub algorithm, \cite{Jenkins_1970a,Jenkins_1970b} the entire set of solutions (with real cluster amplitudes) associated with the system of polynomial equations \eqref{eq:dEVCC}.
These VpCDD solutions are represented as thin solid lines in Fig.~\ref{fig:H4RHF},
while the thick parts of the curves correspond to the energies that we have been able to obtain using the Newton-Raphson algorithm described earlier [see Sec.~\ref{sec:OO4ES}].
Figure~\ref{fig:H4RHF} also shows the TpCCD energies (dashed lines) which are also determined with the Jenkins-Traub algorithm applied to Eq.~\eqref{eq:T2_eq}.
In addition, the difference between TpCCD (VpCCD) and DOCI energies are also plotted in the top (bottom) right panel of Fig.~\ref{fig:H4RHF}.

Considering the ground-state RHF determinant as reference wave function, the convergence towards the VpCCD ground state, $(\sigma_g)^2 (\sigma_u)^2$, is numerically straightforward all along the potential energy curve (PEC). 
On the other hand, converging excited states with the Newton-Raphson algorithm has been found to be much more challenging.
We have not been able to converge the two lowest VpCCD excited states, $(\sigma_g)^2 (\sigma_g^*)^2$ and $(\sigma_u)^2 (\sigma_g^*)^2$, further than $R=\SI{2.0}{\bohr}$ for this set of orbitals. 
Even worse, the two other doubly-excited states, $(\sigma_g)^2 (\sigma_u^*)^2$ and $(\sigma_u)^2 (\sigma_u^*)^2$, have been reached only for $R=\SI{1.0}{\bohr}$.
This is not the case for the $(\sigma_g^*)^2 (\sigma_u^*)^2$ quadruply-excited state for which one can converge VpCCD calculations fairly easily all along the PEC with the Newton-Raphson algorithm. 
This might be due to the fact that the corresponding stationary points are maxima for the quadruply-excited state whereas doubly-excited states correspond to saddle points (see below).
Despite such numerical difficulties, the complete set of solutions could be obtained thanks to the Jenkins-Traub algorithm.

Overall, the VpCCD method provides a fairly good approximation to the DOCI energies. 
At small $R$ (\ie, in the weak correlation regime), VpCCD is in much closer agreement with DOCI than TpCCD, most noticeably for the $(\sigma_g^*)^2 (\sigma_u^*)^2$ quadruply-excited state.
At large $R$ (\ie, in the strong correlation regime), this comparison is trickier.
Yet, the difference between VpCCD and DOCI seems more regular (see the bottom-right panel of Fig.~\ref{fig:H4RHF}) whereas the behavior of TpCCD is more erratic (top-right panel).
Thus, one can state that, if one considers the ground-state RHF determinant as reference wave function, the main difficulties associated with VpCCD calculations concern its convergence, as the energies compare very favourably with the DOCI reference (at least in the weak correlation regime).
At large $R$, the agreement between VpCCD and DOCI is less obvious as we shall see below.

Thanks to previous investigations, we know that some of the TpCCD excited-state solutions can be labeled as non-physical. \cite{Jankowski_1994,Jankowski_1994a,Jankowski_1995,Jankowski_1999,Jankowski_1999a,Jankowski_1999b,Jankowski_1999c,Piecuch_2000,Kossoski_2021}
For example, in the case of the linear \ce{H4} molecule using the ground-state RHF determinant as reference wave function, the lowest-lying DOCI excited state (blue markers in Fig.~\ref{fig:H4RHF}) can be represented by two TpCCD solutions (dashed blue curves). \cite{Kossoski_2021}
These solutions eventually merge for $R>\SI{1.7}{\bohr}$ and become a complex conjugate pair of solutions with real components represented as black dashed lines in Fig.~\ref{fig:H4RHF}. 
The same phenomenon occurs for the fourth doubly-excited state, but the complex conjugate pair of solutions exists up to $R = \SI{3.4}{\bohr}$ before splitting into two real solutions (dashed green curves).
In the case of VpCCD, there are only six real-amplitude solutions in the weak correlation regime. 
However, for $R\gtrsim \SI{3.5}{\bohr}$, two additional real solutions appear as one can see in the inset of Fig.~\ref{fig:H4RHF}.
The fact that these spurious solutions appear as a pair indicate that they may exist for smaller $R$ as a pair of solutions with complex conjugate amplitudes.
Because all the solutions are energetically close in this region of the PECs, it is hard to tell whether a solution is unphysical or not, and which one models better the corresponding DOCI solution.
This is why the curve corresponding to the difference between VpCCD and DOCI for the fourth doubly-excited state stops at $R=\SI{3.2}{\bohr}$ in the bottom right panel of Fig.~\ref{fig:H4RHF}.
The same unpredictability occurs between the green and purple TpCCD curves in the strong correlation regime.
Therefore, in the weak correlation regime, the problems caused by unphysical solutions seem to be less severe in VpCCD than in TpCCD. 
Yet, when the correlation becomes strong, VpCCD is also plagued by unphysical solutions.
Note that unphysical solutions at the VpCCD level are due to the approximate nature of the method which originates from the truncation of the cluster operator $\hT$. 
On the other hand, unphysical TpCCD solutions can originate from this same truncation and/or from the projection step of Eq.~\eqref{eq:TCCnrj}.

The stability analysis of the various stationary points via the computation of the eigenvalues of the Jacobian matrix \eqref{eq:jacobian} provides useful information on the presence of additional solutions. \cite{Szakacs_2008,Surjan_2010}
For example, a change in the number of negative eigenvalues (the saddle point index) indicates the appearance of additional solutions. \cite{Burton_2021}
For $R<\SI{3.4}{\bohr}$, the index of the VpCCD solutions, in ascending energies, increases smoothly (0, 1, 2, 2, 3, and 4) up to the cyan curve which is an index-4 stationary point (\ie, a maximum).
At $R=\SI{3.4}{\bohr}$, the index associated to the green VpCCD solution decreases by one unit, this solution becoming an index-1 saddle point.
We see in the inset of Fig.~\ref{fig:H4RHF} that two additional solutions appear right after this index variation, these two spurious solutions being index-3 saddle points.

Because the agreement between DOCI and both TpCCD and VpCCD ground-state energies is very satisfying when one employs as reference the ground-state RHF wave function, one can reasonably wonder if the same similarity holds in the case of excited states by considering excited-state RHF wave functions as state-specific references.
We have recently shown that this holds for TpCCD when one uses state-specific orbitals optimized at this correlated level, \cite{Kossoski_2021}
but it remains to be seen whether this still applies with state-specific mean-field orbitals.
Using IMOM, \cite{Barca_2018} we have obtained five additional restricted solutions of the RHF equations, corresponding to the five possible non-Aufbau closed-shell determinants (see Fig.~\ref{fig:H4MOM}).
Note that each excited-state solution corresponds to a different set of orthogonal orbitals, but these sets are, a priori, not orthogonal to each others because they originate from distinct Fock operators. 
Of course, spatially symmetry-broken RHF solutions do exist but we have not considered them here.
For $R > \SI{2.4}{\bohr}$, an additional solution, plotted as a black line in Fig.~\ref{fig:H4MOM}, has been found by systematically occupying the two highest-energy molecular orbitals at each SCF cycle.
The molecular orbitals associated with this \ce{H+}~\ce{H-}~\ce{H-}~\ce{H+} ionic configuration have a more localized character than the orbitals constituting the $(\sigma_g^*)^2 (\sigma_u^*)^2$ determinant (the orbitals associated with these two solutions are available in the \hyperlink{SI}{supplementary material}).
A stability analysis of these RHF solutions \cite{Seeger_1977,Fukutome_1981,Stuber_2003} shows that the cyan curve is a maxima at small $R$ but, for $R>\SI{2.4}{\bohr}$, it becomes a saddle point whereas the ionic configuration (\ie, the black curve) corresponds to a maxima.

The MOM excited states represented in Fig.~\ref{fig:H4MOM} can be used as reference wave functions at both the TCC (see Ref.~\onlinecite{Lee_2019}) and VCC levels.
The energies at the three different correlated levels (namely, DOCI, TpCCD, and VpCCD) using these state-specific excited-state RHF reference wave functions are plotted in Fig.~\ref{fig:H4Correlated} and are labeled as MOM-DOCI, MOM-TpCCD, and MOM-VpCCD in the following.
As one can see, these energies are visually indistinguishable, except at large $R$ in the case of the $(\sigma_g^*)^2(\sigma_u^*)^2$ state.
Still, the right panel of Fig.~\ref{fig:H4Correlated} shows that MOM-VpCCD is closer to MOM-DOCI than MOM-TpCCD by roughly one order of magnitude all along the PEC.
Also, we can see in the top-right panel that the difference between MOM-TpCCD and MOM-DOCI is less erratic than its analog using ground-state RHF orbitals (see Fig.~\ref{fig:H4RHF}).

As expected, using state-specific RHF determinants as reference rather than the ground-state one significantly improves the description of excited states at the TpCCD level.
Therefore, if one wants to target excited states at the TpCCD level, it is worth investing in the design of proper state-specific references in order to make the key projection step in Eq.~\eqref{eq:TCCnrj} more effective.
Even if it is less pronounced at the VpCCD level, state-specific RHF reference determinants also improve the accuracy of the excited-state energies (with respect to DOCI).
The most noticeable positive side effect of these state-specific references on VpCCD is the greater ease of convergence.
Indeed, as shown in Fig.~\ref{fig:H4Correlated}, we have been able to converge almost all the states up to $R \simeq \SI{3.5}{\bohr}$.
Therefore, we argue that using state-specific references enlarge the basin of attraction of the associated solution and consequently facilitates the convergence towards it.

The logical next step is to compare DOCI, TpCCD, and VpCCD at the orbital-optimized (oo) level (as described in Sec.~\ref{sec:OO4ES}).
For ground states, DOCI and pCCD have already been shown to perform best when one relaxes the spatial symmetry constraint as it allows the orbitals to be fully localized. \cite{Limacher_2014,Boguslawski_2014b,Boguslawski_2014c} 
On the other hand, relaxing this constraint also considerably increases, in principle, the number of attainable solutions. 
For example, multiple solutions corresponding to the ground state have already been observed. \cite{Limacher_2014,Boguslawski_2014b,Boguslawski_2014c}
However, in the case of excited states, we have only obtained symmetry-adapted solutions even if the orbital optimization algorithm could, in principle, target symmetry-broken solutions. \cite{Henderson_2014a}
It may be due to the lack of flexibility associated with the minimal basis. 
Indeed, we have shown, at the TpCCD level, that for larger molecules in larger basis sets one could also break the spatial symmetry to improve the description of excited states. \cite{Kossoski_2021}
We expect analog symmetry-broken excited-state solutions for larger molecules in the case of VpCCD.

As shown by Henderson \textit{et al.}, \cite{Henderson_2014a} DOCI and TpCCD optimized orbitals are virtually indistinguishable in molecular systems.
Here, we have observed that the VpCCD optimized orbitals are also virtually indistinguishable from the two other sets.
Therefore, as expected, oo-TpCCD and oo-VpCCD energies are also highly similar, so that we only report oo-VpCCD energies in Fig.~\ref{fig:H4ooCorrelated}.
The right panel of Fig.~\ref{fig:H4ooCorrelated} evidences that the accuracy of oo-TpCCD and oo-VpCCD is similar to their MOM-TpCCD and MOM-VpCCD counterparts (see Fig.~\ref{fig:H4Correlated}), at least in the weak correlation regime (always having DOCI as the reference results). 
However, in the strong correlation regime, the scenario is rather different. 
The orbital optimization at the correlated level allows them to strongly localize when the bond is stretched, hence the PECs exhibit the correct dissociation limits.
As a direct consequence, the agreement between VpCCD (and TpCCD) and DOCI is improved at large $R$, as compared to MOM-VpCCD (and MOM-TpCCD).
We can then conclude that, in the absence of strong correlation effects, state-specific RHF determinants provide robust and cheaper alternatives to determinants made of optimized orbitals at the correlated level. 
To further illustrate this, we provide the VpCCD optimized orbitals as well as the MOM orbitals in the \hyperlink{SI}{supplementary material}.

%=========================================================================
\subsection{A strong correlation model: the ring \ce{H4} molecule}
%=========================================================================

%%% FIG 5 %%%
\begin{figure}
  \includegraphics[width=\linewidth]{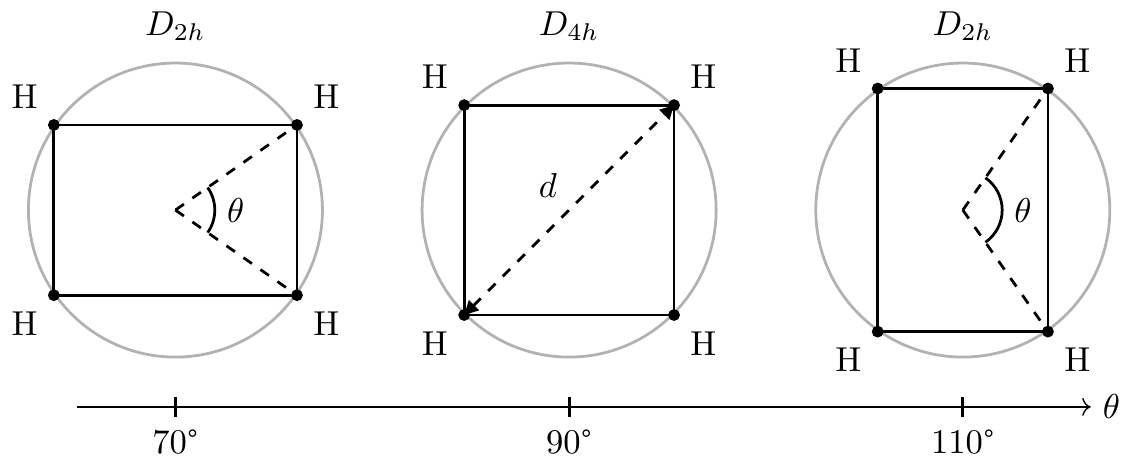}
  \caption{The ring \ce{H4} model.
  Here the diameter of the circle $d$ is set to $\SI{6.569}{\bohr}$.}
  \label{fig:dhmodel}
\end{figure}
%%% %%% %%% %%%

%%% FIG 6 %%%
\begin{figure*}
  \includegraphics[width=0.295\textwidth]{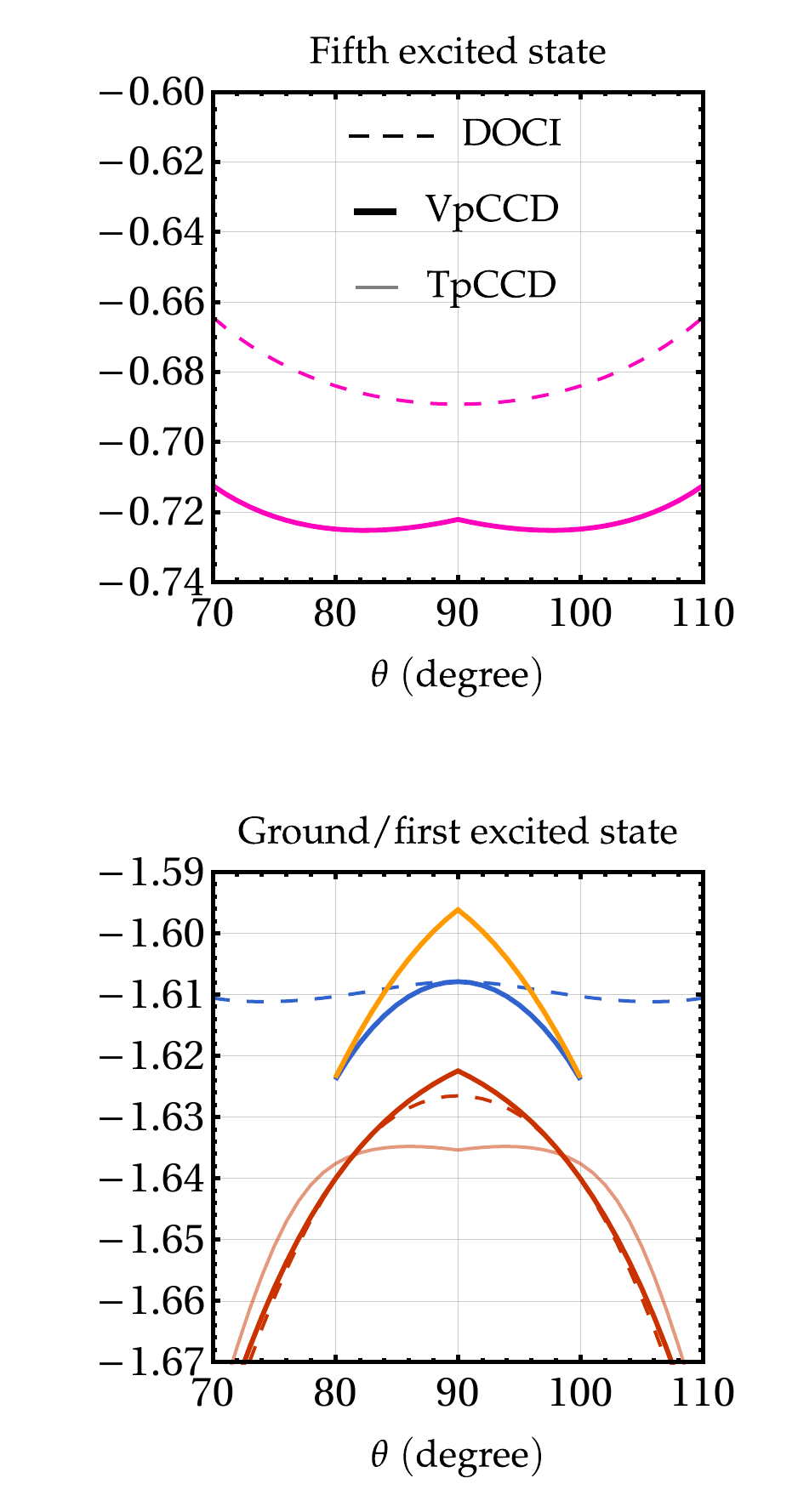}
  \includegraphics[width=0.39\textwidth]{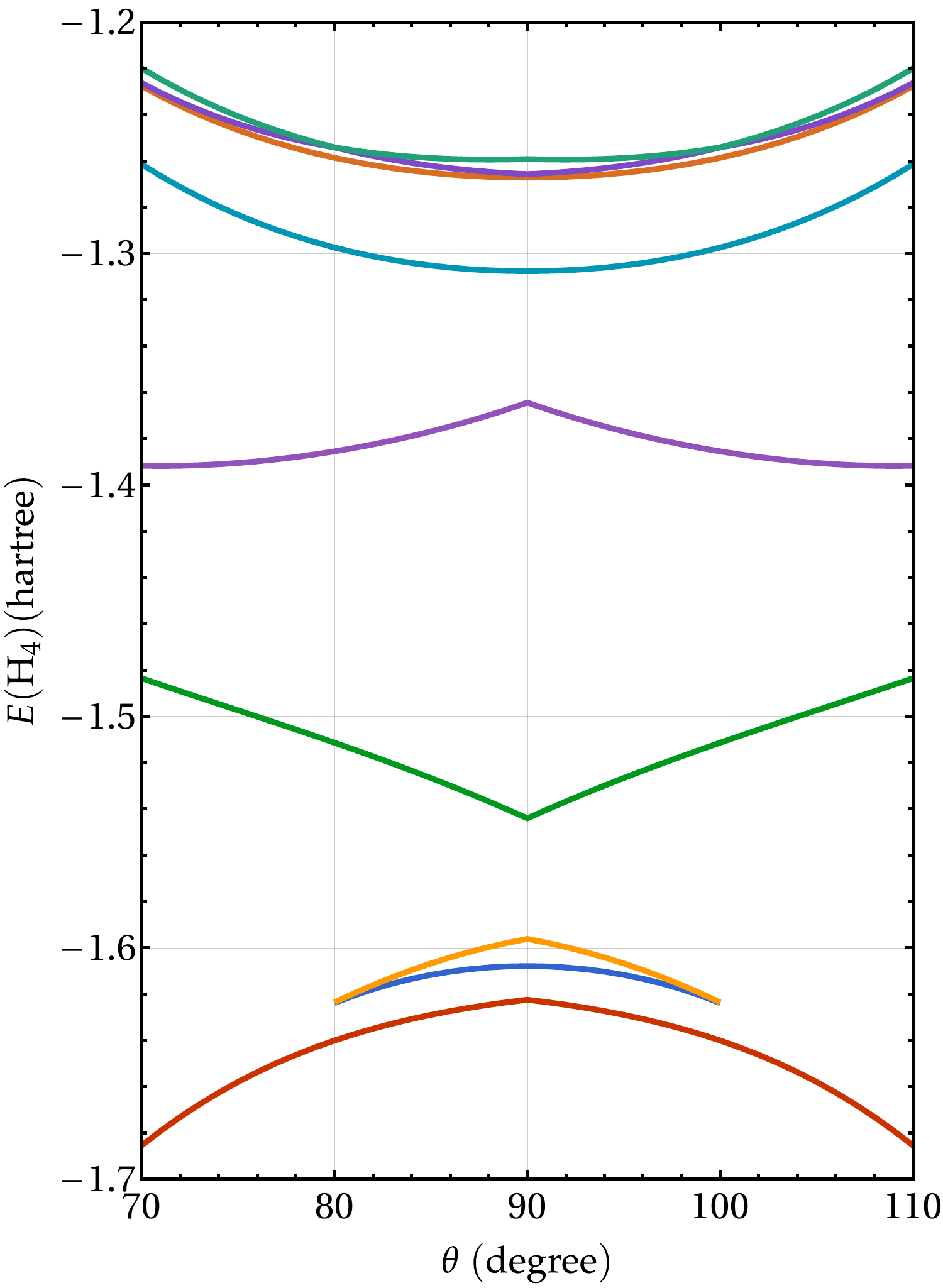}
  \includegraphics[width=0.295\textwidth]{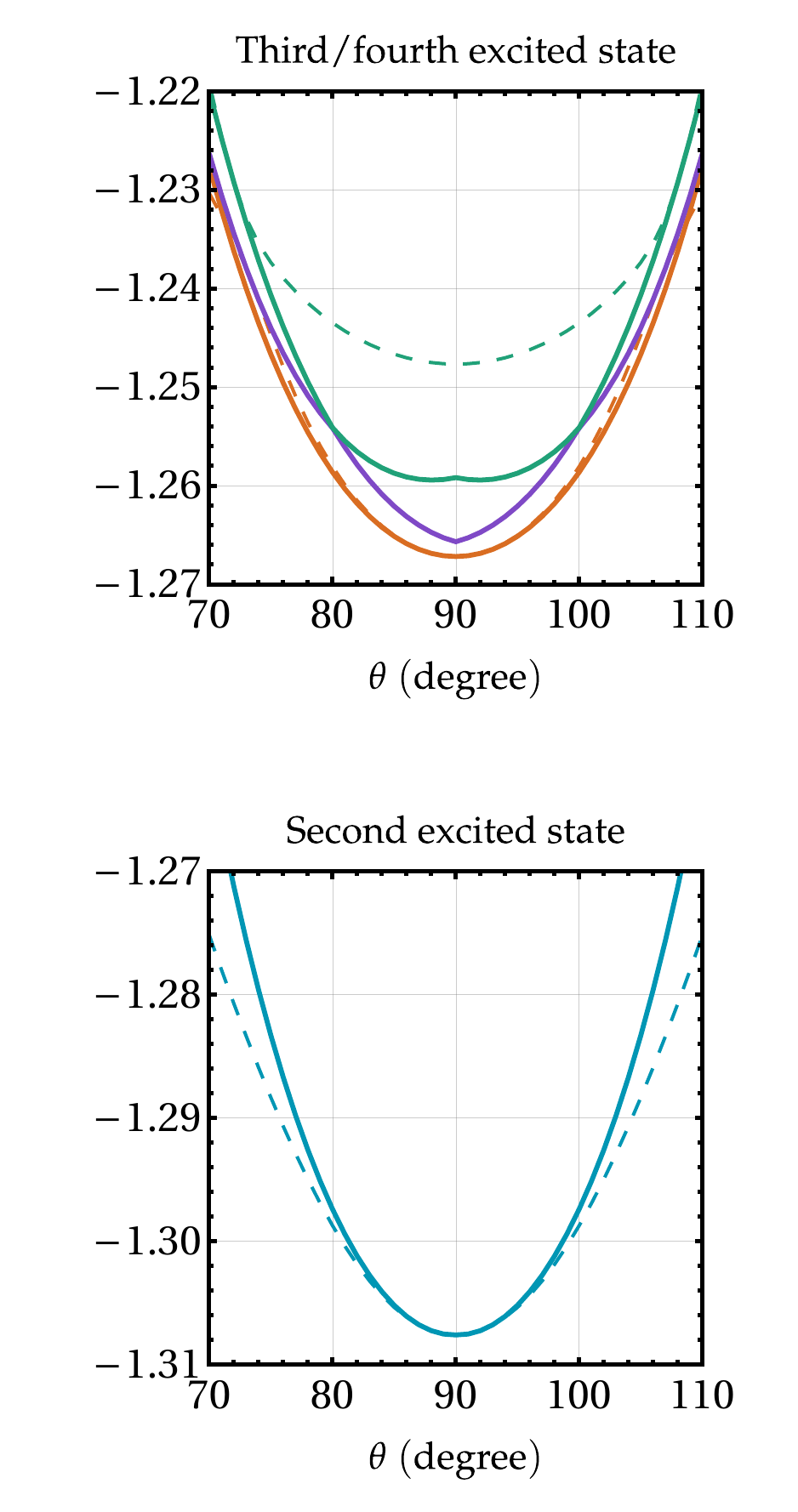}
  \caption{
    Center: Energies (in hartree) of the ring \ce{H4} model in the STO-6G basis set for the VpCCD method as functions of $\theta$ (in degree) using the ground-state RHF determinant of configuration  ($a_{1g}^2b_{2u}^2$) as reference (symmetry-adapted orbitals are considered).
  Bottom-left and bottom-right: VpCCD (thick solid), TpCCD (thin solid), and DOCI (dashed) energies of the ground and first two excited states. 
  Top-right and left panels provide the energies of the quadruply-excited state and of the highest-lying doubly-excited states, respectively.}
    \label{fig:D2hH4}
\end{figure*}
%%% %%% %%% %%%

We now turn our attention to another widely known model for strong correlation, namely the ring \ce{H4} molecule where the four hydrogen atoms lie on a circle of diameter $d = \SI{6.569}{\bohr}$.
As represented in Fig.~\ref{fig:dhmodel}, varying the angle $\theta$ with respect to $\theta = \SI{90}{\degree}$ connects two equivalent $D_{2h}$ rectangular geometries with non-degenerate molecular orbitals. 
At $\theta = \SI{90}{\degree}$ though, the $D_{4h}$ square-planar geometry has strictly degenerate orbitals and strong multi-reference effects are at play. 
Therefore, giving an accurate description of this system as a function of $\theta$ has been shown to be a real challenge for CC methods. \cite{VanVoorhis_2000,Robinson_2012a,Robinson_2012b,Robinson_2012c,Kats_2013,Limacher_2014,Burton_2016,Qiu_2017}

In the following, we restrict ourselves to the minimal STO-6G basis set in which the $D_{2h}$ symmetry-adapted molecular orbitals are determined solely by symmetry.
The resulting four molecular orbitals, ordered by ascending energy, have the symmetry $a_{1g}$, $b_{2u}$, $b_{3u}$, and $b_{1g}$.
At $\theta=\SI{90}{\degree}$, the $b_{2u}$ and $b_{3u}$ orbitals are degenerate and form a pair of orbitals of $e_{g}$ symmetry in the $D_{4h}$ point group.
Although one can choose to break spatial symmetry to gain flexibility, in a first stage, we restrict ourselves to the symmetry-adapted RHF molecular orbitals.
In such situation, excited-state RHF wave functions correspond to non-Aufbau determinants built with this set of symmetry-pure orbitals, hence freeing ourselves from the orbital optimization issue to focus solely on the optimization of the cluster amplitudes.
Because we deal with the seniority-zero subspace, the RHF determinants are made, by definition, of two doubly-occupied orbitals. 
For example, for the ground-state determinant at $\theta=\SI{90}{\degree}$, the lowest-energy $a_{1g}$ orbital and one of the doubly-degenerate $b_{2u}$ and $b_{3u}$ orbitals are doubly-occupied.
Of course, in this case, the seniority-zero determinant is a poor approximation of the exact wave function as it tries to model an inherently multi-reference wave function with a single Slater determinant.

We start by looking at the description of the ground state using the symmetry-adapted orbitals. 
The DOCI (dashed lines), VpCCD (thick solid lines), and TpCCD (thin solid lines) energies are plotted in the bottom-left panel of Fig.~\ref{fig:D2hH4}.
The configuration of the reference determinant is $a_{1g}^2b_{2u}^2$ for $\theta < \SI{90}{\degree}$ and $a_{1g}^2b_{3u}^2$ for $\theta > \SI{90}{\degree}$.
Hereafter, the electronic configuration of the RHF wave function is given for $\theta < \SI{90}{\degree}$; the corresponding configuration for $\theta > \SI{90}{\degree}$ is obtained by simply swapping $b_{2u}$ and $b_{3u}$.
The first interesting fact is that TpCCD does not closely match DOCI for this system.
On the other hand, VpCCD provides energies in fairly good agreement with DOCI. 
Therefore, the hermiticity property of VpCCD leads to a notable improvement over TpCCD.
Yet, VpCCD exhibits a cusp at $\theta = \SI{90}{\degree}$ which is not present in DOCI. 
The derivative of the TpCCD PEC is also discontinuous at $\theta =\SI{90}{\degree}$ with an upside-down cusp compared to VpCCD.
The comparison of the two pCCD variants and DOCI for the ground-state PEC provides similar insights to those reported in Ref.~\onlinecite{VanVoorhis_2000} in the case of VCCD, TCCD, and FCI.
At the RHF level, the ground state and the lowest-lying excited state form a conical intersection. 
This is a drawback of the HF approximation as FCI produces an avoided crossing (HF and FCI energies are given in the \hyperlink{SI}{supplementary material}).
In the seniority-zero subspace, the avoided crossing between these two states is not present.
Indeed, as shown in the bottom-left panel of Fig.~\ref{fig:D2hH4}, the DOCI dashed curves are smooth. 
Yet, they do not form an avoided crossing.

Then, we turn to the description of the excited states.
The simplicity of the ring \ce{H4} model in a minimal basis allows us to access the entire set of solutions using the Jenkins-Traub algorithm (see Sec.~\ref{subsec:linearH4}).
All the VpCCD solutions (with real cluster amplitudes) obtained with a ground-state RHF reference made of symmetry-adapted orbitals are represented in the center panel of Fig.~\ref{fig:D2hH4}, except for the quadruply-excited state which is plotted in the top-left panel.
The convergence towards the ground state and the quadruply-excited state is fairly straightforward using the Newton-Raphson algorithm presented in Sec.~\ref{sec:ES}.
However, the VpCCD solutions represented by the blue and cyan curves are the only two other solutions that we have been able to get for all $\theta$ values using the Newton-Raphson algorithm.
In addition, we have obtained some parts of the three highest VpCCD PECs of Fig.~\ref{fig:D2hH4}, but the iterative algorithm was highly oscillatory and we have not been able to get any of these solutions for all values of $\theta$.

The agreement between the VpCCD and DOCI excited states is less evident than for the linear \ce{H4} model studied in Sec.~\ref{subsec:linearH4}.
The first important point to mention here is that there are more VpCCD than DOCI solutions, for all values of $\theta$.
More importantly, as we shall discuss below, it is challenging to tell which of these solutions are unphysical. 
We believe that three VpCCD solutions can be assigned to DOCI states with certainty: the ground state as well as the pink (quadruply-) and cyan (doubly-)excited states.
However, even if the VpCCD solution corresponding to the quadruply-excited state (top-left panel) is attainable for all $\theta$, it is a poor approximation to its DOCI counterpart, exhibiting a cusp and a local maxima at $\theta=\SI{90}{\degree}$.
Meanwhile, the two highest-lying DOCI doubly-excited states could correspond to some of the three VpCCD solutions (see top-right panel of Fig.~\ref{fig:D2hH4}). 
One could argue that the brown curve should be associated with the dashed brown curve, but for the two other VpCCD solutions it is hard to tell which one is unphysical.
Finally, one can see in the bottom-left panel of Fig.~\ref{fig:D2hH4} that the lowest-lying doubly-excited state could be associated with two VpCCD solutions.
However, these solutions eventually disappear around $\theta=\SI{80}{\degree}$ and $\theta=\SI{100}{\degree}$.
Similarly to the spurious solutions in the linear \ce{H4} model, it is possible that beyond this region the two solutions acquire complex-valued cluster amplitudes.
Even for the part of the PECs where the solutions are real, the agreement with DOCI is quite poor (except around $\theta=\SI{90}{\degree}$ for the lower VpCCD solution), the PECs having the wrong topology.
Alternatively, one could argue that the green VpCCD solution is associated with the lowest-energy DOCI excited state because it has the same topology as the first RHF excited state.
Moreover, this solution exists for all geometry in contrast with the blue and yellow ones.
We think that, at this stage, it would be arbitrary to assign a particular VpCCD solution to this DOCI state.

%%% FIG 7 %%%
\begin{figure}
  \includegraphics[width=\linewidth]{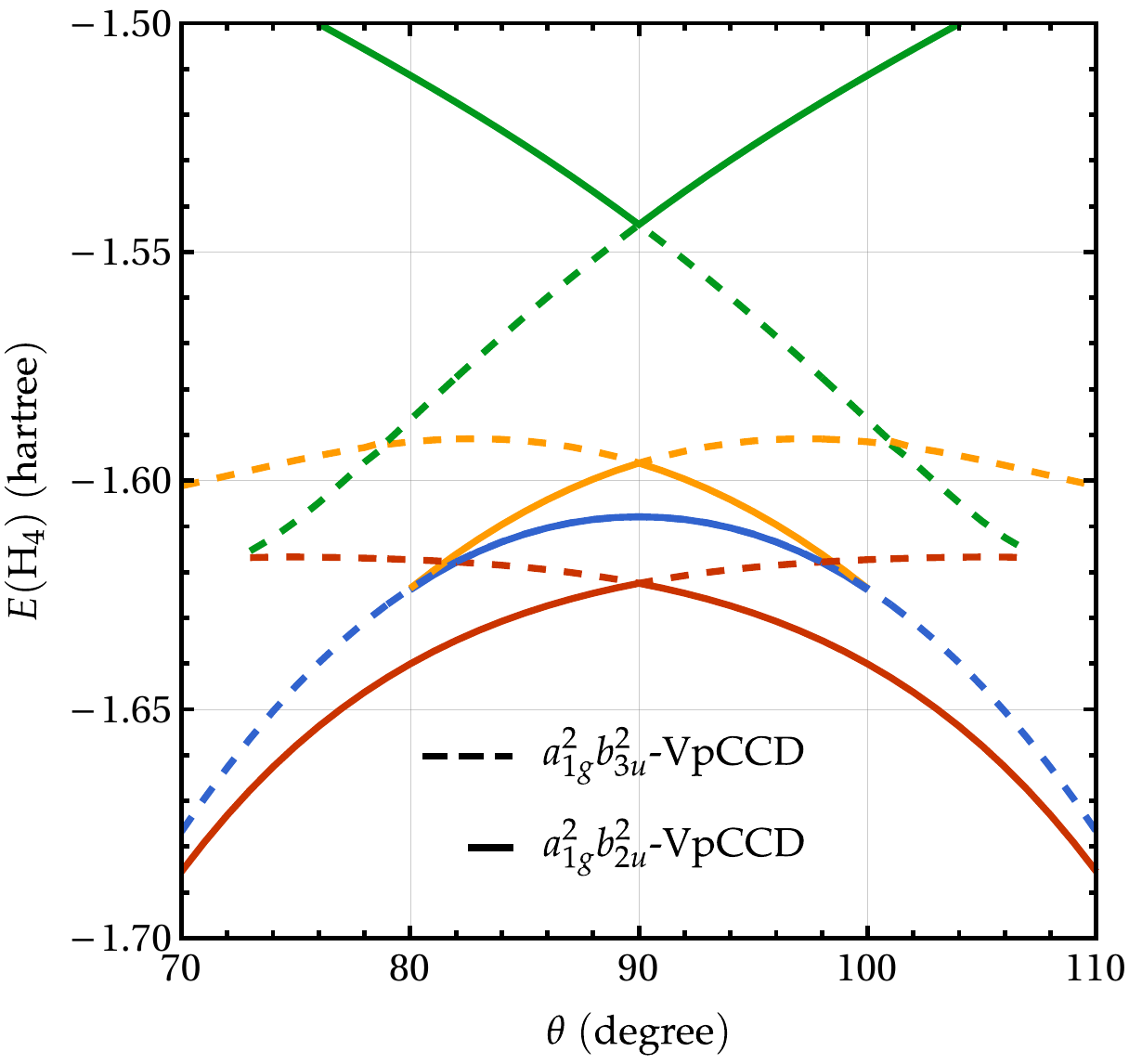}
  \caption{Energies (in hartree) of the ring \ce{H4} model in the STO-6G basis set as functions of $\theta$ (in degree) for the four lowest VpCCD solutions obtained with two symmetry-pure RHF references: the ground-state determinant of configuration $a_{1g}^2 b_{2u}^2$ and the lowest excited-state determinant of configuration $a_{1g}^2 b_{3u}^2$. 
  Note that the dashed lines are used only for readability.}
  \label{fig:MOMring}
\end{figure}
%%% %%% %%% %%%

We now compare the previous set of VpCCD solutions with the ones obtained using non-Aufbau reference determinants made of the same set of symmetry-adapted orbitals.
More specifically, we consider the lowest-lying RHF excited state of configuration $a_{1g}^2b_{3u}^2$, \ie, the other adiabatic state involved in the conical intersection with the $a_{1g}^2b_{2u}^2$ RHF ground state (see the \hyperlink{SI}{supplementary material}).
At $\theta=\SI{90}{\degree}$, these two configurations become degenerate, and the choice of the $e_{g}$ orbital to occupy is arbitrary. 
Therefore, it seems logic to compare these two closely related references as function of $\theta$. 
This may shed light on the meaning of the VpCCD solutions observed in Fig.~\ref{fig:D2hH4}.
The four lowest VpCCD solutions of Fig.~\ref{fig:D2hH4}, \ie, the solutions obtained using the ground-state RHF determinant of configuration $a_{1g}^2b_{2u}^2$ as reference, are also reported in Fig.~\ref{fig:MOMring} alongside the four lowest solutions obtained using the lowest-lying RHF excited state ($a_{1g}^2b_{3u}^2$) as reference.
One can see that three $a_{1g}^2b_{3u}^2$-VpCCD solutions (dashed) are connected with three $a_{1g}^2b_{2u}^2$-VpCCD solutions (solid) at $\theta=\SI{90}{\degree}$, 
while the remaining pair of solutions coincide between $\theta=\SI{80}{\degree}$ and $\theta=\SI{100}{\degree}$. 
For other $\theta$ values, the $a_{1g}^2b_{2u}^2$-VpCCD solution disappears.

As already mentioned earlier, the two lowest diabatic RHF states, $a_{1g}^2b_{2u}^2$ and $a_{1g}^2b_{3u}^2$, intersect at $\theta = \SI{90}{\degree}$.
Hence, the lowest adiabatic RHF state has a cusp at $\theta = \SI{90}{\degree}$.
Cusps observed in ground-state CC PECs are often claimed to be unphysical. \cite{VanVoorhis_2000,Bulik_2015}
However, it has been pointed out by Burton and Thom that these cusps are not unphysical but are consequences of the RHF reference used to construct the corresponding CC wave functions. \cite{Burton_2016}
They further argued that these cusps indicate crossing of solutions at the CC level. 
This is indeed what we observe in Fig.~\ref{fig:MOMring} where the cusps on the VpCCD PECs are actually formed by two VpCCD solutions obtained with distinct reference RHF wave functions.
In short, the inherent single-reference character of pCCD prevents it from correctly describing the FCI avoided crossing.
On the other hand, a non-orthogonal CI (which is inherently multi-reference) between the two RHF states reproduces the correct shape of the PEC. \cite{Burton_2016}
We should also mention that the projected CC method introduced by Qiu \textit{et al.}, in which one constructs a CCSD wave function on top of a projected HF reference, does not exhibit such a cusp. \cite{Qiu_2017}

%%% FIG 8 %%%
\begin{figure*}
  \includegraphics[width=0.48\textwidth]{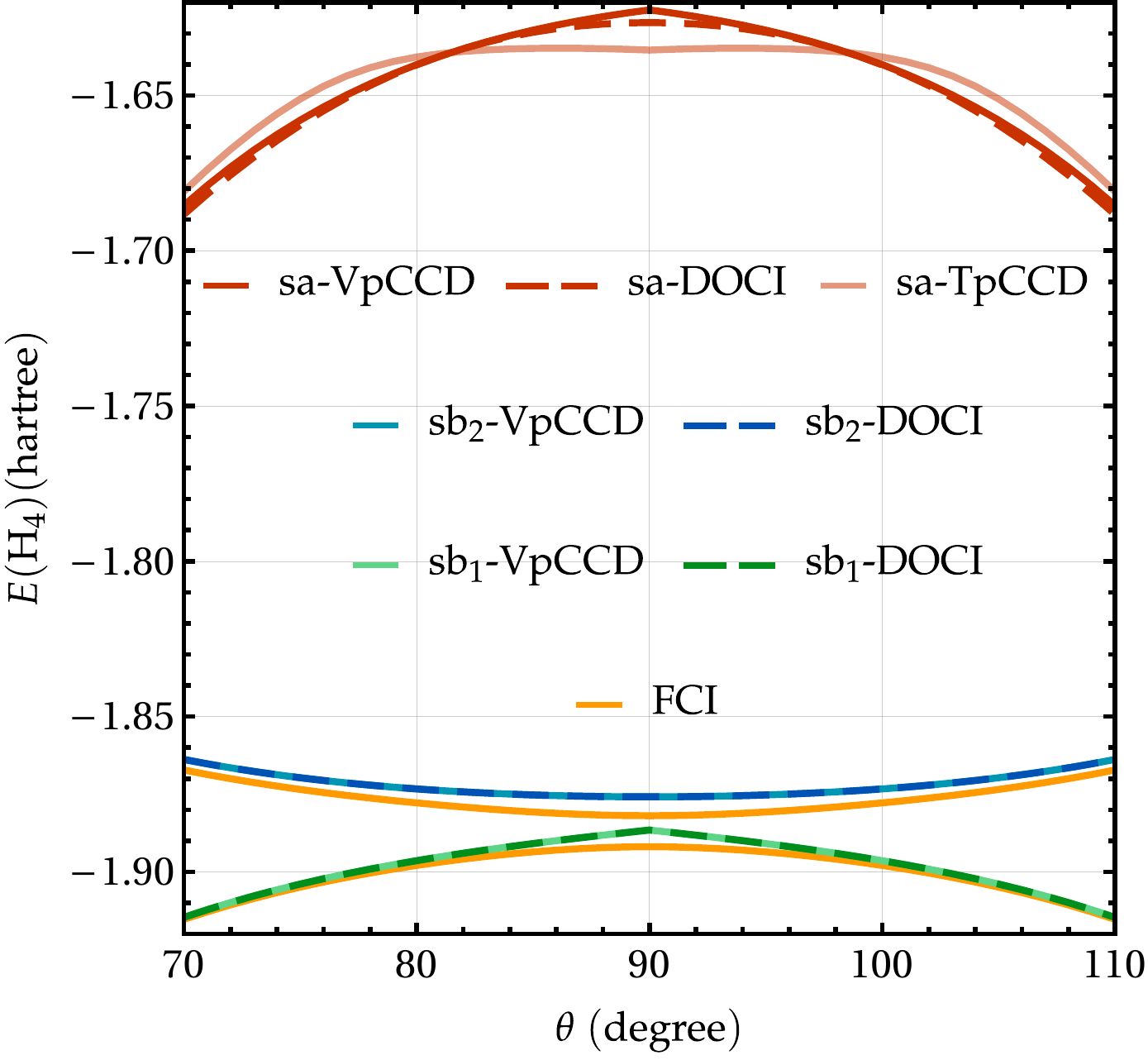}
  \includegraphics[width=0.5\textwidth]{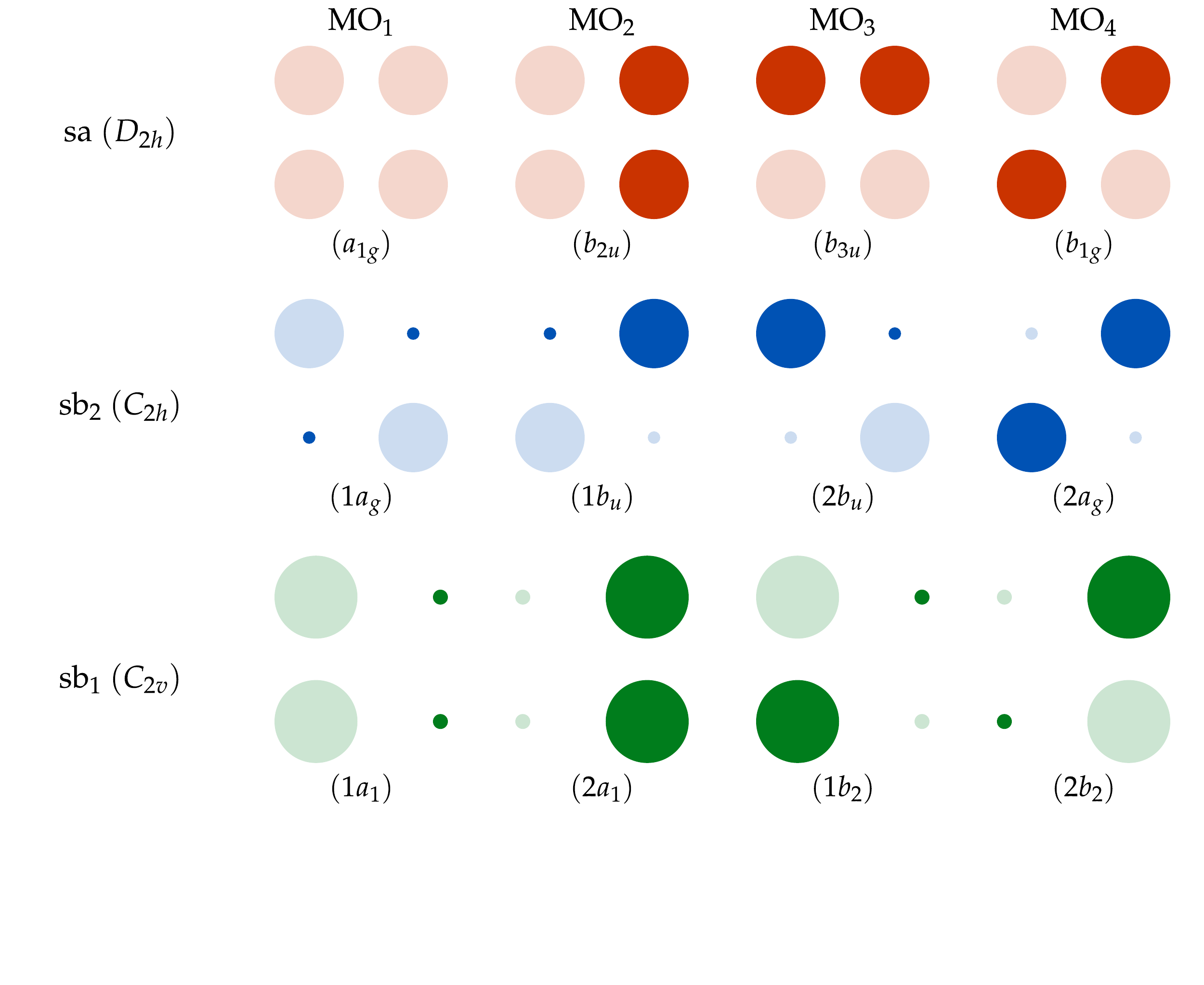}
  \caption{Left: Energies (in hartree) of the ring \ce{H4} model in the STO-6G basis set at various correlated levels (VpCCD, TpCCD, and DOCI) and for various orbital sets (see right panel) as functions of $\theta$ (in degree).
    Right: Orbital representation of the set of symmetry-adapted (sa) orbitals and the two sets of symmetry-broken (sb) orbitals.
    For each set of orbitals, the reference determinant is built from the two leftmost orbitals. 
    The irreducible representation of each orbital in the corresponding point group of the electron density is given in parenthesis.}
  \label{fig:sbring}
\end{figure*}
%%% %%% %%% %%%

As stated earlier, the $D_{2h}$ molecular orbitals are fully determined by the spatial symmetry of the system.
The corresponding set of symmetry-adapted (sa) orbitals is represented in Fig.~\ref{fig:sbring} and ordered by ascending energies.
However, one may wonder if there exists solutions associated with (spatial) symmetry-broken orbital sets.
A stability analysis in the space of real RHF solutions reveals that the symmetry-pure ground-state RHF solution is a minimum with respect to occupied-virtual rotations. \cite{Seeger_1977,Fukutome_1981,Stuber_2003}
Thus, there is, a priori, no spatially symmetry-broken RHF solution lower in energy.

Next, we study the influence of orbital rotations on the VpCCD energy for the ground state of the ring \ce{H4} model.
The diagonalization of the orbital Hessian [see Eq.~\eqref{eq:orbGradHess}] associated with this solution shows that this stationary point is an index-2 saddle point.
Therefore, there is at least one additional state below the sa-VpCCD ground-state PEC.
Indeed, by following the direction provided by the eigenvectors associated with the two distinct negative eigenvalues, we have been able to locate two additional VpCCD solutions.
The first one, labeled as ``sb$_{1}$'' (symmetry-broken) in Fig.~\ref{fig:sbring}, results from occupied-occupied and virtual-virtual rotations.
Let us recall that, although the HF energy is invariant under occupied-occupied and virtual-virtual rotations, it is not the case for pCCD as the seniority-zero subspace depends on the orbital basis used to define it. \cite{Bytautas_2011}
The direction associated with the second negative eigenvalue involves occupied-virtual rotations.
Going downhill following the eigenvector associated with this eigenvalue leads to a different spatially symmetry-broken solution (see the ``sb$_{2}$'' orbital set in Fig.~\ref{fig:sbring}).
The point group of the electron density associated with each solution and the irreducible representation of each orbital are also given in the right panel of Fig.~\ref{fig:sbring}. 
A stability analysis of these two additional solutions shows that they are minima with respect to orbital rotations.
Note that, for each set of symmetry-broken orbitals, four of the six possible reference determinants yield the same correlated energies.
The two other references, $1a_{1}^21b_{2}^2$ and $2a_{1}^22b_{2}^2$ for the sb$_{1}$ set and $1a_{g}^22b_{u}^2$ and $1b_{u}^22a_{g}^2$ for the sb$_{2}$ set, yield energies close to the quadruply-excited state obtained with symmetry-adapted orbitals.

The left panel of Fig.~\ref{fig:sbring} shows that the agreement between VpCCD and DOCI is much better when one considers the two sets of symmetry-broken orbitals. 
In addition, we emphasize that the sb$_{1}$-DOCI/VpCCD PECs (which exhibit a cusp at $\theta = \SI{90}{\degree}$) are fairly good approximations of the ground-state FCI PEC while containing only seniority-zero determinants.
Likewise, the cuspless sb$_{2}$-DOCI/VpCCD PECs are close in energy to the lowest-lying FCI excited-state one.
The downside is that the corresponding wave functions do not possess the correct spatial symmetry.
This is the famous L\"owdin symmetry dilemma. \cite{Lowdin_1963,Lykos_1963,Lowdin_1969}
Moreover, for these two orbital sets, the TpCCD energies are in good agreement with DOCI (see the \hyperlink{SI}{supplementary material}). 
Yet, VpCCD is closer to DOCI than TpCCD as already seen in the linear \ce{H4} case [see Sec.~\ref{subsec:linearH4}].
The cusps exhibited by the sb$_1$-DOCI/VpCCD PECs are also due to a crossing between two diabatic states.
More specifically, it originates from a change of axis along which the $D_{2h}$ symmetry breaks, leading to two different $C_{2v}$ subgroups related by a rotation of $\pi/2$.
Unfortunately, we have not been able to converge to the higher-lying symmetry-broken $C_{2v}$ state.

One would have noticed that we have not plotted the excited-state TpCCD energies in Fig.~\ref{fig:D2hH4}.
In fact, TpCCD suffers from the same issues related to additional solutions and their physical meaning.
Similarly to VpCCD, projection on non-Aufbau references leads to moderate improvements. 
Yet, the TpCCD energy landscapes remain plagued by unphysical solutions.
Consequently, it is hardly possible to assign a TpCCD solution to a given DOCI excited state, as discussed in the case of VpCCD.

Finally, we would like to mention that the improvement of VpCCD/TpCCD brought by state-specific reference wave functions is mitigated in comparison to the case of the linear \ce{H4} molecule.
Therefore, it seems that state-specific (MOM or oo-pCCD) references provide a very significant improvement for weak correlation, but does not help much in the presence of strong correlation.
In such a case, if one is willing to sacrifice the spatial symmetry of the wave function, the description of the ground state (at least) can be improved.
Symmetry-broken excited-state wave functions also exist at the VpCCD level but we have struggled to systematically converge towards these solutions.
Hence, their performance still need to be properly assessed.
This is left for future work.

%%%%%%%%%%%%%%%%%%%%%%%%%%%%%%%%
\section{Conclusion}
\label{sec:ccl}
%%%%%%%%%%%%%%%%%%%%%%%%%%%%%%%%

Recently, there has been a renewed interest in single-reference methods for excited states in the context of Hartree-Fock, density-functional, and coupled-cluster theories.\cite{Gilbert_2008,Thom_2008,Barca_2014,Barca_2018a,Barca_2018b,Mayhall_2010,Lee_2019,Zhao_2016a,Ye_2017,Shea_2018,Thompson_2018,Ye_2019,Tran_2019,Burton_2019c,Zhao_2020,Hait_2020,Hait_2020b,Levi_2020a,Levi_2020b,Dong_2020,Hait_2021,Burton_2021,Kossoski_2021}
This has been made possible thanks to the development of new algorithms specifically designed to target higher-energy solutions of these non-linear equations.
These so-called non-standard solutions provide genuine alternatives to the usual linear response and equation-of-motion formalisms (which are naturally biased towards the reference ground state) for the determination of accurate excited-state energies in molecular systems. 
This is especially true for double excitations which are known to be difficult to model with the two latter formalisms. \cite{Hirata_2000,Sundstrom_2014,Watson_2012,Loos_2018b,Loos_2019c,Loos_2020c,Loos_2020d,Veril_2021}
There is, therefore, a real need for a better understanding of the structure of the energy landscape associated with these methods. 
In this study, we have focused on the case of CC.

Due to the non-linearity of the CC equations, the topology of its energy landscape from which multiple solutions emerge is still far from being thoroughly understood. 
During the last decades though, several groups have been tackling this formidable problem. \cite{Piecuch_2000,Mayhall_2010,Lee_2019,Kossoski_2021,Csirik_2021}
In a recent study, \cite{Kossoski_2021} we have pursued along these lines by investigating the structure of the CC energy surface and the comparison between DOCI and TpCCD for excited states. 
More specifically, we have shown that the agreement which has been observed for ground-state energies \cite{Limacher_2013,Limacher_2014,Henderson_2014a,Henderson_2014b,Henderson_2015,Shepherd_2016} remains in the case of excited states only if one minimizes the TpCCD energy with respect to the orbital coefficients.
In the present study, we have investigated the solution structure of the VCC method, a version of CC where the cluster amplitudes and the energy are determined variationally instead of the usual projective way.
To the best of our knowledge, VCC excited states have never been investigated before.

Restricting ourselves to the case of pCCD (in which the cluster operator includes only pair excitations), we have looked at the VpCCD solution structure of two model systems, namely the linear and ring \ce{H4} molecules, both in the minimal STO-6G basis.
The former system has been used to investigate the influence of the orbital set on the VpCCD and TpCCD energy landscape.
In contrast to TpCCD, VpCCD provides a much better approximation to the DOCI solution structure when one builds the reference determinant with ground-state RHF orbitals, at least in the weak correlation regime. 
When the correlation becomes strong, \ie, when the hydrogen chain is stretched, additional spurious solutions appear due to the truncation of the excitation operator $\hT$,
and VpCCD does not seem to improve with respect to TpCCD.
In either regime, however, these excited-state solutions are hardly attainable using an iterative Newton-Raphson algorithm. 
Therefore, we replaced the ground-state RHF reference wave function by state-specific excited-state RHF references computed with MOM and targeted the corresponding VpCCD solution for each of them.
We have observed that these state-specific references enlarge the basin of attraction of their associated solution, hence easing the convergence of the Newton-Raphson algorithm towards the targeted VpCCD solution.
In addition, considering state-specific RHF orbitals greatly improves the TpCCD results for excited states. 
However, the difference between TpCCD and DOCI energies remains roughly one order of magnitude larger than the one between VpCCD and DOCI.

Then, we have turned our attention to the situation where the reference orbitals are optimized at the correlated level.
In the weak correlation regime, the agreement between DOCI and the two variants of pCCD (TpCCD and VpCCD) is only slightly better than with MOM orbitals. 
However, in the strong correlation regime, the orbital optimization procedure allows the orbitals to localize further (while keeping their spatial symmetry), hence improving the accuracy of VpCCD and TpCCD (with respect to DOCI) at large internuclear separation.
The take-home message of this first part is that TpCCD energies computed with state-specific RHF orbitals provide a good balance between robustness and computational cost to describe excited states, at least in the weak correlation regime.
Of course, further studies on real molecules are required to assess the accuracy of these methods. 

In a second stage, we have studied the ring \ce{H4} molecule to investigate the influence of strong correlation on the energy landscape. 
We have seen that spurious VpCCD solutions, due to the truncation of the cluster operator, seem unavoidable in the presence of strong static correlation.
Therefore, the description of excited states is much less accurate than in the weak correlation regime.
Even worse, these spurious solutions prevent an unambiguous assignment of (some of) the excited states.
This problem remains if one considers state-specific references at the VpCCD level.
TpCCD suffers from the same issues, but in a more severe way.
In addition, inspired by Burton and Thom, \cite{Burton_2016} we have investigated the physical origin of the cusps of the PEC at the VpCCD level.
In agreement with Burton and Thom, \cite{Burton_2016} we have shown that, at the VpCCD level, these cusps are due to crossing of diabatic states obtained with distinct reference determinants.

Finally, we have investigated spatially symmetry-broken VpCCD solutions of the ring \ce{H4} molecule.
In a minimal basis set, the symmetry-adapted molecular orbitals are completely determined by the $D_{2h}$ symmetry of the system.
Yet, one can deliberately break this symmetry to relax the constraints imposed on the molecular orbitals.
Doing so, we have shown that it is possible to locate two symmetry-broken VpCCD ground-state wave functions with energies in much better agreement with FCI, improving in the process the agreement between DOCI and VpCCD.

%%%%%%%%%%%%%%%%%%%%%%%%%%%%%%%%
\section*{Supplementary Material}
%%%%%%%%%%%%%%%%%%%%%%%%%%%%%%%%
Included in the supplementary material are a standalone \textsc{mathematica} notebook gathering modules for the computational methods investigated here (HF, MOM, TpCCD, VpCCD, DOCI, and FCI), raw data for each figure, orbitals obtained at various levels of theory, and files containing the one- and two-electron integrals for the systems studied in the present manuscript.

%%%%%%%%%%%%%%%%%%%%%%%%%%%%%%%%
\begin{acknowledgements}
The authors thank Hugh G.~A.~Burton for insightful discussions on the energy landscape of coupled-cluster methods.
This project has received funding from the European Research Council (ERC) under the European Union's Horizon 2020 research and innovation programme (Grant agreement No.~863481).
\end{acknowledgements}
%%%%%%%%%%%%%%%%%%%%%%%%%%%%%%%%

%%%%%%%%%%%%%%%%%%%%%%%%%%%%%%%%
\section*{Data availability statement}
%%%%%%%%%%%%%%%%%%%%%%%%%%%%%%%%
The \hypertarget{SI}{data} that support the findings of this study are openly available in Zenodo at \href{http://doi.org/10.5281/zenodo.4971904}{http://doi.org/10.5281/zenodo.4971904}.

\appendix

%%%%%%%%%%%%%%%%%%%%%%%%%%%%%%%%%%%%%%
\section{VCCD residual and Jacobian matrices}
\label{app:appendixA}
%%%%%%%%%%%%%%%%%%%%%%%%%%%%%%%%%%%%%%
In this appendix, we provide all the equations required to implement the Newton-Raphson algorithm in order to optimize the VCCD amplitudes.
The corresponding equations for VpCCD are reported in Appendix \ref{app:appendixB}.
For the sake of clarity, hereafter we denote the CC wave function and its energy as $\ket{\PsiCC} \equiv \ket{\Psi}$ and $\EVCC \equiv E$.
In this case, the derivative of the VCC energy functional with respect to the cluster amplitudes reads
\begin{equation}
  \label{eq:derivationVCCampEq}
  \begin{split}
    r_{ij}^{ab} 
    & = \pdv{E(\boldsymbol{t})}{\ta{ij}{ab}} 
    = \pdv{\ta{ij}{ab}}\qty(\frac{\mel{\Psi}{\hH}{\Psi}}{\braket{\Psi}{\Psi}}) 
    \\
    & = \pdv{\mel{\Psi}{\hH}{\Psi}}{\ta{ij}{ab}}\frac{\braket{\Psi}{\Psi}}{\braket{\Psi}{\Psi}^2} - \frac{\mel{\Psi}{\hH}{\Psi}}{\braket{\Psi}{\Psi}^2}\pdv{\braket{\Psi}{\Psi}}{\ta{ij}{ab}} 
    \\
    & = 2\frac{\Re\qty[\mel{\Psi}{\hH \cre{a}\cre{b}\ani{i}\ani{j}}{\Psi}]}{\braket{\Psi}{\Psi}} - 2 E \frac{\Re\qty[\mel{\Psi}{\cre{a}\cre{b}\ani{i}\ani{j}}{\Psi}]}{\braket{\Psi}{\Psi}}
    \\
    &  = 2\frac{\mel{\Psi}{\hH \cre{a}\cre{b}\ani{i}\ani{j}}{\Psi}}{\braket{\Psi}{\Psi}} - 2 E \frac{\mel{\Psi}{\cre{a}\cre{b}\ani{i}\ani{j}}{\Psi}}{\braket{\Psi}{\Psi}},
  \end{split} 
\end{equation}
where we have used the assumptions that both the cluster amplitudes and orbital coefficients are real to simplify the second to last equality.
Then, we differentiate the residual with respect to the cluster amplitudes to obtain the Jacobian matrix elements required to target saddle points. 
This yields the following formula:
\begin{equation}
  \label{eq:matrixElemHessian}
  \begin{split}
    J_{ij,kl}^{ab,cd}
    & = \pdv{r_{ij}^{ab}}{\ta{kl}{cd}} 
    = \pdv{E(\boldsymbol{t})}{\ta{ij}{ab}}{\ta{kl}{cd}} 
    \\
    & = 4\frac{\mel{\Psi}{\hH \cre{a}\cre{b}\ani{i}\ani{j} \cre{c}\cre{d}\ani{k}\ani{l}}{\Psi}}{\braket{\Psi}{\Psi}}
	\\
	& - 4 \frac{\mel{\Psi}{\hH \cre{a}\cre{b}\ani{i}\ani{j}}{\Psi}}{\braket{\Psi}{\Psi}} \frac{\mel{\Psi}{\cre{c}\cre{d}\ani{k}\ani{l}}{\Psi}}{\braket{\Psi}{\Psi}} 
	\\
	& - 4 \frac{\mel{\Psi}{\hH \cre{c}\cre{d}\ani{k}\ani{l}}{\Psi}}{\braket{\Psi}{\Psi}} \frac{\mel{\Psi}{\cre{a}\cre{b}\ani{i}\ani{j}}{\Psi}}{\braket{\Psi}{\Psi}} 
	\\
    & + 8 E \frac{\mel{\Psi}{\cre{c}\cre{d}\ani{k}\ani{l}}{\Psi}}{\braket{\Psi}{\Psi}}\frac{\mel{\Psi}{\cre{a}\cre{b}\ani{i}\ani{j}}{\Psi}}{\braket{\Psi}{\Psi}}
    \\
    & - 4 E \frac{\mel{\Psi}{\cre{a}\cre{b}\ani{i}\ani{j} \cre{c}\cre{d}\ani{k}\ani{l}}{\Psi}}{\braket{\Psi}{\Psi}}.
  \end{split}
\end{equation}

%%%%%%%%%%%%%%%%%%%%%%%%%%%%%%%%%%%%%%
\section{VpCCD residual, Jacobian and density matrices}
\label{app:appendixB}
%%%%%%%%%%%%%%%%%%%%%%%%%%%%%%%%%%%%%%

For the sake of completeness, we also provide the equations in the case of VpCCD. 
These can be easily derived from their VCCD analogs gathered in Appendix \ref{app:appendixA} by setting $j_{\sigma}=i_{\bar{\sigma}}, b_{\sigma}=a_{\bar{\sigma}}, l_{\sigma}=k_{\bar{\sigma}}$, and $d_{\sigma}=c_{\bar{\sigma}}$, where the indexes now refer to spatial orbitals and $\bar{\sigma}$ is the opposite spin of $\sigma$.
Dropping the spin indexes, the matrix elements of the VpCCD residual are 
\begin{equation}
  \label{eq:pair_derivationVpCCDampEq}
    r_{ii}^{aa}
    = 2\frac{\mel{\Psi}{\hH \cre{a}\cre{a}\ani{i}\ani{i}}{\Psi}}{\braket{\Psi}{\Psi}} 
    - 2 E \frac{\mel{\Psi}{\cre{a}\cre{a}\ani{i}\ani{i}}{\Psi}}{\braket{\Psi}{\Psi}},
\end{equation}
and the matrix elements of the VpCCD Jacobian matrix are 
\begin{equation}
  \label{eq:pair_matrixElemHessian}
  \begin{split}
    J_{ii,kk}^{aa,cc} 
    & = 4\frac{\mel{\Psi}{\hH \cre{a}\cre{a}\ani{i}\ani{i} \cre{c}\cre{c}\ani{k}\ani{k}}{\Psi}}{\braket{\Psi}{\Psi}}
	\\
	& - 4\frac{\mel{\Psi}{\hH \cre{a}\cre{a}\ani{i}\ani{i}}{\Psi}}{\braket{\Psi}{\Psi}} \frac{\mel{\Psi}{\cre{c}\cre{c}\ani{k}\ani{k}}{\Psi}}{\braket{\Psi}{\Psi}} 
	\\ 
	& - 4\frac{\mel{\Psi}{\hH \cre{c}\cre{c}\ani{k}\ani{k}}{\Psi}}{\braket{\Psi}{\Psi}} \frac{\mel{\Psi}{\cre{a}\cre{a}\ani{i}\ani{i}}{\Psi}}{\braket{\Psi}{\Psi}}
	\\
    & + 8 E \frac{\mel{\Psi}{\cre{c}\cre{c}\ani{k}\ani{k}}{\Psi}}{\braket{\Psi}{\Psi}}\frac{\mel{\Psi}{\cre{a}\cre{a}\ani{i}\ani{i}}{\Psi}}{\braket{\Psi}{\Psi}}
    \\
    & - 4 E \frac{\mel{\Psi}{\cre{a}\cre{a}\ani{i}\ani{i} \cre{c}\cre{c}\ani{k}\ani{k}}{\Psi}}{\braket{\Psi}{\Psi}}.
  \end{split}
\end{equation}

Following Ref.~\onlinecite{Henderson_2014a}, in the case of pCCD, one can take advantage of the fact that $\ket{\Psi}$ is a linear combination of seniority-zero determinants to compute the one-body density matrix $\boldsymbol{\gamma}$ [see Eq.~\eqref{eq:onebody}] and the two-body density matrix $\boldsymbol{\Gamma}$ [see Eq.~\eqref{eq:twobody}].
Indeed, many elements of $\boldsymbol{\gamma}$ and $\boldsymbol{\Gamma}$ are zero due to seniority considerations. 
Therefore, one only needs to compute the ``diagonal'' elements $\gamma_{pp}$ of the one-body density matrix, while for the two-body density matrix the only non-zero elements are $\Gamma_{pp,qq}$, $\Gamma_{pq,pq}$, and $\Gamma_{pq,qp} = -\frac{1}{2} \Gamma_{pq,pq}$.
  
%%%%%%%%%%%%%%%%%%%%%%%%%%%%%%%% 
\bibliography{VpCCD}

\begin{thebibliography}{182}%
\makeatletter
\providecommand \@ifxundefined [1]{%
 \@ifx{#1\undefined}
}%
\providecommand \@ifnum [1]{%
 \ifnum #1\expandafter \@firstoftwo
 \else \expandafter \@secondoftwo
 \fi
}%
\providecommand \@ifx [1]{%
 \ifx #1\expandafter \@firstoftwo
 \else \expandafter \@secondoftwo
 \fi
}%
\providecommand \natexlab [1]{#1}%
\providecommand \enquote  [1]{``#1''}%
\providecommand \bibnamefont  [1]{#1}%
\providecommand \bibfnamefont [1]{#1}%
\providecommand \citenamefont [1]{#1}%
\providecommand \href@noop [0]{\@secondoftwo}%
\providecommand \href [0]{\begingroup \@sanitize@url \@href}%
\providecommand \@href[1]{\@@startlink{#1}\@@href}%
\providecommand \@@href[1]{\endgroup#1\@@endlink}%
\providecommand \@sanitize@url [0]{\catcode `\\12\catcode `\$12\catcode
  `\&12\catcode `\#12\catcode `\^12\catcode `\_12\catcode `\%12\relax}%
\providecommand \@@startlink[1]{}%
\providecommand \@@endlink[0]{}%
\providecommand \url  [0]{\begingroup\@sanitize@url \@url }%
\providecommand \@url [1]{\endgroup\@href {#1}{\urlprefix }}%
\providecommand \urlprefix  [0]{URL }%
\providecommand \Eprint [0]{\href }%
\providecommand \doibase [0]{http://dx.doi.org/}%
\providecommand \selectlanguage [0]{\@gobble}%
\providecommand \bibinfo  [0]{\@secondoftwo}%
\providecommand \bibfield  [0]{\@secondoftwo}%
\providecommand \translation [1]{[#1]}%
\providecommand \BibitemOpen [0]{}%
\providecommand \bibitemStop [0]{}%
\providecommand \bibitemNoStop [0]{.\EOS\space}%
\providecommand \EOS [0]{\spacefactor3000\relax}%
\providecommand \BibitemShut  [1]{\csname bibitem#1\endcsname}%
\let\auto@bib@innerbib\@empty
%</preamble>
\bibitem [{\citenamefont {{\v C}{\'\i}{\v z}ek}(1966)}]{Cizek_1966}%
  \BibitemOpen
  \bibfield  {author} {\bibinfo {author} {\bibfnamefont {J.}~\bibnamefont {{\v
  C}{\'\i}{\v z}ek}},\ }\href {\doibase 10.1063/1.1727484} {\bibfield
  {journal} {\bibinfo  {journal} {J. Chem. Phys.}\ }\textbf {\bibinfo {volume}
  {45}},\ \bibinfo {pages} {4256} (\bibinfo {year} {1966})}\BibitemShut
  {NoStop}%
\bibitem [{\citenamefont {Paldus}, \citenamefont {\ifmmode \check{C}\else
  \v{C}\fi{}\'{\i}\ifmmode~\check{z}\else \v{z}\fi{}ek},\ and\ \citenamefont
  {Shavitt}(1972)}]{Paldus_1972}%
  \BibitemOpen
  \bibfield  {author} {\bibinfo {author} {\bibfnamefont {J.}~\bibnamefont
  {Paldus}}, \bibinfo {author} {\bibfnamefont {J.}~\bibnamefont {\ifmmode
  \check{C}\else \v{C}\fi{}\'{\i}\ifmmode~\check{z}\else \v{z}\fi{}ek}}, \ and\
  \bibinfo {author} {\bibfnamefont {I.}~\bibnamefont {Shavitt}},\ }\href
  {\doibase 10.1103/PhysRevA.5.50} {\bibfield  {journal} {\bibinfo  {journal}
  {Phys. Rev. A}\ }\textbf {\bibinfo {volume} {5}},\ \bibinfo {pages} {50}
  (\bibinfo {year} {1972})}\BibitemShut {NoStop}%
\bibitem [{\citenamefont {Crawford}\ and\ \citenamefont
  {Schaefer}(2000)}]{Crawford_2000}%
  \BibitemOpen
  \bibfield  {author} {\bibinfo {author} {\bibfnamefont {T.~D.}\ \bibnamefont
  {Crawford}}\ and\ \bibinfo {author} {\bibfnamefont {H.~F.}\ \bibnamefont
  {Schaefer}},\ }in\ \href {\doibase 10.1002/9780470125915.ch2} {\emph
  {\bibinfo {booktitle} {Reviews in {{Computational Chemistry}}}}}\ (\bibinfo
  {publisher} {{John Wiley \& Sons, Ltd}},\ \bibinfo {year} {2000})\ pp.\
  \bibinfo {pages} {33--136}\BibitemShut {NoStop}%
\bibitem [{\citenamefont {Bartlett}\ and\ \citenamefont
  {Musia{\l}}(2007)}]{Bartlett_2007}%
  \BibitemOpen
  \bibfield  {author} {\bibinfo {author} {\bibfnamefont {R.~J.}\ \bibnamefont
  {Bartlett}}\ and\ \bibinfo {author} {\bibfnamefont {M.}~\bibnamefont
  {Musia{\l}}},\ }\href {\doibase 10.1103/RevModPhys.79.291} {\bibfield
  {journal} {\bibinfo  {journal} {Rev. Mod. Phys.}\ }\textbf {\bibinfo {volume}
  {79}},\ \bibinfo {pages} {291} (\bibinfo {year} {2007})}\BibitemShut
  {NoStop}%
\bibitem [{\citenamefont {Shavitt}\ and\ \citenamefont
  {Bartlett}(2009)}]{Shavitt_2009}%
  \BibitemOpen
  \bibfield  {author} {\bibinfo {author} {\bibfnamefont {I.}~\bibnamefont
  {Shavitt}}\ and\ \bibinfo {author} {\bibfnamefont {R.~J.}\ \bibnamefont
  {Bartlett}},\ }\href {\doibase 10.1017/CBO9780511596834} {\emph {\bibinfo
  {title} {Many-{{Body Methods}} in {{Chemistry}} and {{Physics}}: {{MBPT}} and
  {{Coupled}}-{{Cluster Theory}}}}},\ Cambridge {{Molecular Science}}\
  (\bibinfo  {publisher} {{Cambridge University Press}},\ \bibinfo {address}
  {{Cambridge}},\ \bibinfo {year} {2009})\BibitemShut {NoStop}%
\bibitem [{\citenamefont {Purvis}\ and\ \citenamefont
  {Bartlett}(1982)}]{Purvis_1982}%
  \BibitemOpen
  \bibfield  {author} {\bibinfo {author} {\bibfnamefont {G.~D.}\ \bibnamefont
  {Purvis}}\ and\ \bibinfo {author} {\bibfnamefont {R.~J.}\ \bibnamefont
  {Bartlett}},\ }\href {\doibase 10.1063/1.443164} {\bibfield  {journal}
  {\bibinfo  {journal} {J. Chem. Phys.}\ }\textbf {\bibinfo {volume} {76}},\
  \bibinfo {pages} {1910} (\bibinfo {year} {1982})}\BibitemShut {NoStop}%
\bibitem [{\citenamefont {Raghavachari}\ \emph {et~al.}(1989)\citenamefont
  {Raghavachari}, \citenamefont {Trucks}, \citenamefont {Pople},\ and\
  \citenamefont {{Head-Gordon}}}]{Raghavachari_1989}%
  \BibitemOpen
  \bibfield  {author} {\bibinfo {author} {\bibfnamefont {K.}~\bibnamefont
  {Raghavachari}}, \bibinfo {author} {\bibfnamefont {G.~W.}\ \bibnamefont
  {Trucks}}, \bibinfo {author} {\bibfnamefont {J.~A.}\ \bibnamefont {Pople}}, \
  and\ \bibinfo {author} {\bibfnamefont {M.}~\bibnamefont {{Head-Gordon}}},\
  }\href {\doibase 10.1016/S0009-2614(89)87395-6} {\bibfield  {journal}
  {\bibinfo  {journal} {Chem. Phys. Lett.}\ }\textbf {\bibinfo {volume}
  {157}},\ \bibinfo {pages} {479} (\bibinfo {year} {1989})}\BibitemShut
  {NoStop}%
\bibitem [{\citenamefont {Jeziorski}\ and\ \citenamefont
  {Monkhorst}(1981)}]{Jeziorski_1981}%
  \BibitemOpen
  \bibfield  {author} {\bibinfo {author} {\bibfnamefont {B.}~\bibnamefont
  {Jeziorski}}\ and\ \bibinfo {author} {\bibfnamefont {H.~J.}\ \bibnamefont
  {Monkhorst}},\ }\href {\doibase 10.1103/PhysRevA.24.1668} {\bibfield
  {journal} {\bibinfo  {journal} {Phys. Rev. A}\ }\textbf {\bibinfo {volume}
  {24}},\ \bibinfo {pages} {1668} (\bibinfo {year} {1981})}\BibitemShut
  {NoStop}%
\bibitem [{\citenamefont {Mahapatra}, \citenamefont {Datta},\ and\
  \citenamefont {Mukherjee}(1998)}]{Mahapatra_1998}%
  \BibitemOpen
  \bibfield  {author} {\bibinfo {author} {\bibfnamefont {U.~S.}\ \bibnamefont
  {Mahapatra}}, \bibinfo {author} {\bibfnamefont {B.}~\bibnamefont {Datta}}, \
  and\ \bibinfo {author} {\bibfnamefont {D.}~\bibnamefont {Mukherjee}},\ }\href
  {\doibase 10.1080/002689798168448} {\bibfield  {journal} {\bibinfo  {journal}
  {Mol. Phys.}\ }\textbf {\bibinfo {volume} {94}},\ \bibinfo {pages} {157}
  (\bibinfo {year} {1998})}\BibitemShut {NoStop}%
\bibitem [{\citenamefont {Mahapatra}, \citenamefont {Datta},\ and\
  \citenamefont {Mukherjee}(1999)}]{Mahapatra_1999}%
  \BibitemOpen
  \bibfield  {author} {\bibinfo {author} {\bibfnamefont {U.~S.}\ \bibnamefont
  {Mahapatra}}, \bibinfo {author} {\bibfnamefont {B.}~\bibnamefont {Datta}}, \
  and\ \bibinfo {author} {\bibfnamefont {D.}~\bibnamefont {Mukherjee}},\ }\href
  {\doibase 10.1063/1.478523} {\bibfield  {journal} {\bibinfo  {journal} {J.
  Chem. Phys.}\ }\textbf {\bibinfo {volume} {110}},\ \bibinfo {pages} {6171}
  (\bibinfo {year} {1999})}\BibitemShut {NoStop}%
\bibitem [{\citenamefont {Lyakh}\ \emph {et~al.}(2012)\citenamefont {Lyakh},
  \citenamefont {Musia{\l}}, \citenamefont {Lotrich},\ and\ \citenamefont
  {Bartlett}}]{Lyakh_2012}%
  \BibitemOpen
  \bibfield  {author} {\bibinfo {author} {\bibfnamefont {D.~I.}\ \bibnamefont
  {Lyakh}}, \bibinfo {author} {\bibfnamefont {M.}~\bibnamefont {Musia{\l}}},
  \bibinfo {author} {\bibfnamefont {V.~F.}\ \bibnamefont {Lotrich}}, \ and\
  \bibinfo {author} {\bibfnamefont {R.~J.}\ \bibnamefont {Bartlett}},\ }\href
  {\doibase 10.1021/cr2001417} {\bibfield  {journal} {\bibinfo  {journal}
  {Chem. Rev.}\ }\textbf {\bibinfo {volume} {112}},\ \bibinfo {pages} {182}
  (\bibinfo {year} {2012})}\BibitemShut {NoStop}%
\bibitem [{\citenamefont {K{\"o}hn}\ \emph {et~al.}(2013)\citenamefont
  {K{\"o}hn}, \citenamefont {Hanauer}, \citenamefont {M{\"u}ck}, \citenamefont
  {Jagau},\ and\ \citenamefont {Gauss}}]{Kohn_2013}%
  \BibitemOpen
  \bibfield  {author} {\bibinfo {author} {\bibfnamefont {A.}~\bibnamefont
  {K{\"o}hn}}, \bibinfo {author} {\bibfnamefont {M.}~\bibnamefont {Hanauer}},
  \bibinfo {author} {\bibfnamefont {L.~A.}\ \bibnamefont {M{\"u}ck}}, \bibinfo
  {author} {\bibfnamefont {T.-C.}\ \bibnamefont {Jagau}}, \ and\ \bibinfo
  {author} {\bibfnamefont {J.}~\bibnamefont {Gauss}},\ }\href {\doibase
  https://doi.org/10.1002/wcms.1120} {\bibfield  {journal} {\bibinfo  {journal}
  {WIREs Comput. Mol. Sci.}\ }\textbf {\bibinfo {volume} {3}},\ \bibinfo
  {pages} {176} (\bibinfo {year} {2013})}\BibitemShut {NoStop}%
\bibitem [{\citenamefont {Limacher}\ \emph {et~al.}(2013)\citenamefont
  {Limacher}, \citenamefont {Ayers}, \citenamefont {Johnson}, \citenamefont
  {De~Baerdemacker}, \citenamefont {Van~Neck},\ and\ \citenamefont
  {Bultinck}}]{Limacher_2013}%
  \BibitemOpen
  \bibfield  {author} {\bibinfo {author} {\bibfnamefont {P.~A.}\ \bibnamefont
  {Limacher}}, \bibinfo {author} {\bibfnamefont {P.~W.}\ \bibnamefont {Ayers}},
  \bibinfo {author} {\bibfnamefont {P.~A.}\ \bibnamefont {Johnson}}, \bibinfo
  {author} {\bibfnamefont {S.}~\bibnamefont {De~Baerdemacker}}, \bibinfo
  {author} {\bibfnamefont {D.}~\bibnamefont {Van~Neck}}, \ and\ \bibinfo
  {author} {\bibfnamefont {P.}~\bibnamefont {Bultinck}},\ }\href {\doibase
  10.1021/ct300902c} {\bibfield  {journal} {\bibinfo  {journal} {J. Chem.
  Theory Comput.}\ }\textbf {\bibinfo {volume} {9}},\ \bibinfo {pages} {1394}
  (\bibinfo {year} {2013})}\BibitemShut {NoStop}%
\bibitem [{\citenamefont {Limacher}\ \emph {et~al.}(2014)\citenamefont
  {Limacher}, \citenamefont {Kim}, \citenamefont {Ayers}, \citenamefont
  {Johnson}, \citenamefont {Baerdemacker}, \citenamefont {Neck},\ and\
  \citenamefont {Bultinck}}]{Limacher_2014}%
  \BibitemOpen
  \bibfield  {author} {\bibinfo {author} {\bibfnamefont {P.~A.}\ \bibnamefont
  {Limacher}}, \bibinfo {author} {\bibfnamefont {T.~D.}\ \bibnamefont {Kim}},
  \bibinfo {author} {\bibfnamefont {P.~W.}\ \bibnamefont {Ayers}}, \bibinfo
  {author} {\bibfnamefont {P.~A.}\ \bibnamefont {Johnson}}, \bibinfo {author}
  {\bibfnamefont {S.~D.}\ \bibnamefont {Baerdemacker}}, \bibinfo {author}
  {\bibfnamefont {D.~V.}\ \bibnamefont {Neck}}, \ and\ \bibinfo {author}
  {\bibfnamefont {P.}~\bibnamefont {Bultinck}},\ }\href {\doibase
  10.1080/00268976.2013.874600} {\bibfield  {journal} {\bibinfo  {journal}
  {Mol. Phys.}\ }\textbf {\bibinfo {volume} {112}},\ \bibinfo {pages} {853}
  (\bibinfo {year} {2014})}\BibitemShut {NoStop}%
\bibitem [{\citenamefont {Henderson}\ \emph
  {et~al.}(2014{\natexlab{a}})\citenamefont {Henderson}, \citenamefont {Bulik},
  \citenamefont {Stein},\ and\ \citenamefont {Scuseria}}]{Henderson_2014a}%
  \BibitemOpen
  \bibfield  {author} {\bibinfo {author} {\bibfnamefont {T.~M.}\ \bibnamefont
  {Henderson}}, \bibinfo {author} {\bibfnamefont {I.~W.}\ \bibnamefont
  {Bulik}}, \bibinfo {author} {\bibfnamefont {T.}~\bibnamefont {Stein}}, \ and\
  \bibinfo {author} {\bibfnamefont {G.~E.}\ \bibnamefont {Scuseria}},\ }\href
  {\doibase 10.1063/1.4904384} {\bibfield  {journal} {\bibinfo  {journal} {J.
  Chem. Phys.}\ }\textbf {\bibinfo {volume} {141}},\ \bibinfo {pages} {244104}
  (\bibinfo {year} {2014}{\natexlab{a}})}\BibitemShut {NoStop}%
\bibitem [{\citenamefont {Henderson}\ \emph
  {et~al.}(2014{\natexlab{b}})\citenamefont {Henderson}, \citenamefont
  {Scuseria}, \citenamefont {Dukelsky}, \citenamefont {Signoracci},\ and\
  \citenamefont {Duguet}}]{Henderson_2014b}%
  \BibitemOpen
  \bibfield  {author} {\bibinfo {author} {\bibfnamefont {T.~M.}\ \bibnamefont
  {Henderson}}, \bibinfo {author} {\bibfnamefont {G.~E.}\ \bibnamefont
  {Scuseria}}, \bibinfo {author} {\bibfnamefont {J.}~\bibnamefont {Dukelsky}},
  \bibinfo {author} {\bibfnamefont {A.}~\bibnamefont {Signoracci}}, \ and\
  \bibinfo {author} {\bibfnamefont {T.}~\bibnamefont {Duguet}},\ }\href
  {\doibase 10.1103/PhysRevC.89.054305} {\bibfield  {journal} {\bibinfo
  {journal} {Phys. Rev. C}\ }\textbf {\bibinfo {volume} {89}},\ \bibinfo
  {pages} {054305} (\bibinfo {year} {2014}{\natexlab{b}})}\BibitemShut
  {NoStop}%
\bibitem [{\citenamefont {Stein}, \citenamefont {Henderson},\ and\
  \citenamefont {Scuseria}(2014)}]{Stein_2014}%
  \BibitemOpen
  \bibfield  {author} {\bibinfo {author} {\bibfnamefont {T.}~\bibnamefont
  {Stein}}, \bibinfo {author} {\bibfnamefont {T.~M.}\ \bibnamefont
  {Henderson}}, \ and\ \bibinfo {author} {\bibfnamefont {G.~E.}\ \bibnamefont
  {Scuseria}},\ }\href {\doibase 10.1063/1.4880819} {\bibfield  {journal}
  {\bibinfo  {journal} {J. Chem. Phys.}\ }\textbf {\bibinfo {volume} {140}},\
  \bibinfo {pages} {214113} (\bibinfo {year} {2014})}\BibitemShut {NoStop}%
\bibitem [{\citenamefont {Shepherd}, \citenamefont {Henderson},\ and\
  \citenamefont {Scuseria}(2016)}]{Shepherd_2016}%
  \BibitemOpen
  \bibfield  {author} {\bibinfo {author} {\bibfnamefont {J.~J.}\ \bibnamefont
  {Shepherd}}, \bibinfo {author} {\bibfnamefont {T.~M.}\ \bibnamefont
  {Henderson}}, \ and\ \bibinfo {author} {\bibfnamefont {G.~E.}\ \bibnamefont
  {Scuseria}},\ }\href {\doibase 10.1063/1.4942770} {\bibfield  {journal}
  {\bibinfo  {journal} {J. Chem. Phys.}\ }\textbf {\bibinfo {volume} {144}},\
  \bibinfo {pages} {094112} (\bibinfo {year} {2016})}\BibitemShut {NoStop}%
\bibitem [{\citenamefont {Boguslawski}\ and\ \citenamefont
  {Tecmer}(2017)}]{Boguslawski_2017a}%
  \BibitemOpen
  \bibfield  {author} {\bibinfo {author} {\bibfnamefont {K.}~\bibnamefont
  {Boguslawski}}\ and\ \bibinfo {author} {\bibfnamefont {P.}~\bibnamefont
  {Tecmer}},\ }\href {\doibase 10.1021/acs.jctc.6b01134} {\bibfield  {journal}
  {\bibinfo  {journal} {J. Chem. Theory Comput.}\ }\textbf {\bibinfo {volume}
  {13}},\ \bibinfo {pages} {5966} (\bibinfo {year} {2017})}\BibitemShut
  {NoStop}%
\bibitem [{\citenamefont {Boguslawski}(2017)}]{Boguslawski_2017b}%
  \BibitemOpen
  \bibfield  {author} {\bibinfo {author} {\bibfnamefont {K.}~\bibnamefont
  {Boguslawski}},\ }\href {\doibase 10.1063/1.5006124} {\bibfield  {journal}
  {\bibinfo  {journal} {J. Chem. Phys.}\ }\textbf {\bibinfo {volume} {147}},\
  \bibinfo {pages} {139901} (\bibinfo {year} {2017})}\BibitemShut {NoStop}%
\bibitem [{\citenamefont {Johnson}\ \emph {et~al.}(2017)\citenamefont
  {Johnson}, \citenamefont {Limacher}, \citenamefont {Kim}, \citenamefont
  {Richer}, \citenamefont {Miranda-Quintana}, \citenamefont {Heidar-Zadeh},
  \citenamefont {Ayers}, \citenamefont {Bultinck}, \citenamefont {{De
  Baerdemacker}},\ and\ \citenamefont {{Van Neck}}}]{Johnson_2017}%
  \BibitemOpen
  \bibfield  {author} {\bibinfo {author} {\bibfnamefont {P.~A.}\ \bibnamefont
  {Johnson}}, \bibinfo {author} {\bibfnamefont {P.~A.}\ \bibnamefont
  {Limacher}}, \bibinfo {author} {\bibfnamefont {T.~D.}\ \bibnamefont {Kim}},
  \bibinfo {author} {\bibfnamefont {M.}~\bibnamefont {Richer}}, \bibinfo
  {author} {\bibfnamefont {R.~A.}\ \bibnamefont {Miranda-Quintana}}, \bibinfo
  {author} {\bibfnamefont {F.}~\bibnamefont {Heidar-Zadeh}}, \bibinfo {author}
  {\bibfnamefont {P.~W.}\ \bibnamefont {Ayers}}, \bibinfo {author}
  {\bibfnamefont {P.}~\bibnamefont {Bultinck}}, \bibinfo {author}
  {\bibfnamefont {S.}~\bibnamefont {{De Baerdemacker}}}, \ and\ \bibinfo
  {author} {\bibfnamefont {D.}~\bibnamefont {{Van Neck}}},\ }\href {\doibase
  https://doi.org/10.1016/j.comptc.2017.05.010} {\bibfield  {journal} {\bibinfo
   {journal} {Comput. Theor. Chem.}\ }\textbf {\bibinfo {volume} {1116}},\
  \bibinfo {pages} {207} (\bibinfo {year} {2017})}\BibitemShut {NoStop}%
\bibitem [{\citenamefont {Boguslawski}(2019)}]{Boguslawski_2019}%
  \BibitemOpen
  \bibfield  {author} {\bibinfo {author} {\bibfnamefont {K.}~\bibnamefont
  {Boguslawski}},\ }\href {\doibase 10.1021/acs.jctc.8b01053} {\bibfield
  {journal} {\bibinfo  {journal} {J. Chem. Theory Comput.}\ }\textbf {\bibinfo
  {volume} {15}},\ \bibinfo {pages} {18} (\bibinfo {year} {2019})}\BibitemShut
  {NoStop}%
\bibitem [{\citenamefont {Bulik}, \citenamefont {Henderson},\ and\
  \citenamefont {Scuseria}(2015)}]{Bulik_2015}%
  \BibitemOpen
  \bibfield  {author} {\bibinfo {author} {\bibfnamefont {I.~W.}\ \bibnamefont
  {Bulik}}, \bibinfo {author} {\bibfnamefont {T.~M.}\ \bibnamefont
  {Henderson}}, \ and\ \bibinfo {author} {\bibfnamefont {G.~E.}\ \bibnamefont
  {Scuseria}},\ }\href {\doibase 10.1021/acs.jctc.5b00422} {\bibfield
  {journal} {\bibinfo  {journal} {J. Chem. Theory Comput.}\ }\textbf {\bibinfo
  {volume} {11}},\ \bibinfo {pages} {3171} (\bibinfo {year}
  {2015})}\BibitemShut {NoStop}%
\bibitem [{\citenamefont {Gomez}, \citenamefont {Henderson},\ and\
  \citenamefont {Scuseria}(2016)}]{Gomez_2016}%
  \BibitemOpen
  \bibfield  {author} {\bibinfo {author} {\bibfnamefont {J.~A.}\ \bibnamefont
  {Gomez}}, \bibinfo {author} {\bibfnamefont {T.~M.}\ \bibnamefont
  {Henderson}}, \ and\ \bibinfo {author} {\bibfnamefont {G.~E.}\ \bibnamefont
  {Scuseria}},\ }\href {\doibase 10.1063/1.4954891} {\bibfield  {journal}
  {\bibinfo  {journal} {J. Chem. Phys.}\ }\textbf {\bibinfo {volume} {144}},\
  \bibinfo {pages} {244117} (\bibinfo {year} {2016})}\BibitemShut {NoStop}%
\bibitem [{\citenamefont {Kats}\ and\ \citenamefont {Manby}(2013)}]{Kats_2013}%
  \BibitemOpen
  \bibfield  {author} {\bibinfo {author} {\bibfnamefont {D.}~\bibnamefont
  {Kats}}\ and\ \bibinfo {author} {\bibfnamefont {F.~R.}\ \bibnamefont
  {Manby}},\ }\href {\doibase 10.1063/1.4813481} {\bibfield  {journal}
  {\bibinfo  {journal} {J. Chem. Phys.}\ }\textbf {\bibinfo {volume} {139}},\
  \bibinfo {pages} {021102} (\bibinfo {year} {2013})}\BibitemShut {NoStop}%
\bibitem [{\citenamefont {Kats}(2014)}]{Kats_2014}%
  \BibitemOpen
  \bibfield  {author} {\bibinfo {author} {\bibfnamefont {D.}~\bibnamefont
  {Kats}},\ }\href {\doibase 10.1063/1.4892792} {\bibfield  {journal} {\bibinfo
   {journal} {J. Chem. Phys.}\ }\textbf {\bibinfo {volume} {141}},\ \bibinfo
  {pages} {061101} (\bibinfo {year} {2014})}\BibitemShut {NoStop}%
\bibitem [{\citenamefont {Kats}\ \emph {et~al.}(2015)\citenamefont {Kats},
  \citenamefont {Kreplin}, \citenamefont {Werner},\ and\ \citenamefont
  {Manby}}]{Kats_2015}%
  \BibitemOpen
  \bibfield  {author} {\bibinfo {author} {\bibfnamefont {D.}~\bibnamefont
  {Kats}}, \bibinfo {author} {\bibfnamefont {D.}~\bibnamefont {Kreplin}},
  \bibinfo {author} {\bibfnamefont {H.-J.}\ \bibnamefont {Werner}}, \ and\
  \bibinfo {author} {\bibfnamefont {F.~R.}\ \bibnamefont {Manby}},\ }\href
  {\doibase 10.1063/1.4907591} {\bibfield  {journal} {\bibinfo  {journal} {J.
  Chem. Phys.}\ }\textbf {\bibinfo {volume} {142}},\ \bibinfo {pages} {064111}
  (\bibinfo {year} {2015})}\BibitemShut {NoStop}%
\bibitem [{\citenamefont {Kats}(2016)}]{Kats_2016}%
  \BibitemOpen
  \bibfield  {author} {\bibinfo {author} {\bibfnamefont {D.}~\bibnamefont
  {Kats}},\ }\href {\doibase 10.1063/1.4940398} {\bibfield  {journal} {\bibinfo
   {journal} {J. Chem. Phys.}\ }\textbf {\bibinfo {volume} {144}},\ \bibinfo
  {pages} {044102} (\bibinfo {year} {2016})}\BibitemShut {NoStop}%
\bibitem [{\citenamefont {Kats}(2018)}]{Kats_2018}%
  \BibitemOpen
  \bibfield  {author} {\bibinfo {author} {\bibfnamefont {D.}~\bibnamefont
  {Kats}},\ }\href {\doibase 10.1080/00268976.2017.1417646} {\bibfield
  {journal} {\bibinfo  {journal} {Mol. Phys.}\ }\textbf {\bibinfo {volume}
  {116}},\ \bibinfo {pages} {1435} (\bibinfo {year} {2018})}\BibitemShut
  {NoStop}%
\bibitem [{\citenamefont {Kats}\ and\ \citenamefont
  {K{\"o}hn}(2019)}]{Kats_2019}%
  \BibitemOpen
  \bibfield  {author} {\bibinfo {author} {\bibfnamefont {D.}~\bibnamefont
  {Kats}}\ and\ \bibinfo {author} {\bibfnamefont {A.}~\bibnamefont
  {K{\"o}hn}},\ }\href {\doibase 10.1063/1.5096343} {\bibfield  {journal}
  {\bibinfo  {journal} {J. Chem. Phys.}\ }\textbf {\bibinfo {volume} {150}},\
  \bibinfo {pages} {151101} (\bibinfo {year} {2019})}\BibitemShut {NoStop}%
\bibitem [{\citenamefont {Kats}\ and\ \citenamefont {Tew}(2019)}]{Kats_2019a}%
  \BibitemOpen
  \bibfield  {author} {\bibinfo {author} {\bibfnamefont {D.}~\bibnamefont
  {Kats}}\ and\ \bibinfo {author} {\bibfnamefont {D.~P.}\ \bibnamefont {Tew}},\
  }\href {\doibase 10.1021/acs.jctc.8b01047} {\bibfield  {journal} {\bibinfo
  {journal} {J. Chem. Theory Comput.}\ }\textbf {\bibinfo {volume} {15}},\
  \bibinfo {pages} {13} (\bibinfo {year} {2019})}\BibitemShut {NoStop}%
\bibitem [{\citenamefont {Rishi}, \citenamefont {Perera},\ and\ \citenamefont
  {Bartlett}(2016)}]{Rishi_2016}%
  \BibitemOpen
  \bibfield  {author} {\bibinfo {author} {\bibfnamefont {V.}~\bibnamefont
  {Rishi}}, \bibinfo {author} {\bibfnamefont {A.}~\bibnamefont {Perera}}, \
  and\ \bibinfo {author} {\bibfnamefont {R.~J.}\ \bibnamefont {Bartlett}},\
  }\href {\doibase 10.1063/1.4944087} {\bibfield  {journal} {\bibinfo
  {journal} {J. Chem. Phys.}\ }\textbf {\bibinfo {volume} {144}},\ \bibinfo
  {pages} {124117} (\bibinfo {year} {2016})}\BibitemShut {NoStop}%
\bibitem [{\citenamefont {Rishi}, \citenamefont {Perera},\ and\ \citenamefont
  {Bartlett}(2019)}]{Rishi_2019}%
  \BibitemOpen
  \bibfield  {author} {\bibinfo {author} {\bibfnamefont {V.}~\bibnamefont
  {Rishi}}, \bibinfo {author} {\bibfnamefont {A.}~\bibnamefont {Perera}}, \
  and\ \bibinfo {author} {\bibfnamefont {R.~J.}\ \bibnamefont {Bartlett}},\
  }\href {\doibase 10.1080/00268976.2018.1492748} {\bibfield  {journal}
  {\bibinfo  {journal} {Mol. Phys.}\ }\textbf {\bibinfo {volume} {117}},\
  \bibinfo {pages} {2201} (\bibinfo {year} {2019})}\BibitemShut {NoStop}%
\bibitem [{\citenamefont {Rishi}\ and\ \citenamefont
  {Valeev}(2019)}]{Rishi_2019a}%
  \BibitemOpen
  \bibfield  {author} {\bibinfo {author} {\bibfnamefont {V.}~\bibnamefont
  {Rishi}}\ and\ \bibinfo {author} {\bibfnamefont {E.~F.}\ \bibnamefont
  {Valeev}},\ }\href {\doibase 10.1063/1.5097150} {\bibfield  {journal}
  {\bibinfo  {journal} {J. Chem. Phys.}\ }\textbf {\bibinfo {volume} {151}},\
  \bibinfo {pages} {064102} (\bibinfo {year} {2019})}\BibitemShut {NoStop}%
\bibitem [{\citenamefont {Scuseria}, \citenamefont {Henderson},\ and\
  \citenamefont {Sorensen}(2008)}]{Scuseria_2008}%
  \BibitemOpen
  \bibfield  {author} {\bibinfo {author} {\bibfnamefont {G.~E.}\ \bibnamefont
  {Scuseria}}, \bibinfo {author} {\bibfnamefont {T.~M.}\ \bibnamefont
  {Henderson}}, \ and\ \bibinfo {author} {\bibfnamefont {D.~C.}\ \bibnamefont
  {Sorensen}},\ }\href {\doibase 10.1063/1.3043729} {\bibfield  {journal}
  {\bibinfo  {journal} {J. Chem. Phys.}\ }\textbf {\bibinfo {volume} {129}},\
  \bibinfo {pages} {231101} (\bibinfo {year} {2008})}\BibitemShut {NoStop}%
\bibitem [{\citenamefont {Peng}\ \emph {et~al.}(2013)\citenamefont {Peng},
  \citenamefont {Steinmann}, \citenamefont {{van Aggelen}},\ and\ \citenamefont
  {Yang}}]{Peng_2013}%
  \BibitemOpen
  \bibfield  {author} {\bibinfo {author} {\bibfnamefont {D.}~\bibnamefont
  {Peng}}, \bibinfo {author} {\bibfnamefont {S.~N.}\ \bibnamefont {Steinmann}},
  \bibinfo {author} {\bibfnamefont {H.}~\bibnamefont {{van Aggelen}}}, \ and\
  \bibinfo {author} {\bibfnamefont {W.}~\bibnamefont {Yang}},\ }\href {\doibase
  10.1063/1.4820556} {\bibfield  {journal} {\bibinfo  {journal} {J. Chem.
  Phys.}\ }\textbf {\bibinfo {volume} {139}},\ \bibinfo {pages} {104112}
  (\bibinfo {year} {2013})}\BibitemShut {NoStop}%
\bibitem [{\citenamefont {Scuseria}, \citenamefont {Henderson},\ and\
  \citenamefont {Bulik}(2013)}]{Scuseria_2013}%
  \BibitemOpen
  \bibfield  {author} {\bibinfo {author} {\bibfnamefont {G.~E.}\ \bibnamefont
  {Scuseria}}, \bibinfo {author} {\bibfnamefont {T.~M.}\ \bibnamefont
  {Henderson}}, \ and\ \bibinfo {author} {\bibfnamefont {I.~W.}\ \bibnamefont
  {Bulik}},\ }\href {\doibase 10.1063/1.4820557} {\bibfield  {journal}
  {\bibinfo  {journal} {J. Chem. Phys.}\ }\textbf {\bibinfo {volume} {139}},\
  \bibinfo {pages} {104113} (\bibinfo {year} {2013})}\BibitemShut {NoStop}%
\bibitem [{\citenamefont {Shepherd}, \citenamefont {Henderson},\ and\
  \citenamefont {Scuseria}(2014{\natexlab{a}})}]{Shepherd_2014}%
  \BibitemOpen
  \bibfield  {author} {\bibinfo {author} {\bibfnamefont {J.~J.}\ \bibnamefont
  {Shepherd}}, \bibinfo {author} {\bibfnamefont {T.~M.}\ \bibnamefont
  {Henderson}}, \ and\ \bibinfo {author} {\bibfnamefont {G.~E.}\ \bibnamefont
  {Scuseria}},\ }\href {\doibase 10.1063/1.4867783} {\bibfield  {journal}
  {\bibinfo  {journal} {J. Chem. Phys.}\ }\textbf {\bibinfo {volume} {140}},\
  \bibinfo {pages} {124102} (\bibinfo {year} {2014}{\natexlab{a}})}\BibitemShut
  {NoStop}%
\bibitem [{\citenamefont {Shepherd}, \citenamefont {Henderson},\ and\
  \citenamefont {Scuseria}(2014{\natexlab{b}})}]{Shepherd_2014a}%
  \BibitemOpen
  \bibfield  {author} {\bibinfo {author} {\bibfnamefont {J.~J.}\ \bibnamefont
  {Shepherd}}, \bibinfo {author} {\bibfnamefont {T.~M.}\ \bibnamefont
  {Henderson}}, \ and\ \bibinfo {author} {\bibfnamefont {G.~E.}\ \bibnamefont
  {Scuseria}},\ }\href {\doibase 10.1103/PhysRevLett.112.133002} {\bibfield
  {journal} {\bibinfo  {journal} {Phys. Rev. Lett.}\ }\textbf {\bibinfo
  {volume} {112}},\ \bibinfo {pages} {133002} (\bibinfo {year}
  {2014}{\natexlab{b}})}\BibitemShut {NoStop}%
\bibitem [{\citenamefont {Bartlett}\ and\ \citenamefont
  {Musia{\l}}(2006)}]{Bartlett_2006}%
  \BibitemOpen
  \bibfield  {author} {\bibinfo {author} {\bibfnamefont {R.~J.}\ \bibnamefont
  {Bartlett}}\ and\ \bibinfo {author} {\bibfnamefont {M.}~\bibnamefont
  {Musia{\l}}},\ }\href {\doibase 10.1063/1.2387952} {\bibfield  {journal}
  {\bibinfo  {journal} {J. Chem. Phys.}\ }\textbf {\bibinfo {volume} {125}},\
  \bibinfo {pages} {204105} (\bibinfo {year} {2006})}\BibitemShut {NoStop}%
\bibitem [{\citenamefont {Musia{\l}}\ and\ \citenamefont
  {Bartlett}(2007)}]{Musial_2007}%
  \BibitemOpen
  \bibfield  {author} {\bibinfo {author} {\bibfnamefont {M.}~\bibnamefont
  {Musia{\l}}}\ and\ \bibinfo {author} {\bibfnamefont {R.~J.}\ \bibnamefont
  {Bartlett}},\ }\href {\doibase 10.1063/1.2747245} {\bibfield  {journal}
  {\bibinfo  {journal} {J. Chem. Phys.}\ }\textbf {\bibinfo {volume} {127}},\
  \bibinfo {pages} {024106} (\bibinfo {year} {2007})}\BibitemShut {NoStop}%
\bibitem [{\citenamefont {Huntington}\ and\ \citenamefont
  {Nooijen}(2010)}]{Huntington_2010}%
  \BibitemOpen
  \bibfield  {author} {\bibinfo {author} {\bibfnamefont {L.~M.~J.}\
  \bibnamefont {Huntington}}\ and\ \bibinfo {author} {\bibfnamefont
  {M.}~\bibnamefont {Nooijen}},\ }\href {\doibase 10.1063/1.3494113} {\bibfield
   {journal} {\bibinfo  {journal} {J. Chem. Phys.}\ }\textbf {\bibinfo {volume}
  {133}},\ \bibinfo {pages} {184109} (\bibinfo {year} {2010})}\BibitemShut
  {NoStop}%
\bibitem [{\citenamefont {Bartlett}\ and\ \citenamefont
  {Noga}(1988)}]{Bartlett_1988}%
  \BibitemOpen
  \bibfield  {author} {\bibinfo {author} {\bibfnamefont {R.~J.}\ \bibnamefont
  {Bartlett}}\ and\ \bibinfo {author} {\bibfnamefont {J.}~\bibnamefont
  {Noga}},\ }\href {\doibase 10.1016/0009-2614(88)80392-0} {\bibfield
  {journal} {\bibinfo  {journal} {Chemical Physics Letters}\ }\textbf {\bibinfo
  {volume} {150}},\ \bibinfo {pages} {29} (\bibinfo {year} {1988})}\BibitemShut
  {NoStop}%
\bibitem [{\citenamefont {Kutzelnigg}(1991)}]{Kutzelnigg_1991}%
  \BibitemOpen
  \bibfield  {author} {\bibinfo {author} {\bibfnamefont {W.}~\bibnamefont
  {Kutzelnigg}},\ }\href {\doibase 10.1007/BF01117418} {\bibfield  {journal}
  {\bibinfo  {journal} {Theoret. Chim. Acta}\ }\textbf {\bibinfo {volume}
  {80}},\ \bibinfo {pages} {349} (\bibinfo {year} {1991})}\BibitemShut
  {NoStop}%
\bibitem [{\citenamefont {Szalay}, \citenamefont {Nooijen},\ and\ \citenamefont
  {Bartlett}(1995)}]{Szalay_1995}%
  \BibitemOpen
  \bibfield  {author} {\bibinfo {author} {\bibfnamefont {P.~G.}\ \bibnamefont
  {Szalay}}, \bibinfo {author} {\bibfnamefont {M.}~\bibnamefont {Nooijen}}, \
  and\ \bibinfo {author} {\bibfnamefont {R.~J.}\ \bibnamefont {Bartlett}},\
  }\href {\doibase 10.1063/1.469641} {\bibfield  {journal} {\bibinfo  {journal}
  {J. Chem. Phys.}\ }\textbf {\bibinfo {volume} {103}},\ \bibinfo {pages} {281}
  (\bibinfo {year} {1995})}\BibitemShut {NoStop}%
\bibitem [{\citenamefont {Kutzelnigg}(1998)}]{Kutzelnigg_1998}%
  \BibitemOpen
  \bibfield  {author} {\bibinfo {author} {\bibfnamefont {W.}~\bibnamefont
  {Kutzelnigg}},\ }\href {\doibase 10.1080/002689798168358} {\bibfield
  {journal} {\bibinfo  {journal} {Mol. Phys.}\ }\textbf {\bibinfo {volume}
  {94}},\ \bibinfo {pages} {65} (\bibinfo {year} {1998})}\BibitemShut {NoStop}%
\bibitem [{\citenamefont {Kutzelnigg}(2010)}]{Kutzelnigg_2010}%
  \BibitemOpen
  \bibfield  {author} {\bibinfo {author} {\bibfnamefont {W.}~\bibnamefont
  {Kutzelnigg}},\ }in\ \href {\doibase 10.1007/978-90-481-2885-3_12} {\emph
  {\bibinfo {booktitle} {Recent {{Progress}} in {{Coupled Cluster Methods}}:
  {{Theory}} and {{Applications}}}}},\ \bibinfo {series and number} {Challenges
  and {{Advances}} in {{Computational Chemistry}} and {{Physics}}},\ \bibinfo
  {editor} {edited by\ \bibinfo {editor} {\bibfnamefont {P.}~\bibnamefont
  {C{\'a}rsky}}, \bibinfo {editor} {\bibfnamefont {J.}~\bibnamefont {Paldus}},
  \ and\ \bibinfo {editor} {\bibfnamefont {J.}~\bibnamefont {Pittner}}}\
  (\bibinfo  {publisher} {{Springer Netherlands}},\ \bibinfo {address}
  {{Dordrecht}},\ \bibinfo {year} {2010})\ pp.\ \bibinfo {pages}
  {299--356}\BibitemShut {NoStop}%
\bibitem [{\citenamefont {Cooper}\ and\ \citenamefont
  {Knowles}(2010)}]{Cooper_2010}%
  \BibitemOpen
  \bibfield  {author} {\bibinfo {author} {\bibfnamefont {B.}~\bibnamefont
  {Cooper}}\ and\ \bibinfo {author} {\bibfnamefont {P.~J.}\ \bibnamefont
  {Knowles}},\ }\href {\doibase 10.1063/1.3520564} {\bibfield  {journal}
  {\bibinfo  {journal} {J. Chem. Phys.}\ }\textbf {\bibinfo {volume} {133}},\
  \bibinfo {pages} {234102} (\bibinfo {year} {2010})}\BibitemShut {NoStop}%
\bibitem [{\citenamefont {Knowles}\ and\ \citenamefont
  {Cooper}(2010)}]{Knowles_2010}%
  \BibitemOpen
  \bibfield  {author} {\bibinfo {author} {\bibfnamefont {P.~J.}\ \bibnamefont
  {Knowles}}\ and\ \bibinfo {author} {\bibfnamefont {B.}~\bibnamefont
  {Cooper}},\ }\href {\doibase 10.1063/1.3507876} {\bibfield  {journal}
  {\bibinfo  {journal} {J. Chem. Phys.}\ }\textbf {\bibinfo {volume} {133}},\
  \bibinfo {pages} {224106} (\bibinfo {year} {2010})}\BibitemShut {NoStop}%
\bibitem [{\citenamefont {Robinson}\ and\ \citenamefont
  {Knowles}(2011)}]{Robinson_2011}%
  \BibitemOpen
  \bibfield  {author} {\bibinfo {author} {\bibfnamefont {J.~B.}\ \bibnamefont
  {Robinson}}\ and\ \bibinfo {author} {\bibfnamefont {P.~J.}\ \bibnamefont
  {Knowles}},\ }\href {\doibase 10.1063/1.3615060} {\bibfield  {journal}
  {\bibinfo  {journal} {J. Chem. Phys.}\ }\textbf {\bibinfo {volume} {135}},\
  \bibinfo {pages} {044113} (\bibinfo {year} {2011})}\BibitemShut {NoStop}%
\bibitem [{\citenamefont {Harsha}, \citenamefont {Shiozaki},\ and\
  \citenamefont {Scuseria}(2018)}]{Harsha_2018}%
  \BibitemOpen
  \bibfield  {author} {\bibinfo {author} {\bibfnamefont {G.}~\bibnamefont
  {Harsha}}, \bibinfo {author} {\bibfnamefont {T.}~\bibnamefont {Shiozaki}}, \
  and\ \bibinfo {author} {\bibfnamefont {G.~E.}\ \bibnamefont {Scuseria}},\
  }\href {\doibase 10.1063/1.5011033} {\bibfield  {journal} {\bibinfo
  {journal} {J. Chem. Phys.}\ }\textbf {\bibinfo {volume} {148}},\ \bibinfo
  {pages} {044107} (\bibinfo {year} {2018})}\BibitemShut {NoStop}%
\bibitem [{\citenamefont {Van~Voorhis}\ and\ \citenamefont
  {{Head-Gordon}}(2000)}]{VanVoorhis_2000}%
  \BibitemOpen
  \bibfield  {author} {\bibinfo {author} {\bibfnamefont {T.}~\bibnamefont
  {Van~Voorhis}}\ and\ \bibinfo {author} {\bibfnamefont {M.}~\bibnamefont
  {{Head-Gordon}}},\ }\href {\doibase 10.1063/1.1319643} {\bibfield  {journal}
  {\bibinfo  {journal} {J. Chem. Phys.}\ }\textbf {\bibinfo {volume} {113}},\
  \bibinfo {pages} {8873} (\bibinfo {year} {2000})}\BibitemShut {NoStop}%
\bibitem [{\citenamefont {Evangelista}(2011)}]{Evangelista_2011}%
  \BibitemOpen
  \bibfield  {author} {\bibinfo {author} {\bibfnamefont {F.~A.}\ \bibnamefont
  {Evangelista}},\ }\href {\doibase 10.1063/1.3598471} {\bibfield  {journal}
  {\bibinfo  {journal} {J. Chem. Phys.}\ }\textbf {\bibinfo {volume} {134}},\
  \bibinfo {pages} {224102} (\bibinfo {year} {2011})}\BibitemShut {NoStop}%
\bibitem [{\citenamefont {Robinson}\ and\ \citenamefont
  {Knowles}(2012{\natexlab{a}})}]{Robinson_2012}%
  \BibitemOpen
  \bibfield  {author} {\bibinfo {author} {\bibfnamefont {J.~B.}\ \bibnamefont
  {Robinson}}\ and\ \bibinfo {author} {\bibfnamefont {P.~J.}\ \bibnamefont
  {Knowles}},\ }\href {\doibase 10.1021/ct300416b} {\bibfield  {journal}
  {\bibinfo  {journal} {J. Chem. Theory Comput.}\ }\textbf {\bibinfo {volume}
  {8}},\ \bibinfo {pages} {2653} (\bibinfo {year}
  {2012}{\natexlab{a}})}\BibitemShut {NoStop}%
\bibitem [{\citenamefont {Robinson}\ and\ \citenamefont
  {Knowles}(2012{\natexlab{b}})}]{Robinson_2012a}%
  \BibitemOpen
  \bibfield  {author} {\bibinfo {author} {\bibfnamefont {J.~B.}\ \bibnamefont
  {Robinson}}\ and\ \bibinfo {author} {\bibfnamefont {P.~J.}\ \bibnamefont
  {Knowles}},\ }\href {\doibase 10.1063/1.3680560} {\bibfield  {journal}
  {\bibinfo  {journal} {J. Chem. Phys.}\ }\textbf {\bibinfo {volume} {136}},\
  \bibinfo {pages} {054114} (\bibinfo {year} {2012}{\natexlab{b}})}\BibitemShut
  {NoStop}%
\bibitem [{\citenamefont {Robinson}\ and\ \citenamefont
  {Knowles}(2012{\natexlab{c}})}]{Robinson_2012b}%
  \BibitemOpen
  \bibfield  {author} {\bibinfo {author} {\bibfnamefont {J.~B.}\ \bibnamefont
  {Robinson}}\ and\ \bibinfo {author} {\bibfnamefont {P.~J.}\ \bibnamefont
  {Knowles}},\ }\href {\doibase 10.1063/1.4738758} {\bibfield  {journal}
  {\bibinfo  {journal} {J. Chem. Phys.}\ }\textbf {\bibinfo {volume} {137}},\
  \bibinfo {pages} {054301} (\bibinfo {year} {2012}{\natexlab{c}})}\BibitemShut
  {NoStop}%
\bibitem [{\citenamefont {Robinson}\ and\ \citenamefont
  {Knowles}(2012{\natexlab{d}})}]{Robinson_2012c}%
  \BibitemOpen
  \bibfield  {author} {\bibinfo {author} {\bibfnamefont {J.~B.}\ \bibnamefont
  {Robinson}}\ and\ \bibinfo {author} {\bibfnamefont {P.~J.}\ \bibnamefont
  {Knowles}},\ }\href {\doibase 10.1039/C2CP40698E} {\bibfield  {journal}
  {\bibinfo  {journal} {Phys. Chem. Chem. Phys.}\ }\textbf {\bibinfo {volume}
  {14}},\ \bibinfo {pages} {6729} (\bibinfo {year}
  {2012}{\natexlab{d}})}\BibitemShut {NoStop}%
\bibitem [{\citenamefont {Pople}, \citenamefont {Binkley},\ and\ \citenamefont
  {Seeger}(1976)}]{Pople_1976}%
  \BibitemOpen
  \bibfield  {author} {\bibinfo {author} {\bibfnamefont {J.~A.}\ \bibnamefont
  {Pople}}, \bibinfo {author} {\bibfnamefont {J.~S.}\ \bibnamefont {Binkley}},
  \ and\ \bibinfo {author} {\bibfnamefont {R.}~\bibnamefont {Seeger}},\ }\href
  {\doibase https://doi.org/10.1002/qua.560100802} {\bibfield  {journal}
  {\bibinfo  {journal} {Int. J. Quantum Chem.}\ }\textbf {\bibinfo {volume}
  {10}},\ \bibinfo {pages} {1} (\bibinfo {year} {1976})}\BibitemShut {NoStop}%
\bibitem [{\citenamefont {Ring}\ and\ \citenamefont
  {Schuck}(1980)}]{Ring_1980}%
  \BibitemOpen
  \bibfield  {author} {\bibinfo {author} {\bibfnamefont {P.}~\bibnamefont
  {Ring}}\ and\ \bibinfo {author} {\bibfnamefont {P.}~\bibnamefont {Schuck}},\
  }\href@noop {} {\emph {\bibinfo {title} {The {{Nuclear Many}}-{{Body
  Problem}}}}},\ Theoretical and {{Mathematical Physics}}, {{The Nuclear
  Many}}-{{Body Problem}}\ (\bibinfo  {publisher} {{Springer-Verlag}},\
  \bibinfo {address} {{Berlin Heidelberg}},\ \bibinfo {year}
  {1980})\BibitemShut {NoStop}%
\bibitem [{\citenamefont {Bytautas}\ \emph {et~al.}(2011)\citenamefont
  {Bytautas}, \citenamefont {Henderson}, \citenamefont {{Jim{\'e}nez-Hoyos}},
  \citenamefont {Ellis},\ and\ \citenamefont {Scuseria}}]{Bytautas_2011}%
  \BibitemOpen
  \bibfield  {author} {\bibinfo {author} {\bibfnamefont {L.}~\bibnamefont
  {Bytautas}}, \bibinfo {author} {\bibfnamefont {T.~M.}\ \bibnamefont
  {Henderson}}, \bibinfo {author} {\bibfnamefont {C.~A.}\ \bibnamefont
  {{Jim{\'e}nez-Hoyos}}}, \bibinfo {author} {\bibfnamefont {J.~K.}\
  \bibnamefont {Ellis}}, \ and\ \bibinfo {author} {\bibfnamefont {G.~E.}\
  \bibnamefont {Scuseria}},\ }\href {\doibase 10.1063/1.3613706} {\bibfield
  {journal} {\bibinfo  {journal} {J. Chem. Phys.}\ }\textbf {\bibinfo {volume}
  {135}},\ \bibinfo {pages} {044119} (\bibinfo {year} {2011})}\BibitemShut
  {NoStop}%
\bibitem [{\citenamefont {Allen}\ and\ \citenamefont
  {Shull}(1962)}]{Allen_1962}%
  \BibitemOpen
  \bibfield  {author} {\bibinfo {author} {\bibfnamefont {T.~L.}\ \bibnamefont
  {Allen}}\ and\ \bibinfo {author} {\bibfnamefont {H.}~\bibnamefont {Shull}},\
  }\href {\doibase 10.1021/j100818a001} {\bibfield  {journal} {\bibinfo
  {journal} {J. Phys. Chem.}\ }\textbf {\bibinfo {volume} {66}},\ \bibinfo
  {pages} {2281} (\bibinfo {year} {1962})}\BibitemShut {NoStop}%
\bibitem [{\citenamefont {Smith}\ and\ \citenamefont
  {Fogel}(1965)}]{Smith_1965}%
  \BibitemOpen
  \bibfield  {author} {\bibinfo {author} {\bibfnamefont {D.~W.}\ \bibnamefont
  {Smith}}\ and\ \bibinfo {author} {\bibfnamefont {S.~J.}\ \bibnamefont
  {Fogel}},\ }\href {\doibase 10.1063/1.1701519} {\bibfield  {journal}
  {\bibinfo  {journal} {J. Chem. Phys.}\ }\textbf {\bibinfo {volume} {43}},\
  \bibinfo {pages} {S91} (\bibinfo {year} {1965})}\BibitemShut {NoStop}%
\bibitem [{\citenamefont {Veillard}\ and\ \citenamefont
  {Clementi}(1967)}]{Veillard_1967}%
  \BibitemOpen
  \bibfield  {author} {\bibinfo {author} {\bibfnamefont {A.}~\bibnamefont
  {Veillard}}\ and\ \bibinfo {author} {\bibfnamefont {E.}~\bibnamefont
  {Clementi}},\ }\href {\doibase 10.1007/BF01151915} {\bibfield  {journal}
  {\bibinfo  {journal} {Theoret. Chim. Acta}\ }\textbf {\bibinfo {volume}
  {7}},\ \bibinfo {pages} {133} (\bibinfo {year} {1967})}\BibitemShut {NoStop}%
\bibitem [{\citenamefont {Weinhold}\ and\ \citenamefont
  {Wilson}(1967)}]{Weinhold_1967}%
  \BibitemOpen
  \bibfield  {author} {\bibinfo {author} {\bibfnamefont {F.}~\bibnamefont
  {Weinhold}}\ and\ \bibinfo {author} {\bibfnamefont {E.~B.}\ \bibnamefont
  {Wilson}},\ }\href {\doibase 10.1063/1.1841109} {\bibfield  {journal}
  {\bibinfo  {journal} {J. Chem. Phys.}\ }\textbf {\bibinfo {volume} {46}},\
  \bibinfo {pages} {2752} (\bibinfo {year} {1967})}\BibitemShut {NoStop}%
\bibitem [{\citenamefont {Couty}\ and\ \citenamefont
  {Hall}(1997)}]{Couty_1997}%
  \BibitemOpen
  \bibfield  {author} {\bibinfo {author} {\bibfnamefont {M.}~\bibnamefont
  {Couty}}\ and\ \bibinfo {author} {\bibfnamefont {M.~B.}\ \bibnamefont
  {Hall}},\ }\href {\doibase 10.1021/jp963953l} {\bibfield  {journal} {\bibinfo
   {journal} {J. Phys. Chem. A}\ }\textbf {\bibinfo {volume} {101}},\ \bibinfo
  {pages} {6936} (\bibinfo {year} {1997})}\BibitemShut {NoStop}%
\bibitem [{\citenamefont {Kollmar}\ and\ \citenamefont
  {He{\ss}}(2003)}]{Kollmar_2003}%
  \BibitemOpen
  \bibfield  {author} {\bibinfo {author} {\bibfnamefont {C.}~\bibnamefont
  {Kollmar}}\ and\ \bibinfo {author} {\bibfnamefont {B.~A.}\ \bibnamefont
  {He{\ss}}},\ }\href {\doibase 10.1063/1.1590635} {\bibfield  {journal}
  {\bibinfo  {journal} {J. Chem. Phys.}\ }\textbf {\bibinfo {volume} {119}},\
  \bibinfo {pages} {4655} (\bibinfo {year} {2003})}\BibitemShut {NoStop}%
\bibitem [{\citenamefont {Henderson}, \citenamefont {Bulik},\ and\
  \citenamefont {Scuseria}(2015)}]{Henderson_2015}%
  \BibitemOpen
  \bibfield  {author} {\bibinfo {author} {\bibfnamefont {T.~M.}\ \bibnamefont
  {Henderson}}, \bibinfo {author} {\bibfnamefont {I.~W.}\ \bibnamefont
  {Bulik}}, \ and\ \bibinfo {author} {\bibfnamefont {G.~E.}\ \bibnamefont
  {Scuseria}},\ }\href {\doibase 10.1063/1.4921986} {\bibfield  {journal}
  {\bibinfo  {journal} {J. Chem. Phys.}\ }\textbf {\bibinfo {volume} {142}},\
  \bibinfo {pages} {214116} (\bibinfo {year} {2015})}\BibitemShut {NoStop}%
\bibitem [{\citenamefont {Tecmer}\ \emph {et~al.}(2014)\citenamefont {Tecmer},
  \citenamefont {Boguslawski}, \citenamefont {Johnson}, \citenamefont
  {Limacher}, \citenamefont {Chan}, \citenamefont {Verstraelen},\ and\
  \citenamefont {Ayers}}]{Tecmer_2014}%
  \BibitemOpen
  \bibfield  {author} {\bibinfo {author} {\bibfnamefont {P.}~\bibnamefont
  {Tecmer}}, \bibinfo {author} {\bibfnamefont {K.}~\bibnamefont {Boguslawski}},
  \bibinfo {author} {\bibfnamefont {P.~A.}\ \bibnamefont {Johnson}}, \bibinfo
  {author} {\bibfnamefont {P.~A.}\ \bibnamefont {Limacher}}, \bibinfo {author}
  {\bibfnamefont {M.}~\bibnamefont {Chan}}, \bibinfo {author} {\bibfnamefont
  {T.}~\bibnamefont {Verstraelen}}, \ and\ \bibinfo {author} {\bibfnamefont
  {P.~W.}\ \bibnamefont {Ayers}},\ }\href {\doibase 10.1021/jp502127v}
  {\bibfield  {journal} {\bibinfo  {journal} {J. Phys. Chem. A}\ }\textbf
  {\bibinfo {volume} {118}},\ \bibinfo {pages} {9058} (\bibinfo {year}
  {2014})}\BibitemShut {NoStop}%
\bibitem [{\citenamefont {Boguslawski}\ \emph
  {et~al.}(2014{\natexlab{a}})\citenamefont {Boguslawski}, \citenamefont
  {Tecmer}, \citenamefont {Ayers}, \citenamefont {Bultinck}, \citenamefont
  {De~Baerdemacker},\ and\ \citenamefont {Van~Neck}}]{Boguslawski_2014a}%
  \BibitemOpen
  \bibfield  {author} {\bibinfo {author} {\bibfnamefont {K.}~\bibnamefont
  {Boguslawski}}, \bibinfo {author} {\bibfnamefont {P.}~\bibnamefont {Tecmer}},
  \bibinfo {author} {\bibfnamefont {P.~W.}\ \bibnamefont {Ayers}}, \bibinfo
  {author} {\bibfnamefont {P.}~\bibnamefont {Bultinck}}, \bibinfo {author}
  {\bibfnamefont {S.}~\bibnamefont {De~Baerdemacker}}, \ and\ \bibinfo {author}
  {\bibfnamefont {D.}~\bibnamefont {Van~Neck}},\ }\href {\doibase
  10.1103/PhysRevB.89.201106} {\bibfield  {journal} {\bibinfo  {journal} {Phys.
  Rev. B}\ }\textbf {\bibinfo {volume} {89}},\ \bibinfo {pages} {201106}
  (\bibinfo {year} {2014}{\natexlab{a}})}\BibitemShut {NoStop}%
\bibitem [{\citenamefont {Boguslawski}\ \emph
  {et~al.}(2014{\natexlab{b}})\citenamefont {Boguslawski}, \citenamefont
  {Tecmer}, \citenamefont {Bultinck}, \citenamefont {De~Baerdemacker},
  \citenamefont {Van~Neck},\ and\ \citenamefont {Ayers}}]{Boguslawski_2014b}%
  \BibitemOpen
  \bibfield  {author} {\bibinfo {author} {\bibfnamefont {K.}~\bibnamefont
  {Boguslawski}}, \bibinfo {author} {\bibfnamefont {P.}~\bibnamefont {Tecmer}},
  \bibinfo {author} {\bibfnamefont {P.}~\bibnamefont {Bultinck}}, \bibinfo
  {author} {\bibfnamefont {S.}~\bibnamefont {De~Baerdemacker}}, \bibinfo
  {author} {\bibfnamefont {D.}~\bibnamefont {Van~Neck}}, \ and\ \bibinfo
  {author} {\bibfnamefont {P.~W.}\ \bibnamefont {Ayers}},\ }\href {\doibase
  10.1021/ct500759q} {\bibfield  {journal} {\bibinfo  {journal} {J. Chem.
  Theory Comput.}\ }\textbf {\bibinfo {volume} {10}},\ \bibinfo {pages} {4873}
  (\bibinfo {year} {2014}{\natexlab{b}})}\BibitemShut {NoStop}%
\bibitem [{\citenamefont {Boguslawski}\ \emph
  {et~al.}(2014{\natexlab{c}})\citenamefont {Boguslawski}, \citenamefont
  {Tecmer}, \citenamefont {Limacher}, \citenamefont {Johnson}, \citenamefont
  {Ayers}, \citenamefont {Bultinck}, \citenamefont {De~Baerdemacker},\ and\
  \citenamefont {Van~Neck}}]{Boguslawski_2014c}%
  \BibitemOpen
  \bibfield  {author} {\bibinfo {author} {\bibfnamefont {K.}~\bibnamefont
  {Boguslawski}}, \bibinfo {author} {\bibfnamefont {P.}~\bibnamefont {Tecmer}},
  \bibinfo {author} {\bibfnamefont {P.~A.}\ \bibnamefont {Limacher}}, \bibinfo
  {author} {\bibfnamefont {P.~A.}\ \bibnamefont {Johnson}}, \bibinfo {author}
  {\bibfnamefont {P.~W.}\ \bibnamefont {Ayers}}, \bibinfo {author}
  {\bibfnamefont {P.}~\bibnamefont {Bultinck}}, \bibinfo {author}
  {\bibfnamefont {S.}~\bibnamefont {De~Baerdemacker}}, \ and\ \bibinfo {author}
  {\bibfnamefont {D.}~\bibnamefont {Van~Neck}},\ }\href {\doibase
  10.1063/1.4880820} {\bibfield  {journal} {\bibinfo  {journal} {J. Chem.
  Phys.}\ }\textbf {\bibinfo {volume} {140}},\ \bibinfo {pages} {214114}
  (\bibinfo {year} {2014}{\natexlab{c}})}\BibitemShut {NoStop}%
\bibitem [{\citenamefont {Tecmer}, \citenamefont {Boguslawski},\ and\
  \citenamefont {Ayers}(2015)}]{Tecmer_2015}%
  \BibitemOpen
  \bibfield  {author} {\bibinfo {author} {\bibfnamefont {P.}~\bibnamefont
  {Tecmer}}, \bibinfo {author} {\bibfnamefont {K.}~\bibnamefont {Boguslawski}},
  \ and\ \bibinfo {author} {\bibfnamefont {P.~W.}\ \bibnamefont {Ayers}},\
  }\href {\doibase 10.1039/C4CP05293E} {\bibfield  {journal} {\bibinfo
  {journal} {Phys. Chem. Chem. Phys.}\ }\textbf {\bibinfo {volume} {17}},\
  \bibinfo {pages} {14427} (\bibinfo {year} {2015})}\BibitemShut {NoStop}%
\bibitem [{\citenamefont {Boguslawski}\ and\ \citenamefont
  {Ayers}(2015)}]{Boguslawski_2015}%
  \BibitemOpen
  \bibfield  {author} {\bibinfo {author} {\bibfnamefont {K.}~\bibnamefont
  {Boguslawski}}\ and\ \bibinfo {author} {\bibfnamefont {P.~W.}\ \bibnamefont
  {Ayers}},\ }\href {\doibase 10.1021/acs.jctc.5b00776} {\bibfield  {journal}
  {\bibinfo  {journal} {J. Chem. Theory Comput.}\ }\textbf {\bibinfo {volume}
  {11}},\ \bibinfo {pages} {5252} (\bibinfo {year} {2015})}\BibitemShut
  {NoStop}%
\bibitem [{\citenamefont {Boguslawski}, \citenamefont {Tecmer},\ and\
  \citenamefont {Legeza}(2016)}]{Boguslawski_2016a}%
  \BibitemOpen
  \bibfield  {author} {\bibinfo {author} {\bibfnamefont {K.}~\bibnamefont
  {Boguslawski}}, \bibinfo {author} {\bibfnamefont {P.}~\bibnamefont {Tecmer}},
  \ and\ \bibinfo {author} {\bibfnamefont {{\"O}.}~\bibnamefont {Legeza}},\
  }\href {\doibase 10.1103/PhysRevB.94.155126} {\bibfield  {journal} {\bibinfo
  {journal} {Phys. Rev. B}\ }\textbf {\bibinfo {volume} {94}},\ \bibinfo
  {pages} {155126} (\bibinfo {year} {2016})}\BibitemShut {NoStop}%
\bibitem [{\citenamefont {Boguslawski}(2016)}]{Boguslawski_2016b}%
  \BibitemOpen
  \bibfield  {author} {\bibinfo {author} {\bibfnamefont {K.}~\bibnamefont
  {Boguslawski}},\ }\href {\doibase 10.1063/1.4972053} {\bibfield  {journal}
  {\bibinfo  {journal} {J. Chem. Phys.}\ }\textbf {\bibinfo {volume} {145}},\
  \bibinfo {pages} {234105} (\bibinfo {year} {2016})}\BibitemShut {NoStop}%
\bibitem [{\citenamefont {Fecteau}\ \emph {et~al.}(2020)\citenamefont
  {Fecteau}, \citenamefont {Fortin}, \citenamefont {Cloutier},\ and\
  \citenamefont {Johnson}}]{Fecteau_2020}%
  \BibitemOpen
  \bibfield  {author} {\bibinfo {author} {\bibfnamefont {C.-{\'E}.}\
  \bibnamefont {Fecteau}}, \bibinfo {author} {\bibfnamefont {H.}~\bibnamefont
  {Fortin}}, \bibinfo {author} {\bibfnamefont {S.}~\bibnamefont {Cloutier}}, \
  and\ \bibinfo {author} {\bibfnamefont {P.~A.}\ \bibnamefont {Johnson}},\
  }\href {\doibase 10.1063/5.0027393} {\bibfield  {journal} {\bibinfo
  {journal} {J. Chem. Phys.}\ }\textbf {\bibinfo {volume} {153}},\ \bibinfo
  {pages} {164117} (\bibinfo {year} {2020})}\BibitemShut {NoStop}%
\bibitem [{\citenamefont {Johnson}\ \emph {et~al.}(2020)\citenamefont
  {Johnson}, \citenamefont {Fecteau}, \citenamefont {Berthiaume}, \citenamefont
  {Cloutier}, \citenamefont {Carrier}, \citenamefont {Gratton}, \citenamefont
  {Bultinck}, \citenamefont {De~Baerdemacker}, \citenamefont {Van~Neck},
  \citenamefont {Limacher},\ and\ \citenamefont {Ayers}}]{Johnson_2020}%
  \BibitemOpen
  \bibfield  {author} {\bibinfo {author} {\bibfnamefont {P.~A.}\ \bibnamefont
  {Johnson}}, \bibinfo {author} {\bibfnamefont {C.-{\'E}.}\ \bibnamefont
  {Fecteau}}, \bibinfo {author} {\bibfnamefont {F.}~\bibnamefont {Berthiaume}},
  \bibinfo {author} {\bibfnamefont {S.}~\bibnamefont {Cloutier}}, \bibinfo
  {author} {\bibfnamefont {L.}~\bibnamefont {Carrier}}, \bibinfo {author}
  {\bibfnamefont {M.}~\bibnamefont {Gratton}}, \bibinfo {author} {\bibfnamefont
  {P.}~\bibnamefont {Bultinck}}, \bibinfo {author} {\bibfnamefont
  {S.}~\bibnamefont {De~Baerdemacker}}, \bibinfo {author} {\bibfnamefont
  {D.}~\bibnamefont {Van~Neck}}, \bibinfo {author} {\bibfnamefont
  {P.}~\bibnamefont {Limacher}}, \ and\ \bibinfo {author} {\bibfnamefont
  {P.~W.}\ \bibnamefont {Ayers}},\ }\href {\doibase 10.1063/5.0022189}
  {\bibfield  {journal} {\bibinfo  {journal} {J. Chem. Phys.}\ }\textbf
  {\bibinfo {volume} {153}},\ \bibinfo {pages} {104110} (\bibinfo {year}
  {2020})}\BibitemShut {NoStop}%
\bibitem [{\citenamefont {Coleman}(1963)}]{Coleman_1963}%
  \BibitemOpen
  \bibfield  {author} {\bibinfo {author} {\bibfnamefont {A.~J.}\ \bibnamefont
  {Coleman}},\ }\href {\doibase 10.1103/RevModPhys.35.668} {\bibfield
  {journal} {\bibinfo  {journal} {Rev. Mod. Phys.}\ }\textbf {\bibinfo {volume}
  {35}},\ \bibinfo {pages} {668} (\bibinfo {year} {1963})}\BibitemShut
  {NoStop}%
\bibitem [{\citenamefont {Coleman}(1965)}]{Coleman_1965}%
  \BibitemOpen
  \bibfield  {author} {\bibinfo {author} {\bibfnamefont {A.~J.}\ \bibnamefont
  {Coleman}},\ }\href {\doibase 10.1063/1.1704794} {\bibfield  {journal}
  {\bibinfo  {journal} {J. Math. Phys.}\ }\textbf {\bibinfo {volume} {6}},\
  \bibinfo {pages} {1425} (\bibinfo {year} {1965})}\BibitemShut {NoStop}%
\bibitem [{\citenamefont {Henderson}\ and\ \citenamefont
  {Scuseria}(2019)}]{Henderson_2019}%
  \BibitemOpen
  \bibfield  {author} {\bibinfo {author} {\bibfnamefont {T.~M.}\ \bibnamefont
  {Henderson}}\ and\ \bibinfo {author} {\bibfnamefont {G.~E.}\ \bibnamefont
  {Scuseria}},\ }\href {\doibase 10.1063/1.5116715} {\bibfield  {journal}
  {\bibinfo  {journal} {J. Chem. Phys.}\ }\textbf {\bibinfo {volume} {151}},\
  \bibinfo {pages} {051101} (\bibinfo {year} {2019})}\BibitemShut {NoStop}%
\bibitem [{\citenamefont {Khamoshi}, \citenamefont {Henderson},\ and\
  \citenamefont {Scuseria}(2019)}]{Khamoshi_2019}%
  \BibitemOpen
  \bibfield  {author} {\bibinfo {author} {\bibfnamefont {A.}~\bibnamefont
  {Khamoshi}}, \bibinfo {author} {\bibfnamefont {T.~M.}\ \bibnamefont
  {Henderson}}, \ and\ \bibinfo {author} {\bibfnamefont {G.~E.}\ \bibnamefont
  {Scuseria}},\ }\href {\doibase 10.1063/1.5127850} {\bibfield  {journal}
  {\bibinfo  {journal} {J. Chem. Phys.}\ }\textbf {\bibinfo {volume} {151}},\
  \bibinfo {pages} {184103} (\bibinfo {year} {2019})}\BibitemShut {NoStop}%
\bibitem [{\citenamefont {Henderson}\ and\ \citenamefont
  {Scuseria}(2020)}]{Henderson_2020}%
  \BibitemOpen
  \bibfield  {author} {\bibinfo {author} {\bibfnamefont {T.~M.}\ \bibnamefont
  {Henderson}}\ and\ \bibinfo {author} {\bibfnamefont {G.~E.}\ \bibnamefont
  {Scuseria}},\ }\href {\doibase 10.1063/5.0021144} {\bibfield  {journal}
  {\bibinfo  {journal} {J. Chem. Phys.}\ }\textbf {\bibinfo {volume} {153}},\
  \bibinfo {pages} {084111} (\bibinfo {year} {2020})}\BibitemShut {NoStop}%
\bibitem [{\citenamefont {Dutta}, \citenamefont {Henderson},\ and\
  \citenamefont {Scuseria}(2020)}]{Dutta_2020}%
  \BibitemOpen
  \bibfield  {author} {\bibinfo {author} {\bibfnamefont {R.}~\bibnamefont
  {Dutta}}, \bibinfo {author} {\bibfnamefont {T.~M.}\ \bibnamefont
  {Henderson}}, \ and\ \bibinfo {author} {\bibfnamefont {G.~E.}\ \bibnamefont
  {Scuseria}},\ }\href {\doibase 10.1021/acs.jctc.0c00807} {\bibfield
  {journal} {\bibinfo  {journal} {J. Chem. Theory Comput.}\ }\textbf {\bibinfo
  {volume} {16}},\ \bibinfo {pages} {6358} (\bibinfo {year}
  {2020})}\BibitemShut {NoStop}%
\bibitem [{\citenamefont {Khamoshi}\ \emph {et~al.}(2021)\citenamefont
  {Khamoshi}, \citenamefont {Chen}, \citenamefont {Henderson},\ and\
  \citenamefont {Scuseria}}]{Khamoshi_2021}%
  \BibitemOpen
  \bibfield  {author} {\bibinfo {author} {\bibfnamefont {A.}~\bibnamefont
  {Khamoshi}}, \bibinfo {author} {\bibfnamefont {G.~P.}\ \bibnamefont {Chen}},
  \bibinfo {author} {\bibfnamefont {T.~M.}\ \bibnamefont {Henderson}}, \ and\
  \bibinfo {author} {\bibfnamefont {G.~E.}\ \bibnamefont {Scuseria}},\ }\href
  {\doibase 10.1063/5.0039618} {\bibfield  {journal} {\bibinfo  {journal} {J.
  Chem. Phys.}\ }\textbf {\bibinfo {volume} {154}},\ \bibinfo {pages} {074113}
  (\bibinfo {year} {2021})}\BibitemShut {NoStop}%
\bibitem [{\citenamefont {Dutta}\ \emph {et~al.}(2021)\citenamefont {Dutta},
  \citenamefont {Chen}, \citenamefont {Henderson},\ and\ \citenamefont
  {Scuseria}}]{Dutta_2021}%
  \BibitemOpen
  \bibfield  {author} {\bibinfo {author} {\bibfnamefont {R.}~\bibnamefont
  {Dutta}}, \bibinfo {author} {\bibfnamefont {G.~P.}\ \bibnamefont {Chen}},
  \bibinfo {author} {\bibfnamefont {T.~M.}\ \bibnamefont {Henderson}}, \ and\
  \bibinfo {author} {\bibfnamefont {G.~E.}\ \bibnamefont {Scuseria}},\ }\href
  {\doibase 10.1063/5.0045006} {\bibfield  {journal} {\bibinfo  {journal} {J.
  Chem. Phys.}\ }\textbf {\bibinfo {volume} {154}},\ \bibinfo {pages} {114112}
  (\bibinfo {year} {2021})}\BibitemShut {NoStop}%
\bibitem [{\citenamefont {Rowe}(1968)}]{Rowe_1968}%
  \BibitemOpen
  \bibfield  {author} {\bibinfo {author} {\bibfnamefont {D.~J.}\ \bibnamefont
  {Rowe}},\ }\href {\doibase 10.1103/RevModPhys.40.153} {\bibfield  {journal}
  {\bibinfo  {journal} {Rev. Mod. Phys.}\ }\textbf {\bibinfo {volume} {40}},\
  \bibinfo {pages} {153} (\bibinfo {year} {1968})}\BibitemShut {NoStop}%
\bibitem [{\citenamefont {Monkhorst}(1977)}]{Monkhorst_1977}%
  \BibitemOpen
  \bibfield  {author} {\bibinfo {author} {\bibfnamefont {H.~J.}\ \bibnamefont
  {Monkhorst}},\ }\href {\doibase https://doi.org/10.1002/qua.560120850}
  {\bibfield  {journal} {\bibinfo  {journal} {Int. J. Quantum Chem.}\ }\textbf
  {\bibinfo {volume} {12}},\ \bibinfo {pages} {421} (\bibinfo {year}
  {1977})}\BibitemShut {NoStop}%
\bibitem [{\citenamefont {Koch}\ \emph {et~al.}(1990)\citenamefont {Koch},
  \citenamefont {Jensen}, \citenamefont {Jorgensen},\ and\ \citenamefont
  {Helgaker}}]{Koch_1990}%
  \BibitemOpen
  \bibfield  {author} {\bibinfo {author} {\bibfnamefont {H.}~\bibnamefont
  {Koch}}, \bibinfo {author} {\bibfnamefont {H.~J.~A.}\ \bibnamefont {Jensen}},
  \bibinfo {author} {\bibfnamefont {P.}~\bibnamefont {Jorgensen}}, \ and\
  \bibinfo {author} {\bibfnamefont {T.}~\bibnamefont {Helgaker}},\ }\href
  {\doibase 10.1063/1.458815} {\bibfield  {journal} {\bibinfo  {journal} {J.
  Chem. Phys.}\ }\textbf {\bibinfo {volume} {93}},\ \bibinfo {pages} {3345}
  (\bibinfo {year} {1990})}\BibitemShut {NoStop}%
\bibitem [{\citenamefont {Stanton}\ and\ \citenamefont
  {Bartlett}(1993)}]{Stanton_1993}%
  \BibitemOpen
  \bibfield  {author} {\bibinfo {author} {\bibfnamefont {J.~F.}\ \bibnamefont
  {Stanton}}\ and\ \bibinfo {author} {\bibfnamefont {R.~J.}\ \bibnamefont
  {Bartlett}},\ }\href {\doibase 10.1063/1.464746} {\bibfield  {journal}
  {\bibinfo  {journal} {J. Chem. Phys.}\ }\textbf {\bibinfo {volume} {98}},\
  \bibinfo {pages} {7029} (\bibinfo {year} {1993})}\BibitemShut {NoStop}%
\bibitem [{\citenamefont {Koch}\ \emph {et~al.}(1994)\citenamefont {Koch},
  \citenamefont {Kobayashi}, \citenamefont {Sanchez~de Mer{\'a}s},\ and\
  \citenamefont {Jorgensen}}]{Koch_1994}%
  \BibitemOpen
  \bibfield  {author} {\bibinfo {author} {\bibfnamefont {H.}~\bibnamefont
  {Koch}}, \bibinfo {author} {\bibfnamefont {R.}~\bibnamefont {Kobayashi}},
  \bibinfo {author} {\bibfnamefont {A.}~\bibnamefont {Sanchez~de Mer{\'a}s}}, \
  and\ \bibinfo {author} {\bibfnamefont {P.}~\bibnamefont {Jorgensen}},\ }\href
  {\doibase 10.1063/1.466321} {\bibfield  {journal} {\bibinfo  {journal} {J.
  Chem. Phys.}\ }\textbf {\bibinfo {volume} {100}},\ \bibinfo {pages} {4393}
  (\bibinfo {year} {1994})}\BibitemShut {NoStop}%
\bibitem [{\citenamefont {Loos}\ \emph {et~al.}(2018)\citenamefont {Loos},
  \citenamefont {Scemama}, \citenamefont {Blondel}, \citenamefont {Garniron},
  \citenamefont {Caffarel},\ and\ \citenamefont {Jacquemin}}]{Loos_2018b}%
  \BibitemOpen
  \bibfield  {author} {\bibinfo {author} {\bibfnamefont {P.~F.}\ \bibnamefont
  {Loos}}, \bibinfo {author} {\bibfnamefont {A.}~\bibnamefont {Scemama}},
  \bibinfo {author} {\bibfnamefont {A.}~\bibnamefont {Blondel}}, \bibinfo
  {author} {\bibfnamefont {Y.}~\bibnamefont {Garniron}}, \bibinfo {author}
  {\bibfnamefont {M.}~\bibnamefont {Caffarel}}, \ and\ \bibinfo {author}
  {\bibfnamefont {D.}~\bibnamefont {Jacquemin}},\ }\href {\doibase
  10.1021/acs.jctc.8b00406} {\bibfield  {journal} {\bibinfo  {journal} {J.
  Chem. Theory Comput.}\ }\textbf {\bibinfo {volume} {14}},\ \bibinfo {pages}
  {4360} (\bibinfo {year} {2018})}\BibitemShut {NoStop}%
\bibitem [{\citenamefont {Loos}\ \emph {et~al.}(2020)\citenamefont {Loos},
  \citenamefont {Lipparini}, \citenamefont {Boggio-Pasqua}, \citenamefont
  {Scemama},\ and\ \citenamefont {Jacquemin}}]{Loos_2020c}%
  \BibitemOpen
  \bibfield  {author} {\bibinfo {author} {\bibfnamefont {P.~F.}\ \bibnamefont
  {Loos}}, \bibinfo {author} {\bibfnamefont {F.}~\bibnamefont {Lipparini}},
  \bibinfo {author} {\bibfnamefont {M.}~\bibnamefont {Boggio-Pasqua}}, \bibinfo
  {author} {\bibfnamefont {A.}~\bibnamefont {Scemama}}, \ and\ \bibinfo
  {author} {\bibfnamefont {D.}~\bibnamefont {Jacquemin}},\ }\href {\doibase
  10.1021/acs.jctc.9b01216} {\bibfield  {journal} {\bibinfo  {journal} {J.
  Chem. Theory Comput.}\ }\textbf {\bibinfo {volume} {16}},\ \bibinfo {pages}
  {1711} (\bibinfo {year} {2020})}\BibitemShut {NoStop}%
\bibitem [{\citenamefont {Loos}\ \emph {et~al.}(2019)\citenamefont {Loos},
  \citenamefont {Boggio-Pasqua}, \citenamefont {Scemama}, \citenamefont
  {Caffarel},\ and\ \citenamefont {Jacquemin}}]{Loos_2019c}%
  \BibitemOpen
  \bibfield  {author} {\bibinfo {author} {\bibfnamefont {P.-F.}\ \bibnamefont
  {Loos}}, \bibinfo {author} {\bibfnamefont {M.}~\bibnamefont {Boggio-Pasqua}},
  \bibinfo {author} {\bibfnamefont {A.}~\bibnamefont {Scemama}}, \bibinfo
  {author} {\bibfnamefont {M.}~\bibnamefont {Caffarel}}, \ and\ \bibinfo
  {author} {\bibfnamefont {D.}~\bibnamefont {Jacquemin}},\ }\href {\doibase
  10.1021/acs.jctc.8b01205} {\bibfield  {journal} {\bibinfo  {journal} {J.
  Chem. Theory Comput.}\ }\textbf {\bibinfo {volume} {15}},\ \bibinfo {pages}
  {1939} (\bibinfo {year} {2019})}\BibitemShut {NoStop}%
\bibitem [{\citenamefont {Loos}, \citenamefont {Scemama},\ and\ \citenamefont
  {Jacquemin}(2020)}]{Loos_2020d}%
  \BibitemOpen
  \bibfield  {author} {\bibinfo {author} {\bibfnamefont {P.-F.}\ \bibnamefont
  {Loos}}, \bibinfo {author} {\bibfnamefont {A.}~\bibnamefont {Scemama}}, \
  and\ \bibinfo {author} {\bibfnamefont {D.}~\bibnamefont {Jacquemin}},\ }\href
  {\doibase 10.1021/acs.jpclett.0c00014} {\bibfield  {journal} {\bibinfo
  {journal} {J. Phys. Chem. Lett.}\ }\textbf {\bibinfo {volume} {11}},\
  \bibinfo {pages} {2374} (\bibinfo {year} {2020})}\BibitemShut {NoStop}%
\bibitem [{\citenamefont {Kucharski}\ and\ \citenamefont
  {Bartlett}(1991)}]{Kucharski_1991}%
  \BibitemOpen
  \bibfield  {author} {\bibinfo {author} {\bibfnamefont {S.~A.}\ \bibnamefont
  {Kucharski}}\ and\ \bibinfo {author} {\bibfnamefont {R.~J.}\ \bibnamefont
  {Bartlett}},\ }\href {\doibase 10.1007/BF01117419} {\bibfield  {journal}
  {\bibinfo  {journal} {Theor. Chim. Acta}\ }\textbf {\bibinfo {volume} {80}},\
  \bibinfo {pages} {387} (\bibinfo {year} {1991})}\BibitemShut {NoStop}%
\bibitem [{\citenamefont {Christiansen}, \citenamefont {Koch},\ and\
  \citenamefont {J{\o}rgensen}(1995)}]{Christiansen_1995b}%
  \BibitemOpen
  \bibfield  {author} {\bibinfo {author} {\bibfnamefont {O.}~\bibnamefont
  {Christiansen}}, \bibinfo {author} {\bibfnamefont {H.}~\bibnamefont {Koch}},
  \ and\ \bibinfo {author} {\bibfnamefont {P.}~\bibnamefont {J{\o}rgensen}},\
  }\href {\doibase http://dx.doi.org/10.1063/1.470315} {\bibfield  {journal}
  {\bibinfo  {journal} {J. Chem. Phys.}\ }\textbf {\bibinfo {volume} {103}},\
  \bibinfo {pages} {7429} (\bibinfo {year} {1995})}\BibitemShut {NoStop}%
\bibitem [{\citenamefont {Kucharski}\ \emph {et~al.}(2001)\citenamefont
  {Kucharski}, \citenamefont {W{\l}och}, \citenamefont {Musia{\l}},\ and\
  \citenamefont {Bartlett}}]{Kucharski_2001}%
  \BibitemOpen
  \bibfield  {author} {\bibinfo {author} {\bibfnamefont {S.~A.}\ \bibnamefont
  {Kucharski}}, \bibinfo {author} {\bibfnamefont {M.}~\bibnamefont {W{\l}och}},
  \bibinfo {author} {\bibfnamefont {M.}~\bibnamefont {Musia{\l}}}, \ and\
  \bibinfo {author} {\bibfnamefont {R.~J.}\ \bibnamefont {Bartlett}},\ }\href
  {\doibase 10.1063/1.1416173} {\bibfield  {journal} {\bibinfo  {journal} {J.
  Chem. Phys.}\ }\textbf {\bibinfo {volume} {115}},\ \bibinfo {pages} {8263}
  (\bibinfo {year} {2001})}\BibitemShut {NoStop}%
\bibitem [{\citenamefont {Kowalski}\ and\ \citenamefont
  {Piecuch}(2001{\natexlab{a}})}]{Kowalski_2001}%
  \BibitemOpen
  \bibfield  {author} {\bibinfo {author} {\bibfnamefont {K.}~\bibnamefont
  {Kowalski}}\ and\ \bibinfo {author} {\bibfnamefont {P.}~\bibnamefont
  {Piecuch}},\ }\href@noop {} {\bibfield  {journal} {\bibinfo  {journal} {J.
  Chem. Phys.}\ }\textbf {\bibinfo {volume} {115}},\ \bibinfo {pages} {643}
  (\bibinfo {year} {2001}{\natexlab{a}})}\BibitemShut {NoStop}%
\bibitem [{\citenamefont {Hirata}\ and\ \citenamefont
  {Bartlett}(2000)}]{Hirata_2000}%
  \BibitemOpen
  \bibfield  {author} {\bibinfo {author} {\bibfnamefont {S.}~\bibnamefont
  {Hirata}}\ and\ \bibinfo {author} {\bibfnamefont {R.~J.}\ \bibnamefont
  {Bartlett}},\ }\href {\doibase 10.1016/S0009-2614(00)00387-0} {\bibfield
  {journal} {\bibinfo  {journal} {Chem. Phys. Lett.}\ }\textbf {\bibinfo
  {volume} {321}},\ \bibinfo {pages} {216} (\bibinfo {year}
  {2000})}\BibitemShut {NoStop}%
\bibitem [{\citenamefont {Hirata}(2004)}]{Hirata_2004}%
  \BibitemOpen
  \bibfield  {author} {\bibinfo {author} {\bibfnamefont {S.}~\bibnamefont
  {Hirata}},\ }\href@noop {} {\bibfield  {journal} {\bibinfo  {journal} {J.
  Chem. Phys.}\ }\textbf {\bibinfo {volume} {121}},\ \bibinfo {pages} {51}
  (\bibinfo {year} {2004})}\BibitemShut {NoStop}%
\bibitem [{\citenamefont {Piecuch}\ and\ \citenamefont
  {Kowalski}(2000)}]{Piecuch_2000}%
  \BibitemOpen
  \bibfield  {author} {\bibinfo {author} {\bibfnamefont {P.}~\bibnamefont
  {Piecuch}}\ and\ \bibinfo {author} {\bibfnamefont {K.}~\bibnamefont
  {Kowalski}},\ }in\ \href {\doibase 10.1142/9789812792501_0001} {\emph
  {\bibinfo {booktitle} {Computational {{Chemistry}}: {{Reviews}} of {{Current
  Trends}}}}},\ Vol.~\bibinfo {volume} {5}\ (\bibinfo  {publisher} {World
  Scientfic},\ \bibinfo {year} {2000})\ pp.\ \bibinfo {pages}
  {1--104}\BibitemShut {NoStop}%
\bibitem [{\citenamefont {Adamowicz}\ and\ \citenamefont
  {Bartlett}(1985)}]{Adamowicz_1985}%
  \BibitemOpen
  \bibfield  {author} {\bibinfo {author} {\bibfnamefont {L.}~\bibnamefont
  {Adamowicz}}\ and\ \bibinfo {author} {\bibfnamefont {R.~J.}\ \bibnamefont
  {Bartlett}},\ }\href {\doibase 10.1002/qua.560280821} {\bibfield  {journal}
  {\bibinfo  {journal} {Int. J. Quantum Chem.}\ }\textbf {\bibinfo {volume}
  {28}},\ \bibinfo {pages} {217} (\bibinfo {year} {1985})}\BibitemShut
  {NoStop}%
\bibitem [{\citenamefont {Lee}, \citenamefont {Small},\ and\ \citenamefont
  {{Head-Gordon}}(2019)}]{Lee_2019}%
  \BibitemOpen
  \bibfield  {author} {\bibinfo {author} {\bibfnamefont {J.}~\bibnamefont
  {Lee}}, \bibinfo {author} {\bibfnamefont {D.~W.}\ \bibnamefont {Small}}, \
  and\ \bibinfo {author} {\bibfnamefont {M.}~\bibnamefont {{Head-Gordon}}},\
  }\href {\doibase 10.1063/1.5128795} {\bibfield  {journal} {\bibinfo
  {journal} {J. Chem. Phys.}\ }\textbf {\bibinfo {volume} {151}},\ \bibinfo
  {pages} {214103} (\bibinfo {year} {2019})}\BibitemShut {NoStop}%
\bibitem [{\citenamefont {Gilbert}, \citenamefont {Besley},\ and\ \citenamefont
  {Gill}(2008)}]{Gilbert_2008}%
  \BibitemOpen
  \bibfield  {author} {\bibinfo {author} {\bibfnamefont {A.~T.~B.}\
  \bibnamefont {Gilbert}}, \bibinfo {author} {\bibfnamefont {N.~A.}\
  \bibnamefont {Besley}}, \ and\ \bibinfo {author} {\bibfnamefont {P.~M.~W.}\
  \bibnamefont {Gill}},\ }\href {\doibase 10.1021/jp801738f} {\bibfield
  {journal} {\bibinfo  {journal} {J. Phys. Chem. A}\ }\textbf {\bibinfo
  {volume} {112}},\ \bibinfo {pages} {13164} (\bibinfo {year}
  {2008})}\BibitemShut {NoStop}%
\bibitem [{\citenamefont {Barca}, \citenamefont {Gilbert},\ and\ \citenamefont
  {Gill}(2014)}]{Barca_2014}%
  \BibitemOpen
  \bibfield  {author} {\bibinfo {author} {\bibfnamefont {G.~M.~J.}\
  \bibnamefont {Barca}}, \bibinfo {author} {\bibfnamefont {A.~T.~B.}\
  \bibnamefont {Gilbert}}, \ and\ \bibinfo {author} {\bibfnamefont {P.~M.~W.}\
  \bibnamefont {Gill}},\ }\href {\doibase 10.1063/1.4896182} {\bibfield
  {journal} {\bibinfo  {journal} {J. Chem. Phys.}\ }\textbf {\bibinfo {volume}
  {141}},\ \bibinfo {pages} {111104} (\bibinfo {year} {2014})}\BibitemShut
  {NoStop}%
\bibitem [{\citenamefont {Barca}, \citenamefont {Gilbert},\ and\ \citenamefont
  {Gill}(2018{\natexlab{a}})}]{Barca_2018a}%
  \BibitemOpen
  \bibfield  {author} {\bibinfo {author} {\bibfnamefont {G.~M.~J.}\
  \bibnamefont {Barca}}, \bibinfo {author} {\bibfnamefont {A.~T.~B.}\
  \bibnamefont {Gilbert}}, \ and\ \bibinfo {author} {\bibfnamefont {P.~M.~W.}\
  \bibnamefont {Gill}},\ }\href {\doibase 10.1021/acs.jctc.7b00994} {\bibfield
  {journal} {\bibinfo  {journal} {J. Chem. Theory Comput.}\ }\textbf {\bibinfo
  {volume} {14}},\ \bibinfo {pages} {1501} (\bibinfo {year}
  {2018}{\natexlab{a}})}\BibitemShut {NoStop}%
\bibitem [{\citenamefont {Barca}, \citenamefont {Gilbert},\ and\ \citenamefont
  {Gill}(2018{\natexlab{b}})}]{Barca_2018b}%
  \BibitemOpen
  \bibfield  {author} {\bibinfo {author} {\bibfnamefont {G.~M.~J.}\
  \bibnamefont {Barca}}, \bibinfo {author} {\bibfnamefont {A.~T.~B.}\
  \bibnamefont {Gilbert}}, \ and\ \bibinfo {author} {\bibfnamefont {P.~M.~W.}\
  \bibnamefont {Gill}},\ }\href {\doibase 10.1021/acs.jctc.7b00963} {\bibfield
  {journal} {\bibinfo  {journal} {J. Chem. Theory Comput.}\ }\textbf {\bibinfo
  {volume} {14}},\ \bibinfo {pages} {9} (\bibinfo {year}
  {2018}{\natexlab{b}})}\BibitemShut {NoStop}%
\bibitem [{\citenamefont {Thom}\ and\ \citenamefont
  {{Head-Gordon}}(2008)}]{Thom_2008}%
  \BibitemOpen
  \bibfield  {author} {\bibinfo {author} {\bibfnamefont {A.~J.~W.}\
  \bibnamefont {Thom}}\ and\ \bibinfo {author} {\bibfnamefont {M.}~\bibnamefont
  {{Head-Gordon}}},\ }\href {\doibase 10.1103/PhysRevLett.101.193001}
  {\bibfield  {journal} {\bibinfo  {journal} {Phys. Rev. Lett.}\ }\textbf
  {\bibinfo {volume} {101}},\ \bibinfo {pages} {193001} (\bibinfo {year}
  {2008})}\BibitemShut {NoStop}%
\bibitem [{\citenamefont {Zhao}\ and\ \citenamefont
  {Neuscamman}(2016)}]{Zhao_2016a}%
  \BibitemOpen
  \bibfield  {author} {\bibinfo {author} {\bibfnamefont {L.}~\bibnamefont
  {Zhao}}\ and\ \bibinfo {author} {\bibfnamefont {E.}~\bibnamefont
  {Neuscamman}},\ }\href {\doibase 10.1021/acs.jctc.6b00508} {\bibfield
  {journal} {\bibinfo  {journal} {J. Chem. Theory Comput.}\ }\textbf {\bibinfo
  {volume} {12}},\ \bibinfo {pages} {3436} (\bibinfo {year}
  {2016})}\BibitemShut {NoStop}%
\bibitem [{\citenamefont {Ye}\ \emph {et~al.}(2017)\citenamefont {Ye},
  \citenamefont {Welborn}, \citenamefont {Ricke},\ and\ \citenamefont
  {Van~Voorhis}}]{Ye_2017}%
  \BibitemOpen
  \bibfield  {author} {\bibinfo {author} {\bibfnamefont {H.-Z.}\ \bibnamefont
  {Ye}}, \bibinfo {author} {\bibfnamefont {M.}~\bibnamefont {Welborn}},
  \bibinfo {author} {\bibfnamefont {N.~D.}\ \bibnamefont {Ricke}}, \ and\
  \bibinfo {author} {\bibfnamefont {T.}~\bibnamefont {Van~Voorhis}},\ }\href
  {\doibase 10.1063/1.5001262} {\bibfield  {journal} {\bibinfo  {journal} {J.
  Chem. Phys.}\ }\textbf {\bibinfo {volume} {147}},\ \bibinfo {pages} {214104}
  (\bibinfo {year} {2017})}\BibitemShut {NoStop}%
\bibitem [{\citenamefont {Shea}\ and\ \citenamefont
  {Neuscamman}(2018)}]{Shea_2018}%
  \BibitemOpen
  \bibfield  {author} {\bibinfo {author} {\bibfnamefont {J.~A.~R.}\
  \bibnamefont {Shea}}\ and\ \bibinfo {author} {\bibfnamefont {E.}~\bibnamefont
  {Neuscamman}},\ }\href {\doibase 10.1063/1.5045056} {\bibfield  {journal}
  {\bibinfo  {journal} {J. Chem. Phys.}\ }\textbf {\bibinfo {volume} {149}},\
  \bibinfo {pages} {081101} (\bibinfo {year} {2018})}\BibitemShut {NoStop}%
\bibitem [{\citenamefont {Thompson}(2018)}]{Thompson_2018}%
  \BibitemOpen
  \bibfield  {author} {\bibinfo {author} {\bibfnamefont {L.~M.}\ \bibnamefont
  {Thompson}},\ }\href {\doibase 10.1063/1.5049827} {\bibfield  {journal}
  {\bibinfo  {journal} {J. Chem. Phys.}\ }\textbf {\bibinfo {volume} {149}},\
  \bibinfo {pages} {194106} (\bibinfo {year} {2018})}\BibitemShut {NoStop}%
\bibitem [{\citenamefont {Ye}\ and\ \citenamefont
  {Van~Voorhis}(2019)}]{Ye_2019}%
  \BibitemOpen
  \bibfield  {author} {\bibinfo {author} {\bibfnamefont {H.-Z.}\ \bibnamefont
  {Ye}}\ and\ \bibinfo {author} {\bibfnamefont {T.}~\bibnamefont
  {Van~Voorhis}},\ }\href {\doibase 10.1021/acs.jctc.8b01224} {\bibfield
  {journal} {\bibinfo  {journal} {J. Chem. Theory Comput.}\ }\textbf {\bibinfo
  {volume} {15}},\ \bibinfo {pages} {2954} (\bibinfo {year}
  {2019})}\BibitemShut {NoStop}%
\bibitem [{\citenamefont {Tran}, \citenamefont {Shea},\ and\ \citenamefont
  {Neuscamman}(2019)}]{Tran_2019}%
  \BibitemOpen
  \bibfield  {author} {\bibinfo {author} {\bibfnamefont {L.~N.}\ \bibnamefont
  {Tran}}, \bibinfo {author} {\bibfnamefont {J.~A.~R.}\ \bibnamefont {Shea}}, \
  and\ \bibinfo {author} {\bibfnamefont {E.}~\bibnamefont {Neuscamman}},\
  }\href {\doibase 10.1021/acs.jctc.9b00351} {\bibfield  {journal} {\bibinfo
  {journal} {J. Chem. Theory Comput.}\ }\textbf {\bibinfo {volume} {15}},\
  \bibinfo {pages} {4790} (\bibinfo {year} {2019})}\BibitemShut {NoStop}%
\bibitem [{\citenamefont {Burton}\ and\ \citenamefont
  {Thom}(2019)}]{Burton_2019c}%
  \BibitemOpen
  \bibfield  {author} {\bibinfo {author} {\bibfnamefont {H.~G.~A.}\
  \bibnamefont {Burton}}\ and\ \bibinfo {author} {\bibfnamefont {A.~J.~W.}\
  \bibnamefont {Thom}},\ }\href {\doibase 10.1021/acs.jctc.9b00441} {\bibfield
  {journal} {\bibinfo  {journal} {J. Chem. Theory Comput.}\ }\textbf {\bibinfo
  {volume} {15}},\ \bibinfo {pages} {4851} (\bibinfo {year}
  {2019})}\BibitemShut {NoStop}%
\bibitem [{\citenamefont {Zhao}\ and\ \citenamefont
  {Neuscamman}(2020)}]{Zhao_2020}%
  \BibitemOpen
  \bibfield  {author} {\bibinfo {author} {\bibfnamefont {L.}~\bibnamefont
  {Zhao}}\ and\ \bibinfo {author} {\bibfnamefont {E.}~\bibnamefont
  {Neuscamman}},\ }\href {\doibase 10.1021/acs.jctc.9b00530} {\bibfield
  {journal} {\bibinfo  {journal} {J. Chem. Theory Comput.}\ }\textbf {\bibinfo
  {volume} {16}},\ \bibinfo {pages} {164} (\bibinfo {year} {2020})}\BibitemShut
  {NoStop}%
\bibitem [{\citenamefont {Hait}\ and\ \citenamefont
  {Head-Gordon}(2020)}]{Hait_2020}%
  \BibitemOpen
  \bibfield  {author} {\bibinfo {author} {\bibfnamefont {D.}~\bibnamefont
  {Hait}}\ and\ \bibinfo {author} {\bibfnamefont {M.}~\bibnamefont
  {Head-Gordon}},\ }\href {\doibase 10.1021/acs.jctc.9b01127} {\bibfield
  {journal} {\bibinfo  {journal} {J. Chem. Theory Comput.}\ }\textbf {\bibinfo
  {volume} {16}},\ \bibinfo {pages} {1699} (\bibinfo {year}
  {2020})}\BibitemShut {NoStop}%
\bibitem [{\citenamefont {Hait}\ \emph {et~al.}(2020)\citenamefont {Hait},
  \citenamefont {Haugen}, \citenamefont {Yang}, \citenamefont {Oosterbaan},
  \citenamefont {Leone},\ and\ \citenamefont {Head-Gordon}}]{Hait_2020b}%
  \BibitemOpen
  \bibfield  {author} {\bibinfo {author} {\bibfnamefont {D.}~\bibnamefont
  {Hait}}, \bibinfo {author} {\bibfnamefont {E.~A.}\ \bibnamefont {Haugen}},
  \bibinfo {author} {\bibfnamefont {Z.}~\bibnamefont {Yang}}, \bibinfo {author}
  {\bibfnamefont {K.~J.}\ \bibnamefont {Oosterbaan}}, \bibinfo {author}
  {\bibfnamefont {S.~R.}\ \bibnamefont {Leone}}, \ and\ \bibinfo {author}
  {\bibfnamefont {M.}~\bibnamefont {Head-Gordon}},\ }\href {\doibase
  10.1063/5.0018833} {\bibfield  {journal} {\bibinfo  {journal} {J. Chem.
  Phys.}\ }\textbf {\bibinfo {volume} {153}},\ \bibinfo {pages} {134108}
  (\bibinfo {year} {2020})}\BibitemShut {NoStop}%
\bibitem [{\citenamefont {Levi}, \citenamefont {Ivanov},\ and\ \citenamefont
  {J{\'o}nsson}(2020{\natexlab{a}})}]{Levi_2020a}%
  \BibitemOpen
  \bibfield  {author} {\bibinfo {author} {\bibfnamefont {G.}~\bibnamefont
  {Levi}}, \bibinfo {author} {\bibfnamefont {A.~V.}\ \bibnamefont {Ivanov}}, \
  and\ \bibinfo {author} {\bibfnamefont {H.}~\bibnamefont {J{\'o}nsson}},\
  }\href {\doibase 10.1021/acs.jctc.0c00597} {\bibfield  {journal} {\bibinfo
  {journal} {J. Chem. Theory Comput.}\ }\textbf {\bibinfo {volume} {16}},\
  \bibinfo {pages} {6968} (\bibinfo {year} {2020}{\natexlab{a}})}\BibitemShut
  {NoStop}%
\bibitem [{\citenamefont {Levi}, \citenamefont {Ivanov},\ and\ \citenamefont
  {J{\'o}nsson}(2020{\natexlab{b}})}]{Levi_2020b}%
  \BibitemOpen
  \bibfield  {author} {\bibinfo {author} {\bibfnamefont {G.}~\bibnamefont
  {Levi}}, \bibinfo {author} {\bibfnamefont {A.~V.}\ \bibnamefont {Ivanov}}, \
  and\ \bibinfo {author} {\bibfnamefont {H.}~\bibnamefont {J{\'o}nsson}},\
  }\href {\doibase 10.1039/D0FD00064G} {\bibfield  {journal} {\bibinfo
  {journal} {Faraday Discuss.}\ }\textbf {\bibinfo {volume} {224}},\ \bibinfo
  {pages} {448} (\bibinfo {year} {2020}{\natexlab{b}})}\BibitemShut {NoStop}%
\bibitem [{\citenamefont {Dong}\ \emph {et~al.}(2020)\citenamefont {Dong},
  \citenamefont {Mahler}, \citenamefont {{Kempfer-Robertson}},\ and\
  \citenamefont {Thompson}}]{Dong_2020}%
  \BibitemOpen
  \bibfield  {author} {\bibinfo {author} {\bibfnamefont {X.}~\bibnamefont
  {Dong}}, \bibinfo {author} {\bibfnamefont {A.~D.}\ \bibnamefont {Mahler}},
  \bibinfo {author} {\bibfnamefont {E.~M.}\ \bibnamefont
  {{Kempfer-Robertson}}}, \ and\ \bibinfo {author} {\bibfnamefont {L.~M.}\
  \bibnamefont {Thompson}},\ }\href {\doibase 10.1021/acs.jctc.0c00488}
  {\bibfield  {journal} {\bibinfo  {journal} {J. Chem. Theory Comput.}\
  }\textbf {\bibinfo {volume} {16}},\ \bibinfo {pages} {5635} (\bibinfo {year}
  {2020})}\BibitemShut {NoStop}%
\bibitem [{\citenamefont {Hait}\ and\ \citenamefont
  {Head-Gordon}(2021)}]{Hait_2021}%
  \BibitemOpen
  \bibfield  {author} {\bibinfo {author} {\bibfnamefont {D.}~\bibnamefont
  {Hait}}\ and\ \bibinfo {author} {\bibfnamefont {M.}~\bibnamefont
  {Head-Gordon}},\ }\href {\doibase 10.1021/acs.jpclett.1c00744} {\bibfield
  {journal} {\bibinfo  {journal} {J. Phys. Chem. Lett.}\ }\textbf {\bibinfo
  {volume} {12}},\ \bibinfo {pages} {4517} (\bibinfo {year}
  {2021})}\BibitemShut {NoStop}%
\bibitem [{\citenamefont {Pulay}(1980)}]{Pulay_1980}%
  \BibitemOpen
  \bibfield  {author} {\bibinfo {author} {\bibfnamefont {P.}~\bibnamefont
  {Pulay}},\ }\href {\doibase 10.1016/0009-2614(80)80396-4} {\bibfield
  {journal} {\bibinfo  {journal} {Chem. Phys. Lett.}\ }\textbf {\bibinfo
  {volume} {73}},\ \bibinfo {pages} {393} (\bibinfo {year} {1980})}\BibitemShut
  {NoStop}%
\bibitem [{\citenamefont {Pulay}(1982)}]{Pulay_1982}%
  \BibitemOpen
  \bibfield  {author} {\bibinfo {author} {\bibfnamefont {P.}~\bibnamefont
  {Pulay}},\ }\href {\doibase 10.1002/jcc.540030413} {\bibfield  {journal}
  {\bibinfo  {journal} {J. Comput. Chem.}\ }\textbf {\bibinfo {volume} {3}},\
  \bibinfo {pages} {556} (\bibinfo {year} {1982})}\BibitemShut {NoStop}%
\bibitem [{\citenamefont {Scuseria}, \citenamefont {Lee},\ and\ \citenamefont
  {{Schaefer III}}(1986)}]{Scuseria_1986}%
  \BibitemOpen
  \bibfield  {author} {\bibinfo {author} {\bibfnamefont {G.~E.}\ \bibnamefont
  {Scuseria}}, \bibinfo {author} {\bibfnamefont {T.~J.}\ \bibnamefont {Lee}}, \
  and\ \bibinfo {author} {\bibfnamefont {H.~F.}\ \bibnamefont {{Schaefer
  III}}},\ }\href {\doibase 10.1016/0009-2614(86)80461-4} {\bibfield  {journal}
  {\bibinfo  {journal} {Chem. Phys. Lett.}\ }\textbf {\bibinfo {volume}
  {130}},\ \bibinfo {pages} {236} (\bibinfo {year} {1986})}\BibitemShut
  {NoStop}%
\bibitem [{\citenamefont {Kossoski}\ \emph {et~al.}(0)\citenamefont {Kossoski},
  \citenamefont {Marie}, \citenamefont {Scemama}, \citenamefont {Caffarel},\
  and\ \citenamefont {Loos}}]{Kossoski_2021}%
  \BibitemOpen
  \bibfield  {author} {\bibinfo {author} {\bibfnamefont {F.}~\bibnamefont
  {Kossoski}}, \bibinfo {author} {\bibfnamefont {A.}~\bibnamefont {Marie}},
  \bibinfo {author} {\bibfnamefont {A.}~\bibnamefont {Scemama}}, \bibinfo
  {author} {\bibfnamefont {M.}~\bibnamefont {Caffarel}}, \ and\ \bibinfo
  {author} {\bibfnamefont {P.-F.}\ \bibnamefont {Loos}},\ }\href {\doibase
  10.1021/acs.jctc.1c00348} {\bibfield  {journal} {\bibinfo  {journal} {J.
  Chem. Theory Comput.}\ }\textbf {\bibinfo {volume} {0}},\ \bibinfo {pages}
  {null} (\bibinfo {year} {0})}\BibitemShut {NoStop}%
\bibitem [{\citenamefont {{\v Z}ivkovi{\'c}}(1977)}]{Zivkovic_1977}%
  \BibitemOpen
  \bibfield  {author} {\bibinfo {author} {\bibfnamefont {T.~P.}\ \bibnamefont
  {{\v Z}ivkovi{\'c}}},\ }\href {\doibase 10.1002/qua.560120849} {\bibfield
  {journal} {\bibinfo  {journal} {Int. J. Quantum Chem.}\ }\textbf {\bibinfo
  {volume} {12}},\ \bibinfo {pages} {413} (\bibinfo {year} {1977})}\BibitemShut
  {NoStop}%
\bibitem [{\citenamefont {{\v Z}ivkovi{\'c}}\ and\ \citenamefont
  {Monkhorst}(1978)}]{Zivkovic_1978}%
  \BibitemOpen
  \bibfield  {author} {\bibinfo {author} {\bibfnamefont {T.~P.}\ \bibnamefont
  {{\v Z}ivkovi{\'c}}}\ and\ \bibinfo {author} {\bibfnamefont {H.~J.}\
  \bibnamefont {Monkhorst}},\ }\href {\doibase 10.1063/1.523761} {\bibfield
  {journal} {\bibinfo  {journal} {J. Math. Phys.}\ }\textbf {\bibinfo {volume}
  {19}},\ \bibinfo {pages} {1007} (\bibinfo {year} {1978})}\BibitemShut
  {NoStop}%
\bibitem [{\citenamefont {Jankowski}, \citenamefont {Kowalski},\ and\
  \citenamefont {Jankowski}(1994{\natexlab{a}})}]{Jankowski_1994}%
  \BibitemOpen
  \bibfield  {author} {\bibinfo {author} {\bibfnamefont {J.}~\bibnamefont
  {Jankowski}}, \bibinfo {author} {\bibfnamefont {K.}~\bibnamefont {Kowalski}},
  \ and\ \bibinfo {author} {\bibfnamefont {P.}~\bibnamefont {Jankowski}},\
  }\href {\doibase 10.1016/0009-2614(94)00391-2} {\bibfield  {journal}
  {\bibinfo  {journal} {Chem. Phys. Lett.}\ }\textbf {\bibinfo {volume}
  {222}},\ \bibinfo {pages} {608} (\bibinfo {year}
  {1994}{\natexlab{a}})}\BibitemShut {NoStop}%
\bibitem [{\citenamefont {Jankowski}, \citenamefont {Kowalski},\ and\
  \citenamefont {Jankowski}(1994{\natexlab{b}})}]{Jankowski_1994a}%
  \BibitemOpen
  \bibfield  {author} {\bibinfo {author} {\bibfnamefont {K.}~\bibnamefont
  {Jankowski}}, \bibinfo {author} {\bibfnamefont {K.}~\bibnamefont {Kowalski}},
  \ and\ \bibinfo {author} {\bibfnamefont {P.}~\bibnamefont {Jankowski}},\
  }\href {\doibase 10.1002/qua.560500504} {\bibfield  {journal} {\bibinfo
  {journal} {Int. J. Quantum Chem.}\ }\textbf {\bibinfo {volume} {50}},\
  \bibinfo {pages} {353} (\bibinfo {year} {1994}{\natexlab{b}})}\BibitemShut
  {NoStop}%
\bibitem [{\citenamefont {Jankowski}, \citenamefont {Kowalski},\ and\
  \citenamefont {Jankowski}(1995)}]{Jankowski_1995}%
  \BibitemOpen
  \bibfield  {author} {\bibinfo {author} {\bibfnamefont {K.}~\bibnamefont
  {Jankowski}}, \bibinfo {author} {\bibfnamefont {K.}~\bibnamefont {Kowalski}},
  \ and\ \bibinfo {author} {\bibfnamefont {P.}~\bibnamefont {Jankowski}},\
  }\href {\doibase 10.1002/qua.560530507} {\bibfield  {journal} {\bibinfo
  {journal} {Int. J. Quantum Chem.}\ }\textbf {\bibinfo {volume} {53}},\
  \bibinfo {pages} {501} (\bibinfo {year} {1995})}\BibitemShut {NoStop}%
\bibitem [{\citenamefont {Kowalski}\ and\ \citenamefont
  {Jankowski}(1998{\natexlab{a}})}]{Kowalski_1998}%
  \BibitemOpen
  \bibfield  {author} {\bibinfo {author} {\bibfnamefont {K.}~\bibnamefont
  {Kowalski}}\ and\ \bibinfo {author} {\bibfnamefont {K.}~\bibnamefont
  {Jankowski}},\ }\href {\doibase 10.1016/S0009-2614(98)00464-3} {\bibfield
  {journal} {\bibinfo  {journal} {Chem. Phys. Lett.}\ }\textbf {\bibinfo
  {volume} {290}},\ \bibinfo {pages} {180} (\bibinfo {year}
  {1998}{\natexlab{a}})}\BibitemShut {NoStop}%
\bibitem [{\citenamefont {Kowalski}\ and\ \citenamefont
  {Jankowski}(1998{\natexlab{b}})}]{Kowalski_1998a}%
  \BibitemOpen
  \bibfield  {author} {\bibinfo {author} {\bibfnamefont {K.}~\bibnamefont
  {Kowalski}}\ and\ \bibinfo {author} {\bibfnamefont {K.}~\bibnamefont
  {Jankowski}},\ }\href {\doibase 10.1103/PhysRevLett.81.1195} {\bibfield
  {journal} {\bibinfo  {journal} {Phys. Rev. Lett.}\ }\textbf {\bibinfo
  {volume} {81}},\ \bibinfo {pages} {1195} (\bibinfo {year}
  {1998}{\natexlab{b}})}\BibitemShut {NoStop}%
\bibitem [{\citenamefont {Jankowski}\ and\ \citenamefont
  {Kowalski}(1999{\natexlab{a}})}]{Jankowski_1999}%
  \BibitemOpen
  \bibfield  {author} {\bibinfo {author} {\bibfnamefont {K.}~\bibnamefont
  {Jankowski}}\ and\ \bibinfo {author} {\bibfnamefont {K.}~\bibnamefont
  {Kowalski}},\ }\href {\doibase 10.1063/1.478900} {\bibfield  {journal}
  {\bibinfo  {journal} {J. Chem. Phys.}\ }\textbf {\bibinfo {volume} {110}},\
  \bibinfo {pages} {9345} (\bibinfo {year} {1999}{\natexlab{a}})}\BibitemShut
  {NoStop}%
\bibitem [{\citenamefont {Jankowski}\ and\ \citenamefont
  {Kowalski}(1999{\natexlab{b}})}]{Jankowski_1999a}%
  \BibitemOpen
  \bibfield  {author} {\bibinfo {author} {\bibfnamefont {K.}~\bibnamefont
  {Jankowski}}\ and\ \bibinfo {author} {\bibfnamefont {K.}~\bibnamefont
  {Kowalski}},\ }\href {\doibase 10.1063/1.479575} {\bibfield  {journal}
  {\bibinfo  {journal} {J. Chem. Phys.}\ }\textbf {\bibinfo {volume} {111}},\
  \bibinfo {pages} {2940} (\bibinfo {year} {1999}{\natexlab{b}})}\BibitemShut
  {NoStop}%
\bibitem [{\citenamefont {Jankowski}\ and\ \citenamefont
  {Kowalski}(1999{\natexlab{c}})}]{Jankowski_1999b}%
  \BibitemOpen
  \bibfield  {author} {\bibinfo {author} {\bibfnamefont {K.}~\bibnamefont
  {Jankowski}}\ and\ \bibinfo {author} {\bibfnamefont {K.}~\bibnamefont
  {Kowalski}},\ }\href {\doibase 10.1063/1.479576} {\bibfield  {journal}
  {\bibinfo  {journal} {J. Chem. Phys.}\ }\textbf {\bibinfo {volume} {111}},\
  \bibinfo {pages} {2952} (\bibinfo {year} {1999}{\natexlab{c}})}\BibitemShut
  {NoStop}%
\bibitem [{\citenamefont {Jankowski}\ and\ \citenamefont
  {Kowalski}(1999{\natexlab{d}})}]{Jankowski_1999c}%
  \BibitemOpen
  \bibfield  {author} {\bibinfo {author} {\bibfnamefont {K.}~\bibnamefont
  {Jankowski}}\ and\ \bibinfo {author} {\bibfnamefont {K.}~\bibnamefont
  {Kowalski}},\ }\href {\doibase 10.1063/1.478262} {\bibfield  {journal}
  {\bibinfo  {journal} {J. Chem. Phys.}\ }\textbf {\bibinfo {volume} {110}},\
  \bibinfo {pages} {3714} (\bibinfo {year} {1999}{\natexlab{d}})}\BibitemShut
  {NoStop}%
\bibitem [{\citenamefont {Podeszwa}\ and\ \citenamefont
  {Stolarczyk}(2002)}]{Podeszwa_2002}%
  \BibitemOpen
  \bibfield  {author} {\bibinfo {author} {\bibfnamefont {R.}~\bibnamefont
  {Podeszwa}}\ and\ \bibinfo {author} {\bibfnamefont {L.~Z.}\ \bibnamefont
  {Stolarczyk}},\ }\href {\doibase 10.1016/S0009-2614(02)01653-6} {\bibfield
  {journal} {\bibinfo  {journal} {Chem. Phys. Lett.}\ }\textbf {\bibinfo
  {volume} {366}},\ \bibinfo {pages} {426} (\bibinfo {year}
  {2002})}\BibitemShut {NoStop}%
\bibitem [{\citenamefont {Podeszwa}\ \emph {et~al.}(2003)\citenamefont
  {Podeszwa}, \citenamefont {Stolarczyk}, \citenamefont {Jankowski},\ and\
  \citenamefont {Rubiniec}}]{Podeszwa_2003}%
  \BibitemOpen
  \bibfield  {author} {\bibinfo {author} {\bibfnamefont {R.}~\bibnamefont
  {Podeszwa}}, \bibinfo {author} {\bibfnamefont {L.~Z.}\ \bibnamefont
  {Stolarczyk}}, \bibinfo {author} {\bibfnamefont {K.}~\bibnamefont
  {Jankowski}}, \ and\ \bibinfo {author} {\bibfnamefont {K.}~\bibnamefont
  {Rubiniec}},\ }\href {\doibase 10.1007/s00214-003-0434-6} {\bibfield
  {journal} {\bibinfo  {journal} {Theor Chem Acc}\ }\textbf {\bibinfo {volume}
  {109}},\ \bibinfo {pages} {309} (\bibinfo {year} {2003})}\BibitemShut
  {NoStop}%
\bibitem [{\citenamefont {Paldus}\ \emph {et~al.}(1993)\citenamefont {Paldus},
  \citenamefont {Piecuch}, \citenamefont {Pylypow},\ and\ \citenamefont
  {Jeziorski}}]{Paldus_1993}%
  \BibitemOpen
  \bibfield  {author} {\bibinfo {author} {\bibfnamefont {J.}~\bibnamefont
  {Paldus}}, \bibinfo {author} {\bibfnamefont {P.}~\bibnamefont {Piecuch}},
  \bibinfo {author} {\bibfnamefont {L.}~\bibnamefont {Pylypow}}, \ and\
  \bibinfo {author} {\bibfnamefont {B.}~\bibnamefont {Jeziorski}},\ }\href
  {\doibase 10.1103/PhysRevA.47.2738} {\bibfield  {journal} {\bibinfo
  {journal} {Phys. Rev. A}\ }\textbf {\bibinfo {volume} {47}},\ \bibinfo
  {pages} {2738} (\bibinfo {year} {1993})}\BibitemShut {NoStop}%
\bibitem [{\citenamefont {Kowalski}\ and\ \citenamefont
  {Piecuch}(2000{\natexlab{a}})}]{Kowalski_2000}%
  \BibitemOpen
  \bibfield  {author} {\bibinfo {author} {\bibfnamefont {K.}~\bibnamefont
  {Kowalski}}\ and\ \bibinfo {author} {\bibfnamefont {P.}~\bibnamefont
  {Piecuch}},\ }\href {\doibase 10.1103/PhysRevA.61.052506} {\bibfield
  {journal} {\bibinfo  {journal} {Phys. Rev. A}\ }\textbf {\bibinfo {volume}
  {61}},\ \bibinfo {pages} {052506} (\bibinfo {year}
  {2000}{\natexlab{a}})}\BibitemShut {NoStop}%
\bibitem [{\citenamefont {Kowalski}\ and\ \citenamefont
  {Piecuch}(2000{\natexlab{b}})}]{Kowalski_2000a}%
  \BibitemOpen
  \bibfield  {author} {\bibinfo {author} {\bibfnamefont {K.}~\bibnamefont
  {Kowalski}}\ and\ \bibinfo {author} {\bibfnamefont {P.}~\bibnamefont
  {Piecuch}},\ }\href {\doibase
  10.1002/1097-461X(2000)80:4/5<757::AID-QUA25>3.0.CO;2-A} {\bibfield
  {journal} {\bibinfo  {journal} {Int. J. Quantum Chem.}\ }\textbf {\bibinfo
  {volume} {80}},\ \bibinfo {pages} {757} (\bibinfo {year}
  {2000}{\natexlab{b}})}\BibitemShut {NoStop}%
\bibitem [{\citenamefont {Mayhall}\ and\ \citenamefont
  {Raghavachari}(2010)}]{Mayhall_2010}%
  \BibitemOpen
  \bibfield  {author} {\bibinfo {author} {\bibfnamefont {N.~J.}\ \bibnamefont
  {Mayhall}}\ and\ \bibinfo {author} {\bibfnamefont {K.}~\bibnamefont
  {Raghavachari}},\ }\href {\doibase 10.1021/ct100321k} {\bibfield  {journal}
  {\bibinfo  {journal} {J. Chem. Theory Comput.}\ }\textbf {\bibinfo {volume}
  {6}},\ \bibinfo {pages} {2714} (\bibinfo {year} {2010})}\BibitemShut
  {NoStop}%
\bibitem [{\citenamefont {Noga}\ and\ \citenamefont
  {Bartlett}(1987)}]{Noga_1987}%
  \BibitemOpen
  \bibfield  {author} {\bibinfo {author} {\bibfnamefont {J.}~\bibnamefont
  {Noga}}\ and\ \bibinfo {author} {\bibfnamefont {R.~J.}\ \bibnamefont
  {Bartlett}},\ }\href {\doibase 10.1063/1.452353} {\bibfield  {journal}
  {\bibinfo  {journal} {J. Chem. Phys.}\ }\textbf {\bibinfo {volume} {86}},\
  \bibinfo {pages} {7041} (\bibinfo {year} {1987})}\BibitemShut {NoStop}%
\bibitem [{\citenamefont {Scuseria}\ and\ \citenamefont
  {Schaefer}(1988)}]{Scuseria_1988b}%
  \BibitemOpen
  \bibfield  {author} {\bibinfo {author} {\bibfnamefont {G.~E.}\ \bibnamefont
  {Scuseria}}\ and\ \bibinfo {author} {\bibfnamefont {H.~F.}\ \bibnamefont
  {Schaefer}},\ }\href {\doibase https://doi.org/10.1016/0009-2614(88)80110-6}
  {\bibfield  {journal} {\bibinfo  {journal} {Chem. Phys. Lett.}\ }\textbf
  {\bibinfo {volume} {152}},\ \bibinfo {pages} {382} (\bibinfo {year}
  {1988})}\BibitemShut {NoStop}%
\bibitem [{\citenamefont {Kowalski}\ and\ \citenamefont
  {Piecuch}(2001{\natexlab{b}})}]{Kowalski_2001b}%
  \BibitemOpen
  \bibfield  {author} {\bibinfo {author} {\bibfnamefont {K.}~\bibnamefont
  {Kowalski}}\ and\ \bibinfo {author} {\bibfnamefont {P.}~\bibnamefont
  {Piecuch}},\ }\href {\doibase https://doi.org/10.1016/S0009-2614(01)01010-7}
  {\bibfield  {journal} {\bibinfo  {journal} {Chem. Phys. Lett.}\ }\textbf
  {\bibinfo {volume} {347}},\ \bibinfo {pages} {237} (\bibinfo {year}
  {2001}{\natexlab{b}})}\BibitemShut {NoStop}%
\bibitem [{\citenamefont {{\v C}{\'\i}{\v z}ek}, \citenamefont {Paldus},\ and\
  \citenamefont {{\v S}roubkov{\'a}}(1969)}]{Cizek_1969}%
  \BibitemOpen
  \bibfield  {author} {\bibinfo {author} {\bibfnamefont {J.}~\bibnamefont {{\v
  C}{\'\i}{\v z}ek}}, \bibinfo {author} {\bibfnamefont {J.}~\bibnamefont
  {Paldus}}, \ and\ \bibinfo {author} {\bibfnamefont {L.}~\bibnamefont {{\v
  S}roubkov{\'a}}},\ }\href {\doibase https://doi.org/10.1002/qua.560030202}
  {\bibfield  {journal} {\bibinfo  {journal} {Int. J. Quantum Chem.}\ }\textbf
  {\bibinfo {volume} {3}},\ \bibinfo {pages} {149} (\bibinfo {year}
  {1969})}\BibitemShut {NoStop}%
\bibitem [{\citenamefont {Lehtola}\ \emph {et~al.}(2017)\citenamefont
  {Lehtola}, \citenamefont {Tubman}, \citenamefont {Whaley},\ and\
  \citenamefont {{Head-Gordon}}}]{Lehtola_2017}%
  \BibitemOpen
  \bibfield  {author} {\bibinfo {author} {\bibfnamefont {S.}~\bibnamefont
  {Lehtola}}, \bibinfo {author} {\bibfnamefont {N.~M.}\ \bibnamefont {Tubman}},
  \bibinfo {author} {\bibfnamefont {K.~B.}\ \bibnamefont {Whaley}}, \ and\
  \bibinfo {author} {\bibfnamefont {M.}~\bibnamefont {{Head-Gordon}}},\ }\href
  {\doibase 10.1063/1.4996044} {\bibfield  {journal} {\bibinfo  {journal} {J.
  Chem. Phys.}\ }\textbf {\bibinfo {volume} {147}},\ \bibinfo {pages} {154105}
  (\bibinfo {year} {2017})}\BibitemShut {NoStop}%
\bibitem [{\citenamefont {Magoulas}\ \emph {et~al.}(2021)\citenamefont
  {Magoulas}, \citenamefont {Gururangan}, \citenamefont {Piecuch},
  \citenamefont {Deustua},\ and\ \citenamefont {Shen}}]{Magoulas_2021}%
  \BibitemOpen
  \bibfield  {author} {\bibinfo {author} {\bibfnamefont {I.}~\bibnamefont
  {Magoulas}}, \bibinfo {author} {\bibfnamefont {K.}~\bibnamefont
  {Gururangan}}, \bibinfo {author} {\bibfnamefont {P.}~\bibnamefont {Piecuch}},
  \bibinfo {author} {\bibfnamefont {J.~E.}\ \bibnamefont {Deustua}}, \ and\
  \bibinfo {author} {\bibfnamefont {J.}~\bibnamefont {Shen}},\ }\href {\doibase
  10.1021/acs.jctc.1c00181} {\bibfield  {journal} {\bibinfo  {journal} {J.
  Chem. Theory Comput.}\ }\textbf {\bibinfo {volume} {17}},\ \bibinfo {pages}
  {4006} (\bibinfo {year} {2021})}\BibitemShut {NoStop}%
\bibitem [{\citenamefont {Scuseria}\ and\ \citenamefont
  {Schaefer}(1987)}]{Scuseria_1987}%
  \BibitemOpen
  \bibfield  {author} {\bibinfo {author} {\bibfnamefont {G.~E.}\ \bibnamefont
  {Scuseria}}\ and\ \bibinfo {author} {\bibfnamefont {H.~F.}\ \bibnamefont
  {Schaefer}},\ }\href {\doibase 10.1016/0009-2614(87)85122-9} {\bibfield
  {journal} {\bibinfo  {journal} {Chem. Phys. Lett.}\ }\textbf {\bibinfo
  {volume} {142}},\ \bibinfo {pages} {354} (\bibinfo {year}
  {1987})}\BibitemShut {NoStop}%
\bibitem [{\citenamefont {Bozkaya}\ \emph {et~al.}(2011)\citenamefont
  {Bozkaya}, \citenamefont {Turney}, \citenamefont {Yamaguchi}, \citenamefont
  {Schaefer},\ and\ \citenamefont {Sherrill}}]{Bozkaya_2011}%
  \BibitemOpen
  \bibfield  {author} {\bibinfo {author} {\bibfnamefont {U.}~\bibnamefont
  {Bozkaya}}, \bibinfo {author} {\bibfnamefont {J.~M.}\ \bibnamefont {Turney}},
  \bibinfo {author} {\bibfnamefont {Y.}~\bibnamefont {Yamaguchi}}, \bibinfo
  {author} {\bibfnamefont {H.~F.}\ \bibnamefont {Schaefer}}, \ and\ \bibinfo
  {author} {\bibfnamefont {C.~D.}\ \bibnamefont {Sherrill}},\ }\href {\doibase
  10.1063/1.3631129} {\bibfield  {journal} {\bibinfo  {journal} {J. Chem.
  Phys.}\ }\textbf {\bibinfo {volume} {135}},\ \bibinfo {pages} {104103}
  (\bibinfo {year} {2011})}\BibitemShut {NoStop}%
\bibitem [{\citenamefont {{Wolfram Research{,} Inc.}}(2020)}]{Mathematica}%
  \BibitemOpen
  \bibfield  {author} {\bibinfo {author} {\bibnamefont {{Wolfram Research{,}
  Inc.}}},\ }\href@noop {} {\enquote {\bibinfo {title} {Mathematica, {V}ersion
  12.1},}\ }\bibinfo {howpublished} {Champaign, Illinois} (\bibinfo {year}
  {2020})\BibitemShut {NoStop}%
\bibitem [{\citenamefont {Garniron}\ \emph {et~al.}(2017)\citenamefont
  {Garniron}, \citenamefont {Scemama}, \citenamefont {Loos},\ and\
  \citenamefont {Caffarel}}]{Garniron_2017b}%
  \BibitemOpen
  \bibfield  {author} {\bibinfo {author} {\bibfnamefont {Y.}~\bibnamefont
  {Garniron}}, \bibinfo {author} {\bibfnamefont {A.}~\bibnamefont {Scemama}},
  \bibinfo {author} {\bibfnamefont {P.-F.}\ \bibnamefont {Loos}}, \ and\
  \bibinfo {author} {\bibfnamefont {M.}~\bibnamefont {Caffarel}},\ }\href
  {\doibase 10.1063/1.4992127} {\bibfield  {journal} {\bibinfo  {journal} {J.
  Chem. Phys.}\ }\textbf {\bibinfo {volume} {147}},\ \bibinfo {pages} {034101}
  (\bibinfo {year} {2017})}\BibitemShut {NoStop}%
\bibitem [{\citenamefont {Garniron}\ \emph {et~al.}(2018)\citenamefont
  {Garniron}, \citenamefont {Scemama}, \citenamefont {Giner}, \citenamefont
  {Caffarel},\ and\ \citenamefont {Loos}}]{Garniron_2018}%
  \BibitemOpen
  \bibfield  {author} {\bibinfo {author} {\bibfnamefont {Y.}~\bibnamefont
  {Garniron}}, \bibinfo {author} {\bibfnamefont {A.}~\bibnamefont {Scemama}},
  \bibinfo {author} {\bibfnamefont {E.}~\bibnamefont {Giner}}, \bibinfo
  {author} {\bibfnamefont {M.}~\bibnamefont {Caffarel}}, \ and\ \bibinfo
  {author} {\bibfnamefont {P.~F.}\ \bibnamefont {Loos}},\ }\href {\doibase
  10.1063/1.5044503} {\bibfield  {journal} {\bibinfo  {journal} {J. Chem.
  Phys.}\ }\textbf {\bibinfo {volume} {149}},\ \bibinfo {pages} {064103}
  (\bibinfo {year} {2018})}\BibitemShut {NoStop}%
\bibitem [{\citenamefont {Garniron}\ \emph {et~al.}(2019)\citenamefont
  {Garniron}, \citenamefont {Gasperich}, \citenamefont {Applencourt},
  \citenamefont {Benali}, \citenamefont {Fert{\'e}}, \citenamefont {Paquier},
  \citenamefont {Pradines}, \citenamefont {Assaraf}, \citenamefont {Reinhardt},
  \citenamefont {Toulouse}, \citenamefont {Barbaresco}, \citenamefont {Renon},
  \citenamefont {David}, \citenamefont {Malrieu}, \citenamefont {V{\'e}ril},
  \citenamefont {Caffarel}, \citenamefont {Loos}, \citenamefont {Giner},\ and\
  \citenamefont {Scemama}}]{Garniron_2019}%
  \BibitemOpen
  \bibfield  {author} {\bibinfo {author} {\bibfnamefont {Y.}~\bibnamefont
  {Garniron}}, \bibinfo {author} {\bibfnamefont {K.}~\bibnamefont {Gasperich}},
  \bibinfo {author} {\bibfnamefont {T.}~\bibnamefont {Applencourt}}, \bibinfo
  {author} {\bibfnamefont {A.}~\bibnamefont {Benali}}, \bibinfo {author}
  {\bibfnamefont {A.}~\bibnamefont {Fert{\'e}}}, \bibinfo {author}
  {\bibfnamefont {J.}~\bibnamefont {Paquier}}, \bibinfo {author} {\bibfnamefont
  {B.}~\bibnamefont {Pradines}}, \bibinfo {author} {\bibfnamefont
  {R.}~\bibnamefont {Assaraf}}, \bibinfo {author} {\bibfnamefont
  {P.}~\bibnamefont {Reinhardt}}, \bibinfo {author} {\bibfnamefont
  {J.}~\bibnamefont {Toulouse}}, \bibinfo {author} {\bibfnamefont
  {P.}~\bibnamefont {Barbaresco}}, \bibinfo {author} {\bibfnamefont
  {N.}~\bibnamefont {Renon}}, \bibinfo {author} {\bibfnamefont
  {G.}~\bibnamefont {David}}, \bibinfo {author} {\bibfnamefont {J.~P.}\
  \bibnamefont {Malrieu}}, \bibinfo {author} {\bibfnamefont {M.}~\bibnamefont
  {V{\'e}ril}}, \bibinfo {author} {\bibfnamefont {M.}~\bibnamefont {Caffarel}},
  \bibinfo {author} {\bibfnamefont {P.~F.}\ \bibnamefont {Loos}}, \bibinfo
  {author} {\bibfnamefont {E.}~\bibnamefont {Giner}}, \ and\ \bibinfo {author}
  {\bibfnamefont {A.}~\bibnamefont {Scemama}},\ }\href {\doibase
  10.1021/acs.jctc.9b00176} {\bibfield  {journal} {\bibinfo  {journal} {J.
  Chem. Theory Comput.}\ }\textbf {\bibinfo {volume} {15}},\ \bibinfo {pages}
  {3591} (\bibinfo {year} {2019})}\BibitemShut {NoStop}%
\bibitem [{\citenamefont {Hehre}, \citenamefont {Stewart},\ and\ \citenamefont
  {Pople}(1969)}]{Hehre_1969}%
  \BibitemOpen
  \bibfield  {author} {\bibinfo {author} {\bibfnamefont {W.~J.}\ \bibnamefont
  {Hehre}}, \bibinfo {author} {\bibfnamefont {R.~F.}\ \bibnamefont {Stewart}},
  \ and\ \bibinfo {author} {\bibfnamefont {J.~A.}\ \bibnamefont {Pople}},\
  }\href {\doibase 10.1063/1.1672392} {\bibfield  {journal} {\bibinfo
  {journal} {J. Chem. Phys.}\ }\textbf {\bibinfo {volume} {51}},\ \bibinfo
  {pages} {2657} (\bibinfo {year} {1969})}\BibitemShut {NoStop}%
\bibitem [{\citenamefont {Hachmann}, \citenamefont {Cardoen},\ and\
  \citenamefont {Chan}(2006)}]{Hachmann_2006}%
  \BibitemOpen
  \bibfield  {author} {\bibinfo {author} {\bibfnamefont {J.}~\bibnamefont
  {Hachmann}}, \bibinfo {author} {\bibfnamefont {W.}~\bibnamefont {Cardoen}}, \
  and\ \bibinfo {author} {\bibfnamefont {G.~K.-L.}\ \bibnamefont {Chan}},\
  }\href {\doibase 10.1063/1.2345196} {\bibfield  {journal} {\bibinfo
  {journal} {J. Chem. Phys.}\ }\textbf {\bibinfo {volume} {125}},\ \bibinfo
  {pages} {144101} (\bibinfo {year} {2006})}\BibitemShut {NoStop}%
\bibitem [{\citenamefont {Al-Saidi}, \citenamefont {Zhang},\ and\ \citenamefont
  {Krakauer}(2007)}]{Al-Saidi_2007}%
  \BibitemOpen
  \bibfield  {author} {\bibinfo {author} {\bibfnamefont {W.~A.}\ \bibnamefont
  {Al-Saidi}}, \bibinfo {author} {\bibfnamefont {S.}~\bibnamefont {Zhang}}, \
  and\ \bibinfo {author} {\bibfnamefont {H.}~\bibnamefont {Krakauer}},\ }\href
  {\doibase 10.1063/1.2770707} {\bibfield  {journal} {\bibinfo  {journal} {J.
  Chem. Phys.}\ }\textbf {\bibinfo {volume} {127}},\ \bibinfo {pages} {144101}
  (\bibinfo {year} {2007})}\BibitemShut {NoStop}%
\bibitem [{\citenamefont {Sinitskiy}, \citenamefont {Greenman},\ and\
  \citenamefont {Mazziotti}(2010)}]{Sinitskiy_2010}%
  \BibitemOpen
  \bibfield  {author} {\bibinfo {author} {\bibfnamefont {A.~V.}\ \bibnamefont
  {Sinitskiy}}, \bibinfo {author} {\bibfnamefont {L.}~\bibnamefont {Greenman}},
  \ and\ \bibinfo {author} {\bibfnamefont {D.~A.}\ \bibnamefont {Mazziotti}},\
  }\href {\doibase 10.1063/1.3459059} {\bibfield  {journal} {\bibinfo
  {journal} {J. Chem. Phys.}\ }\textbf {\bibinfo {volume} {133}},\ \bibinfo
  {pages} {014104} (\bibinfo {year} {2010})}\BibitemShut {NoStop}%
\bibitem [{\citenamefont {Stella}\ \emph {et~al.}(2011)\citenamefont {Stella},
  \citenamefont {Attaccalite}, \citenamefont {Sorella},\ and\ \citenamefont
  {Rubio}}]{Stella_2011}%
  \BibitemOpen
  \bibfield  {author} {\bibinfo {author} {\bibfnamefont {L.}~\bibnamefont
  {Stella}}, \bibinfo {author} {\bibfnamefont {C.}~\bibnamefont {Attaccalite}},
  \bibinfo {author} {\bibfnamefont {S.}~\bibnamefont {Sorella}}, \ and\
  \bibinfo {author} {\bibfnamefont {A.}~\bibnamefont {Rubio}},\ }\href
  {\doibase 10.1103/PhysRevB.84.245117} {\bibfield  {journal} {\bibinfo
  {journal} {Phys. Rev. B}\ }\textbf {\bibinfo {volume} {84}},\ \bibinfo
  {pages} {245117} (\bibinfo {year} {2011})}\BibitemShut {NoStop}%
\bibitem [{\citenamefont {Motta}\ \emph {et~al.}(2017)\citenamefont {Motta},
  \citenamefont {Ceperley}, \citenamefont {Chan}, \citenamefont {Gomez},
  \citenamefont {Gull}, \citenamefont {Guo}, \citenamefont {Jim{\'e}nez-Hoyos},
  \citenamefont {Lan}, \citenamefont {Li}, \citenamefont {Ma} \emph
  {et~al.}}]{Motta_2017}%
  \BibitemOpen
  \bibfield  {author} {\bibinfo {author} {\bibfnamefont {M.}~\bibnamefont
  {Motta}}, \bibinfo {author} {\bibfnamefont {D.~M.}\ \bibnamefont {Ceperley}},
  \bibinfo {author} {\bibfnamefont {G.~K.-L.}\ \bibnamefont {Chan}}, \bibinfo
  {author} {\bibfnamefont {J.~A.}\ \bibnamefont {Gomez}}, \bibinfo {author}
  {\bibfnamefont {E.}~\bibnamefont {Gull}}, \bibinfo {author} {\bibfnamefont
  {S.}~\bibnamefont {Guo}}, \bibinfo {author} {\bibfnamefont {C.~A.}\
  \bibnamefont {Jim{\'e}nez-Hoyos}}, \bibinfo {author} {\bibfnamefont {T.~N.}\
  \bibnamefont {Lan}}, \bibinfo {author} {\bibfnamefont {J.}~\bibnamefont
  {Li}}, \bibinfo {author} {\bibfnamefont {F.}~\bibnamefont {Ma}},  \emph
  {et~al.},\ }\href {\doibase 10.1103/PhysRevX.7.031059} {\bibfield  {journal}
  {\bibinfo  {journal} {Phys. Rev. X}\ }\textbf {\bibinfo {volume} {7}},\
  \bibinfo {pages} {031059} (\bibinfo {year} {2017})}\BibitemShut {NoStop}%
\bibitem [{\citenamefont {Motta}\ \emph {et~al.}(2020)\citenamefont {Motta},
  \citenamefont {Genovese}, \citenamefont {Ma}, \citenamefont {Cui},
  \citenamefont {Sawaya}, \citenamefont {Chan}, \citenamefont {Chepiga},
  \citenamefont {Helms}, \citenamefont {Jim\'enez-Hoyos}, \citenamefont
  {Millis}, \citenamefont {Ray}, \citenamefont {Ronca}, \citenamefont {Shi},
  \citenamefont {Sorella}, \citenamefont {Stoudenmire}, \citenamefont {White},\
  and\ \citenamefont {Zhang}}]{Motta_2020}%
  \BibitemOpen
  \bibfield  {author} {\bibinfo {author} {\bibfnamefont {M.}~\bibnamefont
  {Motta}}, \bibinfo {author} {\bibfnamefont {C.}~\bibnamefont {Genovese}},
  \bibinfo {author} {\bibfnamefont {F.}~\bibnamefont {Ma}}, \bibinfo {author}
  {\bibfnamefont {Z.-H.}\ \bibnamefont {Cui}}, \bibinfo {author} {\bibfnamefont
  {R.}~\bibnamefont {Sawaya}}, \bibinfo {author} {\bibfnamefont {G.~K.-L.}\
  \bibnamefont {Chan}}, \bibinfo {author} {\bibfnamefont {N.}~\bibnamefont
  {Chepiga}}, \bibinfo {author} {\bibfnamefont {P.}~\bibnamefont {Helms}},
  \bibinfo {author} {\bibfnamefont {C.}~\bibnamefont {Jim\'enez-Hoyos}},
  \bibinfo {author} {\bibfnamefont {A.~J.}\ \bibnamefont {Millis}}, \bibinfo
  {author} {\bibfnamefont {U.}~\bibnamefont {Ray}}, \bibinfo {author}
  {\bibfnamefont {E.}~\bibnamefont {Ronca}}, \bibinfo {author} {\bibfnamefont
  {H.}~\bibnamefont {Shi}}, \bibinfo {author} {\bibfnamefont {S.}~\bibnamefont
  {Sorella}}, \bibinfo {author} {\bibfnamefont {E.~M.}\ \bibnamefont
  {Stoudenmire}}, \bibinfo {author} {\bibfnamefont {S.~R.}\ \bibnamefont
  {White}}, \ and\ \bibinfo {author} {\bibfnamefont {S.}~\bibnamefont {Zhang}}
  (\bibinfo {collaboration} {Simons Collaboration on the Many-Electron
  Problem}),\ }\href {\doibase 10.1103/PhysRevX.10.031058} {\bibfield
  {journal} {\bibinfo  {journal} {Phys. Rev. X}\ }\textbf {\bibinfo {volume}
  {10}},\ \bibinfo {pages} {031058} (\bibinfo {year} {2020})}\BibitemShut
  {NoStop}%
\bibitem [{\citenamefont {Vollhardt}(2020)}]{Vollhard_2020}%
  \BibitemOpen
  \bibfield  {author} {\bibinfo {author} {\bibfnamefont {D.}~\bibnamefont
  {Vollhardt}},\ }\href {\doibase 10.1103/Physics.13.142} {\bibfield  {journal}
  {\bibinfo  {journal} {Physics}\ }\textbf {\bibinfo {volume} {13}},\ \bibinfo
  {pages} {142} (\bibinfo {year} {2020})}\BibitemShut {NoStop}%
\bibitem [{\citenamefont {Giner}\ \emph {et~al.}(2020)\citenamefont {Giner},
  \citenamefont {Scemama}, \citenamefont {Loos},\ and\ \citenamefont
  {Toulouse}}]{Giner_2020}%
  \BibitemOpen
  \bibfield  {author} {\bibinfo {author} {\bibfnamefont {E.}~\bibnamefont
  {Giner}}, \bibinfo {author} {\bibfnamefont {A.}~\bibnamefont {Scemama}},
  \bibinfo {author} {\bibfnamefont {P.-F.}\ \bibnamefont {Loos}}, \ and\
  \bibinfo {author} {\bibfnamefont {J.}~\bibnamefont {Toulouse}},\ }\href
  {\doibase 10.1063/5.0002892} {\bibfield  {journal} {\bibinfo  {journal} {J.
  Chem. Phys.}\ }\textbf {\bibinfo {volume} {152}},\ \bibinfo {pages} {174104}
  (\bibinfo {year} {2020})}\BibitemShut {NoStop}%
\bibitem [{\citenamefont {Jenkins}\ and\ \citenamefont
  {Traub}(1970{\natexlab{a}})}]{Jenkins_1970a}%
  \BibitemOpen
  \bibfield  {author} {\bibinfo {author} {\bibfnamefont {M.~A.}\ \bibnamefont
  {Jenkins}}\ and\ \bibinfo {author} {\bibfnamefont {J.~F.}\ \bibnamefont
  {Traub}},\ }\href {\doibase 10.1137/0707045} {\bibfield  {journal} {\bibinfo
  {journal} {SIAM J. Numer. Anal.}\ }\textbf {\bibinfo {volume} {7}},\ \bibinfo
  {pages} {545} (\bibinfo {year} {1970}{\natexlab{a}})}\BibitemShut {NoStop}%
\bibitem [{\citenamefont {Jenkins}\ and\ \citenamefont
  {Traub}(1970{\natexlab{b}})}]{Jenkins_1970b}%
  \BibitemOpen
  \bibfield  {author} {\bibinfo {author} {\bibfnamefont {M.~A.}\ \bibnamefont
  {Jenkins}}\ and\ \bibinfo {author} {\bibfnamefont {J.~F.}\ \bibnamefont
  {Traub}},\ }\href {\doibase 10.1007/BF02163334} {\bibfield  {journal}
  {\bibinfo  {journal} {Numer. Math.}\ }\textbf {\bibinfo {volume} {14}},\
  \bibinfo {pages} {252} (\bibinfo {year} {1970}{\natexlab{b}})}\BibitemShut
  {NoStop}%
\bibitem [{\citenamefont {Szak{\'a}cs}\ and\ \citenamefont
  {Surj{\'a}n}(2008)}]{Szakacs_2008}%
  \BibitemOpen
  \bibfield  {author} {\bibinfo {author} {\bibfnamefont {P.}~\bibnamefont
  {Szak{\'a}cs}}\ and\ \bibinfo {author} {\bibfnamefont {P.~R.}\ \bibnamefont
  {Surj{\'a}n}},\ }\href {\doibase https://doi.org/10.1002/qua.21723}
  {\bibfield  {journal} {\bibinfo  {journal} {Int. J. Quantum Chem.}\ }\textbf
  {\bibinfo {volume} {108}},\ \bibinfo {pages} {2043} (\bibinfo {year}
  {2008})}\BibitemShut {NoStop}%
\bibitem [{\citenamefont {Surj{\'a}n}\ and\ \citenamefont
  {Szabados}(2010)}]{Surjan_2010}%
  \BibitemOpen
  \bibfield  {author} {\bibinfo {author} {\bibfnamefont {P.~R.}\ \bibnamefont
  {Surj{\'a}n}}\ and\ \bibinfo {author} {\bibfnamefont {{\'A}.}~\bibnamefont
  {Szabados}},\ }in\ \href {\doibase 10.1007/978-90-481-2885-3_19} {\emph
  {\bibinfo {booktitle} {Recent {{Progress}} in {{Coupled Cluster Methods}}:
  {{Theory}} and {{Applications}}}}},\ \bibinfo {series and number} {Challenges
  and {{Advances}} in {{Computational Chemistry}} and {{Physics}}},\ \bibinfo
  {editor} {edited by\ \bibinfo {editor} {\bibfnamefont {P.}~\bibnamefont
  {C{\'a}rsky}}, \bibinfo {editor} {\bibfnamefont {J.}~\bibnamefont {Paldus}},
  \ and\ \bibinfo {editor} {\bibfnamefont {J.}~\bibnamefont {Pittner}}}\
  (\bibinfo  {publisher} {{Springer Netherlands}},\ \bibinfo {address}
  {{Dordrecht}},\ \bibinfo {year} {2010})\ pp.\ \bibinfo {pages}
  {513--534}\BibitemShut {NoStop}%
\bibitem [{\citenamefont {Burton}\ and\ \citenamefont
  {Wales}(2021)}]{Burton_2021}%
  \BibitemOpen
  \bibfield  {author} {\bibinfo {author} {\bibfnamefont {H.~G.~A.}\
  \bibnamefont {Burton}}\ and\ \bibinfo {author} {\bibfnamefont {D.~J.}\
  \bibnamefont {Wales}},\ }\href {\doibase 10.1021/acs.jctc.0c00772} {\bibfield
   {journal} {\bibinfo  {journal} {J. Chem. Theory Comput.}\ }\textbf {\bibinfo
  {volume} {17}},\ \bibinfo {pages} {151} (\bibinfo {year} {2021})}\BibitemShut
  {NoStop}%
\bibitem [{\citenamefont {Barca}\ and\ \citenamefont
  {Loos}(2018)}]{Barca_2018}%
  \BibitemOpen
  \bibfield  {author} {\bibinfo {author} {\bibfnamefont {G.~M.}\ \bibnamefont
  {Barca}}\ and\ \bibinfo {author} {\bibfnamefont {P.-F.}\ \bibnamefont
  {Loos}},\ }in\ \href {\doibase 10.1016/bs.aiq.2017.03.004} {\emph {\bibinfo
  {booktitle} {Adv. Quantum Chem.}}},\ Vol.~\bibinfo {volume} {76}\ (\bibinfo
  {publisher} {{Elsevier}},\ \bibinfo {year} {2018})\ pp.\ \bibinfo {pages}
  {147--165}\BibitemShut {NoStop}%
\bibitem [{\citenamefont {Seeger}\ and\ \citenamefont
  {Pople}(1977)}]{Seeger_1977}%
  \BibitemOpen
  \bibfield  {author} {\bibinfo {author} {\bibfnamefont {R.}~\bibnamefont
  {Seeger}}\ and\ \bibinfo {author} {\bibfnamefont {J.~A.}\ \bibnamefont
  {Pople}},\ }\href {\doibase 10.1063/1.434318} {\bibfield  {journal} {\bibinfo
   {journal} {J. Chem. Phys.}\ }\textbf {\bibinfo {volume} {66}},\ \bibinfo
  {pages} {3045} (\bibinfo {year} {1977})}\BibitemShut {NoStop}%
\bibitem [{\citenamefont {Fukutome}(1981)}]{Fukutome_1981}%
  \BibitemOpen
  \bibfield  {author} {\bibinfo {author} {\bibfnamefont {H.}~\bibnamefont
  {Fukutome}},\ }\href {\doibase 10.1002/qua.560200502} {\bibfield  {journal}
  {\bibinfo  {journal} {Int. J. Quantum Chem.}\ ,\ \bibinfo {pages} {955}}
  (\bibinfo {year} {1981})}\BibitemShut {NoStop}%
\bibitem [{\citenamefont {Stuber}\ and\ \citenamefont
  {Paldus}(2003)}]{Stuber_2003}%
  \BibitemOpen
  \bibfield  {author} {\bibinfo {author} {\bibfnamefont {J.}~\bibnamefont
  {Stuber}}\ and\ \bibinfo {author} {\bibfnamefont {J.}~\bibnamefont
  {Paldus}},\ }\enquote {\bibinfo {title} {{Symmetry Breaking in the
  Independent Particle Model}},}\ in\ \href@noop {} {\emph {\bibinfo
  {booktitle} {Fundamental World of Quantum Chemistry: A Tribute to the Memory
  of Per-Olov L\"{o}wdin}}},\ Vol.~\bibinfo {volume} {1},\ \bibinfo {editor}
  {edited by\ \bibinfo {editor} {\bibfnamefont {E.~J.}\ \bibnamefont
  {Br\"{a}ndas}}\ and\ \bibinfo {editor} {\bibfnamefont {E.~S.}\ \bibnamefont
  {Kryachko}}}\ (\bibinfo  {publisher} {Kluwer Academic},\ \bibinfo {address}
  {Dordrecht},\ \bibinfo {year} {2003})\ p.~\bibinfo {pages} {67}\BibitemShut
  {NoStop}%
\bibitem [{\citenamefont {Burton}\ and\ \citenamefont
  {Thom}(2016)}]{Burton_2016}%
  \BibitemOpen
  \bibfield  {author} {\bibinfo {author} {\bibfnamefont {H.~G.~A.}\
  \bibnamefont {Burton}}\ and\ \bibinfo {author} {\bibfnamefont {A.~J.~W.}\
  \bibnamefont {Thom}},\ }\href {\doibase 10.1021/acs.jctc.5b01005} {\bibfield
  {journal} {\bibinfo  {journal} {J. Chem. Theory Comput.}\ }\textbf {\bibinfo
  {volume} {12}},\ \bibinfo {pages} {167} (\bibinfo {year} {2016})}\BibitemShut
  {NoStop}%
\bibitem [{\citenamefont {Qiu}, \citenamefont {Henderson},\ and\ \citenamefont
  {Scuseria}(2017)}]{Qiu_2017}%
  \BibitemOpen
  \bibfield  {author} {\bibinfo {author} {\bibfnamefont {Y.}~\bibnamefont
  {Qiu}}, \bibinfo {author} {\bibfnamefont {T.~M.}\ \bibnamefont {Henderson}},
  \ and\ \bibinfo {author} {\bibfnamefont {G.~E.}\ \bibnamefont {Scuseria}},\
  }\href {\doibase 10.1063/1.4983065} {\bibfield  {journal} {\bibinfo
  {journal} {J. Chem. Phys.}\ }\textbf {\bibinfo {volume} {146}},\ \bibinfo
  {pages} {184105} (\bibinfo {year} {2017})}\BibitemShut {NoStop}%
\bibitem [{\citenamefont {L{\"o}wdin}(1963)}]{Lowdin_1963}%
  \BibitemOpen
  \bibfield  {author} {\bibinfo {author} {\bibfnamefont {P.-O.}\ \bibnamefont
  {L{\"o}wdin}},\ }\href {\doibase 10.1103/RevModPhys.35.415} {\bibfield
  {journal} {\bibinfo  {journal} {Rev. Mod. Phys.}\ }\textbf {\bibinfo {volume}
  {35}},\ \bibinfo {pages} {496} (\bibinfo {year} {1963})}\BibitemShut
  {NoStop}%
\bibitem [{\citenamefont {Lykos}\ and\ \citenamefont
  {Pratt}(1963)}]{Lykos_1963}%
  \BibitemOpen
  \bibfield  {author} {\bibinfo {author} {\bibfnamefont {P.}~\bibnamefont
  {Lykos}}\ and\ \bibinfo {author} {\bibfnamefont {G.~W.}\ \bibnamefont
  {Pratt}},\ }\href {\doibase 10.1103/RevModPhys.35.496} {\bibfield  {journal}
  {\bibinfo  {journal} {Rev. Mod. Phys.}\ }\textbf {\bibinfo {volume} {35}},\
  \bibinfo {pages} {496} (\bibinfo {year} {1963})}\BibitemShut {NoStop}%
\bibitem [{\citenamefont {L{\"o}wdin}(1969)}]{Lowdin_1969}%
  \BibitemOpen
  \bibfield  {author} {\bibinfo {author} {\bibfnamefont {P.-O.}\ \bibnamefont
  {L{\"o}wdin}},\ }\enquote {\bibinfo {title} {Some aspects on the correlation
  problem and possible extensions of the independent-particle model},}\ in\
  \href {\doibase https://doi.org/10.1002/9780470143599.ch9} {\emph {\bibinfo
  {booktitle} {Adv. Chem. Phys.}}}\ (\bibinfo  {publisher} {John Wiley \& Sons,
  Ltd},\ \bibinfo {year} {1969})\ pp.\ \bibinfo {pages} {283--340}\BibitemShut
  {NoStop}%
\bibitem [{\citenamefont {Sundstrom}\ and\ \citenamefont
  {{Head-Gordon}}(2014)}]{Sundstrom_2014}%
  \BibitemOpen
  \bibfield  {author} {\bibinfo {author} {\bibfnamefont {E.~J.}\ \bibnamefont
  {Sundstrom}}\ and\ \bibinfo {author} {\bibfnamefont {M.}~\bibnamefont
  {{Head-Gordon}}},\ }\href {\doibase 10.1063/1.4868120} {\bibfield  {journal}
  {\bibinfo  {journal} {J. Chem. Phys.}\ }\textbf {\bibinfo {volume} {140}},\
  \bibinfo {pages} {114103} (\bibinfo {year} {2014})}\BibitemShut {NoStop}%
\bibitem [{\citenamefont {Watson}\ and\ \citenamefont
  {Chan}(2012)}]{Watson_2012}%
  \BibitemOpen
  \bibfield  {author} {\bibinfo {author} {\bibfnamefont {M.~A.}\ \bibnamefont
  {Watson}}\ and\ \bibinfo {author} {\bibfnamefont {G.~K.-L.}\ \bibnamefont
  {Chan}},\ }\href {\doibase 10.1021/ct300591z} {\bibfield  {journal} {\bibinfo
   {journal} {J. Chem. Theory Comput.}\ }\textbf {\bibinfo {volume} {8}},\
  \bibinfo {pages} {4013} (\bibinfo {year} {2012})}\BibitemShut {NoStop}%
\bibitem [{\citenamefont {V{\'e}ril}\ \emph {et~al.}(2021)\citenamefont
  {V{\'e}ril}, \citenamefont {Scemama}, \citenamefont {Caffarel}, \citenamefont
  {Lipparini}, \citenamefont {Boggio-Pasqua}, \citenamefont {Jacquemin},\ and\
  \citenamefont {Loos}}]{Veril_2021}%
  \BibitemOpen
  \bibfield  {author} {\bibinfo {author} {\bibfnamefont {M.}~\bibnamefont
  {V{\'e}ril}}, \bibinfo {author} {\bibfnamefont {A.}~\bibnamefont {Scemama}},
  \bibinfo {author} {\bibfnamefont {M.}~\bibnamefont {Caffarel}}, \bibinfo
  {author} {\bibfnamefont {F.}~\bibnamefont {Lipparini}}, \bibinfo {author}
  {\bibfnamefont {M.}~\bibnamefont {Boggio-Pasqua}}, \bibinfo {author}
  {\bibfnamefont {D.}~\bibnamefont {Jacquemin}}, \ and\ \bibinfo {author}
  {\bibfnamefont {P.-F.}\ \bibnamefont {Loos}},\ }\href {\doibase
  10.1002/wcms.1517} {\bibfield  {journal} {\bibinfo  {journal} {WIREs Comput.
  Mol. Sci.}\ ,\ \bibinfo {pages} {e1517}} (\bibinfo {year}
  {2021})}\BibitemShut {NoStop}%
\bibitem [{\citenamefont {Csirik}\ and\ \citenamefont
  {Laestadius}(2021)}]{Csirik_2021}%
  \BibitemOpen
  \bibfield  {author} {\bibinfo {author} {\bibfnamefont {M.~A.}\ \bibnamefont
  {Csirik}}\ and\ \bibinfo {author} {\bibfnamefont {A.}~\bibnamefont
  {Laestadius}},\ }\href@noop {} {\enquote {\bibinfo {title} {Coupled-cluster
  theory revisited},}\ } (\bibinfo {year} {2021}),\ \Eprint
  {http://arxiv.org/abs/2105.13134} {arXiv:2105.13134 [math.NA]} \BibitemShut
  {NoStop}%
\end{thebibliography}%
%%%%%%%%%%%%%%%%%%%%%%%%%%%%%%%%

\end{document}